\documentclass[12pt, a4paper]{article}
\usepackage{amsmath,amssymb}
\usepackage{cancel}
\usepackage{color}
\usepackage{bm}
\usepackage{braket}
\usepackage{slashed}
\usepackage{tikz-feynhand}
\usepackage{cite}

\newcommand{\im}{\textrm{Im}\,}

\makeatletter

\@addtoreset{equation}{section}
\makeatother

\begin{document}

\title{%\vbox{
\baselineskip 14pt
%\hfill \hbox{\normalsize KUNS-???}\\
\vskip 1.7cm
\textbf{
CP-like Symmetry \\
with Discrete and Continuous Groups\\
and CP Violation/Restoration}
\vskip 0.5cm
%}
}
\author{
\centerline{Hiroshi~Ohki$^{1}$, \
Shohei~Uemura$^{2}$\thanks{s\_uemura@cc.nara-wu.ac.jp}, \
}
\\*[20pt]
\\
\textit{
\normalsize
\centerline{${}^{1}$
Department of Physics, Nara Women's University, Nara 630-8506, Japan}}
\\
\textit{\normalsize
\centerline{${}^{2}$ Research organization of Integrative STEAM Education,}}\\
\textit{\normalsize
\centerline{Nara Women's University, Nara 630-8506, Japan}}
\\*[50pt]}

\date{
%\today
%\\
\centerline{\small \textbf{Abstract}}
\begin{minipage}{0.9\linewidth}
\medskip
\medskip
\small
We study physical implications of general CP symmetry including CP-like symmetry.
Various scattering amplitudes of CP asymmetry are calculated in CP-like symmetric models.
We explicitly show that the CP-like transformation leads to a specific relation between different CP asymmetries.
The resultant relation is similar to the one obtained in GUT baryogenesis and sphaleron processes,
where we also obtain a required condition for generating particle number asymmetry in CP-like symmetric models.
In addition, we propose a generalization of a CP-like transformation for continuous symmetry groups.
Since the CP transformation is an outer automorphism, which depends on the internal symmetry group,
it turns out that the physical CP and CP-like symmetries can be mutually converted
through the spontaneous symmetry breaking (SSB) of the internal symmetry.
We investigate properties of physical CP asymmetry in both CP and CP-like symmetric phases,
and find that the spontaneous CP violation and restoration
can be observed even in models with continuous groups.
We demonstrate that CP-like symmetric models with continuous Lie groups can be naturally realized in physical CP symmetric models through the SSB.
\end{minipage}
}

\newpage

\begin{titlepage}
\maketitle
\thispagestyle{empty}
\clearpage
\end{titlepage}

\section{Introduction}

Violation of CP symmetry is a key property of the Standard Model (SM).
CP violation in the SM has been confirmed by several experiments,
and the observed CP asymmetries are consistently described
by the complex phases in the Yukawa couplings~\cite{ParticleDataGroup:2022pth}.
In addition, CP violation is a necessary condition for the matter/antimatter asymmetry of the universe \cite{Sakharov:1967dj}.
However, the CP violation of the SM is not sufficient to reproduce the baryon number density of the nature,
and hence CP violation should be related to new physics beyond the SM.
It is important to study the property of CP transformations
in such an underlying theory.

If an underlying theory has other (flavor) symmetry groups,
a generalized CP transformation is considered as an extension of the CP transformation in the SM
\cite{Neufeld:1987wa, Feruglio:2012cw, Holthausen:2012dk}.
There are vast phenomenological applications of generalized CP.
For instance, generalized CP is introduced as $\mu-\tau$ reflection symmetry 
\cite{Harrison:2002et, Grimus:2003yn}.\footnote{For review of $\mu-\tau$ symmetry see \cite{Xing:2015fdg} and references therein.}
It is also studied in flavor symmetric models with discrete groups such as $S_4$
\cite{Ding:2013hpa, Feruglio:2013hia, Li:2013jya, Li:2014eia, Girardi:2015rwa},
$A_4$ \cite{Girardi:2015rwa, Ahn:2013mva, Ding:2013bpa, Li:2016nap},
$A_5$ \cite{Girardi:2015rwa, Li:2015jxa, DiIura:2015kfa, Ballett:2015wia, Turner:2015uta, DiIura:2018fnk},
$T'$ \cite{Girardi:2015rwa, Girardi:2013sza},
$\Delta(27)$ \cite{Nishi:2013jqa},
$\Delta(48)$ \cite{Ding:2013nsa, Ding:2014hva},
$\Delta(96)$ \cite{Ding:2014ssa},
$\Delta(3n^2)$ \cite{Ding:2015rwa, Hagedorn:2014wha, Hagedorn:2021ldq},
$\Delta(6n^2)$ \cite{King:2014rwa, Hagedorn:2014wha, Ding:2014ora, Lu:2018oxc, Hagedorn:2021ldq},
and other discrete groups \cite{Li:2016ppt, Rong:2016cpk, Yao:2016zev, Chauhan:2022gkz}.\footnote{For review of non-Abelian discrete groups in particle physics,
see \cite{Altarelli:2010gt, Ishimori:2010au, King:2013eh, Branco:2011zb, Kobayashi:2022moq} and references therein.}
Recently generalized CP symmetries (and its spontaneous breaking) in modular symmetric models
have been investigated
\cite{Wang:2019ovr, Liu:2020akv, Novichkov:2020eep, deAnda:2018ecu, Wang:2019xbo, Okada:2020brs, Okada:2021qdf, Tanimoto:2021ehw, Mishra:2023cjc, Ding:2023ynd, Ding:2021iqp, Kobayashi:2019uyt, Kobayashi:2020uaj, Ishiguro:2020nuf, Ishiguro:2020tmo},
where the generalized CP transformation also acts on the moduli space   \cite{Novichkov:2019sqv, Baur:2019kwi}.
String compactification also gives rise to conventional flavor symmetries \cite{Abe:2009vi, Marchesano:2013ega, Berasaluce-Gonzalez:2012abm},
and hence, it provides a unified origin of generalized CP, modular, and discrete flavor symmetries
\cite{Baur:2019kwi, Baur:2019iai, Nilles:2020nnc, Nilles:2020kgo, Ohki:2020bpo, Ishiguro:2021ccl}.

From a theoretical viewpoint, a generalized CP transformation is characterized by its action on the internal symmetry group $G$.
Its action on $G$ is given by $u_{\textrm{CP}}:\, G \to CP^{-1}\circ G \circ CP$,
and $u_{\textrm{CP}}$ must be an (outer) automorphism if the model is invariant under $G$ and CP
\cite{Holthausen:2012dk, Grimus:1995zi}.
This is a consistency condition for generalized CP transformation.
As for a continuous symmetry
it is known that most of semi-simple Lie algebras\footnote{The only exception is $\mathfrak{so}(8)$, which exhibits the triality.}
have an outer automorphism of $\mathbb{Z}_2$ or trivial group,
which corresponds to the complex conjugation CP transformation,
and hence the physical CP is uniquely defined up to inner automorphism \cite{Grimus:1995zi}.

By contrast, outer automorphisms of a discrete group have a more variety, 
for example ${\rm Out}(\Delta(54)) \cong S_4$  \cite{Trautner:2016ezn}.
Among them, proper CP transformation is defined to be a transformation which maps all complex representations
to their own complex conjugate representations.
It is referred to as a class-inverting automorphism (CIA) \cite{Chen:2014tpa}, 
which transforms $g\in G \to u_{\textrm{CP}}(g) = h \sim g^{-1}$. 
On the other hand, a general CP transformation associated with non-CIA is called a CP-like transformation.
In fact, there are discrete groups without CIA, which are referred to as Type I groups in \cite{Chen:2014tpa}.
Thus any of outer automorphisms of Type I group should correspond to a CP-like transformation.
Since the CP-like transformation includes non-trivial interchanges of particles and other (anti)particles
which are not related by complex conjugations,
it has been shown that the physical CP symmetry is violated in models with a Type I symmetry group~\cite{Chen:2014tpa, Ratz:2016scn, Nilles:2018wex}.
Despite its interesting property, phenomenological aspects of the CP-like symmetric models with a general symmetry
group other than type I have not been investigated comprehensively so far.
In addition, the dynamical origins of the CP-like symmetric models and its relation to the proper CP symmetry
for general and continuous symmetry groups have not been carefully investigated in the previous works.

In this paper, we study physical implications of general CP transformations 
with continuous and discrete symmetry groups.
Throughout the paper we only consider the global symmetry $G$. 
To precisely discuss the CP asymmetry of a theory, 
we have to consider a physical CP transformation,
under which all particles are transformed into their own antiparticles. 
For applicability to type I groups, 
it is essential that both the physical CP transformation 
and the associated unitary matrix are defined independently of group automorphisms.
Thus in our study we first define field transformations for general CP transformations, 
and subsequently categorize them based on the symmetry properties of the theory, as well as group automorphisms.
As a result, 
general CP transformations will be classified into three different types: 
proper CP, CP-like, and inconsistent CP transformations. 
Among them we will pay particular attention to CP-like transformations. 
We will provide various constructions of CP-like symmetric models for a general symmetry which 
includes a direct product of groups and continuous groups, 
without resorting to the characteristic property of type I groups. 
We then explicitly show that in CP-like symmetric models there exists a characteristic relation between different CP violating amplitudes.
The resultant relation is similar to the $B-L$ in GUT baryogenesis \cite{Yoshimura:1978ex} and sphaleron process \cite{Kuzmin:1985mm},
so that the CP-like model can provide us with an alternative mechanism for generating matter asymmetry
as well as asymmetric dark matter \cite{Kaplan:2009ag}.
In addition, we find that there exists a CP-like eigenstate which does not have a CP-like partner,
and hence its number conservation is violated too.
We also study the effects of spontaneous symmetry breaking (SSB) of the internal symmetry on the CP(-like) symmetric models.
We show that the proper CP and CP-like symmetries
are mutually changeable through the SSB of the internal symmetry.\footnote{CP $\to$ CP-like symmetry has been already studied in \cite{Ratz:2016scn}.}
We will point out that the emergence of CP-like symmetry is not directly related to the absence of the CIA of the discrete group.
In fact, we demonstrate that various CP-like models with discrete and continuous Lie groups can be naturally realized
in proper CP symmetric models through the SSB.
We show CP-like symmetry appearance from $\Delta(54) \times U(1)$, $SU(3)$ and $SU(2)\times U(1)$ models.
We explicitly show how CP-violation occurs in the broken phase if physical CP becomes CP-like in some toy models.

This paper is organized as follows.
In Sec.~\ref{sec:review_CP},
we briefly review CP transformation and study general CP transformation from the viewpoint of internal symmetry.
We clarify the difference between proper CP symmetry and CP-like symmetry.
We then propose a generalization of CP-like transformation for continuous groups.
In Sec.~\ref{sec:CP-like}, we consider general CP symmetric model with internal symmetry.
We use a toy model with $\Delta(27)$ symmetry for illustrative purposes.
We calculate scattering amplitude in CP-like symmetric models, and study physical implications of the CP-like invariance.
Then we consider general CP transformation with continuous group and introduce CP-like symmetry.
In Sec.~\ref{sec:SSB}, we study CP symmetric model with SSB of internal symmetry.
We consider models with various internal symmetry including discrete and continuous groups.
We show CP-like symmetric model can be originated from physical CP symmetric model and vice versa.
Section~\ref{sec:Conclusion} is devoted to conclusion.
In App.~\ref{sec:A1}, we summarize group property of $\Delta(54)$.
In App.~\ref{app:CG}, we show part of Clebsch-Gordan (CG) coefficients of $\Delta(27)$ and $\Delta(54)$ relevant to our discussions.
In App.~\ref{App:1-loop}, we explicitly compute one-loop amplitude of a CP violating process.
In App.~\ref{sec:trace}, we study sufficient condition for general CP violation.

\section{CP Transformation for General Group}
\label{sec:review_CP}

We first review general CP transformations with internal symmetry group $G$.
Throughout the paper, we only consider global symmetry. 
In order to discuss CP asymmetry of physical quantities, 
it is important to establish the difference between the two physical amplitudes of particles and their antiparticles.
For this purpose, we introduce a physical CP transformation,
which can be defined independently of the internal group property. 
On the other hand, a general CP transformation is characterized by its actions on the internal symmetry, i.e., automorphisms of $G$.
We study a CP-like transformation for general symmetry groups including continuous and non-simple groups.
We will show that general CP transformations can be classified into three classes: proper CP, CP-like, and inconsistent CP transformations. 
Our notation is based on \cite{Chen:2014tpa, Peskin:1995ev} and references therein.

\subsection{General and Physical CP Transformations}

The CP transformation transforms a particle with momentum $\mathbf{p}$ into an antiparticle with inverse momentum.
Therefore the standard definition of the CP transformation for a complex scalar field $\phi(x)$ is given as 
\begin{align}
{\textrm{CP}}: \phi(x) \to \eta \phi^*(\tilde x)
\end{align}
where $\tilde x = (t, -\mathbf{x})^T$ and $\eta$ is a $U(1)$ phase.
CP transformations for fermions and gauge bosons should be modified
depending on its spin structures.
In this section, we concentrate on scalar fields for simplicity.

If there are multiple fields in a theory, CP transformation can be generalized.
In particular, with internal symmetries such as gauge and flavor symmetries,
the definitions of CP transformations as well as CP invariance are extended.
Let us consider a quantum field theory with $N$ fields denoted by $\phi_{i = 1,..., N}$.
If a theory has an internal symmetry associated with a group $G$,
$\boldsymbol{\phi} = (\phi_1, \phi_2, \cdots, \phi_N)^T$ forms $N$ dimensional (in general) reducible representation of $G$;
\begin{align}
  g \in G: \boldsymbol{\phi}(x) \to \rho_\phi(g) \boldsymbol{\phi}(x),
\end{align}
where $\rho_\phi (g)$ is a unitary representation of $g$.
The most general CP transformation for $\boldsymbol{\phi}$ is given by \cite{Holthausen:2012dk, Feruglio:2012cw, Branco:2011zb}
\begin{align}
  \textrm{CP}:&\boldsymbol{\phi}(x) \to  U \boldsymbol{\phi}^*(\tilde x),
  \label{eq:CPtrans}
  &\textrm{CP}&:\boldsymbol{\phi}^*(x) \to  U^* \boldsymbol{\phi}(\tilde x),
\end{align}
where $U$ is an $N\times N$ unitary matrix.
We note that the most general CP transformation in Eq.~\eqref{eq:CPtrans} may in some cases not lead to physical CP conservation.

Here we define a \textit{physical CP transformation}.
We denote $\Phi = (\phi_{\mathbf{r}_{1}}, \cdots, \phi_{\mathbf{r}_{N'}})$ as a direct sum of irreducible representations of $N$-multiplet field $\boldsymbol{\phi}$,
where $\mathbf{r}_{i} \ (i=1,\cdots N')$ is an irreducible representation of $G$.
The standard CP transformation is defined to 
transform particles to their antiparticles with inverse momenta.
Therefore, the physical CP transformation is defined to transform
a field $\phi_{\mathbf{r}_{i}}$ in irreducible representation $\mathbf{r}_{i}$ to its complex conjugate;
\begin{align}
\label{eq:CPphys}
  \textrm{physical CP}:\phi_{\mathbf{r}_i}(x) \to U_{\mathbf{r}_i}\phi_{\mathbf{r}_i}^*(\tilde x), \ \ (i=1,\cdots, N'),
\end{align}
where $U_{\mathbf{r}_i}$ is a unitary matrix associated 
with the physical CP transformation.\footnote{If there is $F$-fold multiplicity in representation $\mathbf{r}_i$,
$U_{\mathbf{r}_i}$ is given as a $(\dim \mathbf{r}_i \times F)^2$ unitary matrix.}
The general CP transformation is not restricted solely to transformations among the same irreducible representations,
so that the group associated with the general CP transformation in Eq.~\eqref{eq:CPtrans}
should include the one corresponding to the physical CP transformation in Eq.~\eqref{eq:CPphys}.
Note that in our terminology, the set of representations $\mathbf{r}_{1}, \cdots , \mathbf{r}_{N'}$ does not necessarily include all the irreducible representations of $G$, 
and hence the physical CP transformation given in Eq.~\eqref{eq:CPphys} can be defined independently of the group (outer) automorphism. 

\subsection{Proper CP, CP-like, and Inconsistent CP Transformations}\label{sec:CPtransformations}

If $CP$ and $G$ are symmetries of the theory, 
the consistency condition $CP^{-1}\circ g \circ CP \in G$ should hold~\cite{Holthausen:2012dk, Grimus:1995zi}.
Thus a \textit{consistent CP transformation} is referred to as a general CP transformation which satisfies
\begin{align}
  CP^{-1}\circ G \circ CP \subset G.
  \label{eq:cc_of_CP}
\end{align}
In this case, for any $g \in G$ we have $h \in G$ satisfying the following relation,
\begin{align}
  U_{ij} \rho(g)^*_{jk} U^{\dagger}_{kl} = \rho(h)_{il}.
  \label{eq:cc_of_CP_matrix}
\end{align}
This means that 
if there is an automorphism $u_{\textrm{CP}}: G \to G$ such that $u_{\textrm{CP}}(g) \equiv h$,
then one can obtain a consistent transformation based on it. 

On the other hand, an \textit{inconsistent CP transformation} is referred to as
a general CP transformation which does not satisfy the above relation, that is
\begin{align}
  CP^{-1}\circ G \circ CP \not\subset G.
\end{align}
It is also obvious that the inconsistent CP symmetry is not compatible with the symmetry group $G$.
It follows that either an inconsistent CP is not a symmetry, or $G$ is enlarged to $\tilde G$ such that $CP^{-1}\circ \tilde{G} \circ CP \subset \tilde{G}$.

Let us consider the consistent CP transformation in detail.
If there is a complex conjugation automorphism which maps all the irreducible complex representations of $G$ to its own conjugates, 
we can obtain a {\it proper CP} transformation, 
which acts on $\phi_{\mathbf{r}_i}$ as 
\begin{align}
\textrm{proper CP}: \phi_{\mathbf{r}_i}(x) \to U_{\mathbf{r}_i}\phi_{\mathbf{r}_i}^*(\tilde x), \quad \forall i.
\label{eq:properCP}
\end{align}
Such a consistent CP transformation can only be ensured 
if and only if the unitary matrix $U_{\mathbf{r}_i}$ satisfies the following relation
\begin{align}
U_{\mathbf{r}_i} \rho_{\mathbf{r}_i}(g)^* U^{\dagger}_{\mathbf{r}_i} = \rho_{\mathbf{r}_i}(h), \quad \forall i.
\label{eq:cc_class_inverting}
\end{align}
Thus all the irreducible representations satisfy Eqs.~\eqref{eq:properCP} and \eqref{eq:cc_class_inverting} 
for proper CP transformation.

It is important to reiterate that there exists a subtle but crucial difference between 
the physical and proper CP transformations in Eqs.~\eqref{eq:CPphys} and \eqref{eq:properCP}.
In our definition, the physical CP transformation acts only on fields existing in a theory, 
which does not imply that the consistency condition in Eq.~\eqref{eq:CPphys} 
is applicable to all irreducible representations of $G$.
By contrast, the proper CP transformation is based on an automorphism of $G$, 
thus Eq.~\eqref{eq:cc_class_inverting} applies to all irreducible representations~\cite{Chen:2014tpa}.
That is, the proper CP transformation should be a subset of the physical CP transformations.
This difference is important for precisely investigating a CP asymmetry of physical quantities 
particularly for models with a symmetry group that does not have any complex conjugation outer automorphisms.

On the other hand, a \textit{CP-like transformation} for general (and continuous) groups should be defined as a consistent but not proper CP transformation. 
Since the CP-like transformation is a consistent CP transformation, 
one can obtain a CP-like transformation only if there exists a non-complex conjugation automorphism,
i.e. there is at least one irreducible representation which is not mapped to its conjugate by the automorphism.
As a result, the CP-like transformation acts like a physical CP transformation on some fields, 
but differently on other fields. 
Therefore, if a model only has fields that are transformed into their complex conjugate fields, 
the CP-like transformation can be a physical CP transformation.  

We should note that the consistency condition in Eq.~\eqref{eq:cc_of_CP}  is a necessary condition for the CP (or CP-like) invariance of a theory,
but not a sufficient condition.
In order to discuss the symmetry properties of models,  
it is important to specify the unitary matrix $U$ accompanying a general CP transformation. 
In fact, the structure of unitary matrix influences the assignment of fundamental quantum numbers to particle states 
that are not determined by the internal symmetry group.
In addition, it is not straightforward to impose CP-like symmetry in a theory, 
particularly when dealing with general (non-simple) groups, such as $U(1)$ and a direct product of groups $G_1 \times G_2$.
We will present such non-trivial examples of CP-like models for continuous groups later in Sec.~\ref{sec:CP-like}.

Finally, we also emphasize that it is important to consider an inconsistent but physical CP transformation.
If $G$ does not have any complex conjugation automorphism,
the physical CP transformation should be an inconsistent CP transformation in general.

\subsection{Classification of General CP transformations}

Based on the previous discussions,
any of general CP transformations in Eq.~\eqref{eq:CPtrans} 
can be classified into one of the following three types: 
\begin{itemize}
  \item[1.] \textbf{Consistent CP transformation}: general CP transformation satisfying consistency condition in Eq.~\eqref{eq:cc_of_CP}.
  This class of CP transformations can be ensured if there is an automorphism of $G$ via $CP^{-1}\circ G \circ CP \subset G$.
  It is possible to impose invariance under this type of CP transformation and $G$ simultaneously.
  There are two types in this class. 
  \begin{itemize}
    \item[a.] \textbf{Proper CP}: physical CP transformation consistent with a complex conjugation automorphism of $G$.
Field transformations are specified with transformation unitary matrices in Eq.~\eqref{eq:properCP}.
In the case of a discrete group, it is consistent with a CIA.\footnote{The $\mathbb{Z}_2$ outer automorphism of $SU(N) \ (N \neq 2)$ is class inverting~\cite{Bischer:2022rvf}.} 
If a model is proper CP invariant, all the fields are transformed to their complex conjugate fields 
under the proper CP transformation.
\item[b.] \textbf{CP-like}: consistent but not proper CP transformation. 
    It includes a transformation of 
    a particle in representation $\mathbf{r}_i$ to an (anti)particle in other representation $\mathbf{r}_j^*$.
    If a model does not have such fields, it can be a physical CP. 
  \end{itemize}
  \item[2.] \textbf{Inconsistent CP}:
  general CP not satisfying the consistency condition Eq.~\eqref{eq:cc_of_CP}.
  We cannot impose invariance under inconsistent CP transformation, or $G$ is enlarged to $\tilde G$ which satisfies Eq.~\eqref{eq:cc_of_CP}.
  If there is no complex conjugation automorphism of $G$, physical CP transformation is inconsistent CP transformation in general.
  CP is not a symmetry in such a model.
\end{itemize}
We also summarize the classification in Tab.~\ref{tab:CP-class}.
It should be emphasized that we classify the generic CP transformations 
in terms of the symmetry property of a theory and the group automorphisms. 
Consequently, some of CP transformations may not be well-defined for a given $G$.
See \cite{Chen:2014tpa} for a classification of (discrete) groups in terms of possible CP transformations.

\begin{table}[thbp]
  \centering
  \begin{tabular}{| c | c c c|}
    \hline
    & consistency with $G$ & automorphism & physical/unphysical \\ \hline
    Proper CP & \begin{tabular}{c} consistent  \\ \end{tabular} & \begin{tabular}{c} complex conjugation \\ (class-inverting) \end{tabular} & \begin{tabular}{c} physical \\ \end{tabular}
    \\ \hline
    CP-like  & consistent  & \begin{tabular}{c} non-complex conjugation \\ (non class-inverting) \end{tabular} & \begin{tabular}{c} physical/unphysical \\ \end{tabular}
    \\ \hline
    inconsistent CP  & inconsistent & no & physical/unphysical \\
    \hline
  \end{tabular}
  \caption{Classification of general CP transformations.
  We also show their properties inside parenthesis () in the case of discrete groups.}
  \label{tab:CP-class}
\end{table}

\subsection{Examples of CP-like Transformation with Order Two Automorphism
}

As shown above, a CP-like symmetry is not directly related to the matter/antimatter symmetry in general.
We note, however, that a CP-like symmetry can be naturally emerge from 
SSB of $G$ in an underlying physical CP symmetric theory~\cite{Ratz:2016scn}. 
More importantly, as will be shown later,
a physical CP symmetry and a CP-like symmetry of order two 
can be mutually converted through SSB.
As a result, certain CP-like transformations could originate 
from a proper CP transformation based on the complex conjugation automorphism of $G$. 
Thus we will pay particular attention to CP-like transformations of order two, 
which we refer to as {\it CP$_{\it 2}$-like transformations}.
In this case the CP$_2$-like transformation of $N$-multiplet field $\Phi(x)$ is specified with 
\begin{align}
&  \textrm{CP$_2$-like} : \Phi(x) \to U_\textrm{CP$_2$-like} \Phi^*(\tilde x),
\label{eq:cc_of_sCPlike}
\end{align}
where the unitary matrix $U_{\textrm{CP$_2$-like}}$ includes anti-diagonal block elements as,
\begin{align}
U_{\textrm{CP$_2$-like}} =
\left(\begin{array}{cc|cc|cc}
U_{\mathbf{r}_1} & \mathbf{0} & \cdots &&& \\
\mathbf{0} & \ddots &&&& \\ \hline
\vdots && \mathbf{0} & U_{\mathbf{r}_{ij}} && \\
&& U_{\mathbf{r}_{ji}} & \mathbf{0} && \vdots  \\ \hline
&&&& \ddots & \mathbf{0} \\
&&& \cdots & \mathbf{0} & U_{\mathbf{r}_{N'}}
\end{array}\right).
\label{eq:U_CPlike}
\end{align}
In order to be consistent with Eq.~\eqref{eq:cc_of_CP_matrix}
the unitary matrices should satisfy
\begin{align}
U_{\mathbf{r}_{ij}} \rho_{\mathbf{r}_j}(g)^* U^\dagger_{\mathbf{r}_{ij}} = \rho_{\mathbf{r}_i}(h),
\quad \quad
U_{\mathbf{r}_{ji}} \rho_{\mathbf{r}_i}(g)^* U^\dagger_{\mathbf{r}_{ji}} = \rho_{\mathbf{r}_j}(h).
\end{align}
From the above conditions,
we see that $\mathbf{r}_i$ and $\mathbf{r}_j$ have the same dimension.
Therefore the CP$_2$-like transformation in Eq.~\eqref{eq:U_CPlike}
exchanges a particle in representation $\mathbf{r}_i$ to an (anti)particles in other representation $\mathbf{r}_j^*$.
Obviously other particles are properly transformed to their complex conjugate particles.\footnote{
If irreducible representations of all the fields contents are restricted to such properly transformed representations,
CP-like transformation can be interpreted as a physical CP.}
As will be demonstrated subsequently, there are significant phenomenological implications in CP$_2$-like symmetric models.
Below, we illustrate some examples where we construct CP$_2$-like symmetric models 
without resorting to the characteristic property of type I groups.
For the sake of simplicity, in the subsequent discussion, 
we will not consider $Z_{N \neq 2}$ CP-like transformations.\footnote{E.g., $S_3$ outer automorphism group of $SO(8)$.
For a study of non-CP ($Z_{N \geq 3}$) transformations, see~\cite{Chen:2014tpa, Henning:2021ctv, Bischer:2022rvf}.}
Therefore ``CP-like'' will generally refers to CP$_2$-like transformation unless explicitly specified otherwise.

One simple example is $G=\mathbb{Z}_3$ group ($a^3=e, a \in \mathbb{Z}_3$), which has three irreducible representations of $\mathbf{1}_{0,1,2}$.
Let us consider a model with a triplet scalar field $\Phi=(\phi_{\mathbf{1}_{0}}, \phi_{\mathbf{1}_{1}}, \phi_{\mathbf{1}_{2}})^T$,
where $\mathbf{1}_{0,1,2}$ are the irreducible representations of $\mathbb{Z}_3$.
A group action of $\mathbb{Z}_3$ on $\Phi$ is given as 
\begin{align}
\rho_\Phi(a)  =
  \begin{pmatrix}
    1 & 0 & 0\\
    0 & \omega & 0\\
    0 & 0 & \omega^2\\
  \end{pmatrix},
\end{align}
where $\omega = e^{\frac{2\pi i}3}$.
We find that the following CP-like transformation,
\begin{align}
\label{eq:Z3}
\textrm{CP-like} : \Phi(x) \to U_{\textrm{CP-like}} \Phi^*(\tilde x),
\quad \quad
  U_{\textrm{CP-like}}=&
  \begin{pmatrix}
    1 & 0 & 0\\
    0 & 0 & 1\\
    0 & 1 & 0\\
  \end{pmatrix}, \ \
  \end{align}
satisfies the consistency condition $U_{\textrm{CP-like}} \rho(a)^* U_{\textrm{CP-like}}^\dagger =  \rho(a)$.
The corresponding automorphism is given as an identity map; $u_{\textrm{CP-like}}^{\mathbb{Z}_3}(g)=g$ for any $g \in \mathbb{Z}_3$.
In this case, it should be noted that
the two exchanged fields $\phi_{\mathbf{1}_1}$ and $\phi_{\mathbf{1}_2}^*$ possess identical charges under $\mathbb{Z}_3$.
In the context of a CP-like symmetric model, this implies that these two fields cannot be distinguished
from a quantum theory perspective. 
Consequently, the particle (energy) eigenstate should also be either a {\it CP-like even} or {\it odd} particle, 
resulting from a linear combination of these two fields,  
in analogous to the standard CP even or odd (neutral) particle with intrinsic parity $\pm 1$. 
Unlike the standard CP transformation, we find that the CP-like eigenstate can be charged under the internal symmetry.
This characteristic (CP-like eigenstate with $\pm 1$ eigenvalue) is a distinctive feature of CP-like transformations with an order of two.
We will carefully study their physical amplitudes based on the energy eigenstates 
in this specific context in Sec.~\ref{sec:CP-like}.

The situation changes if we consider a direct product group of $\mathbb{Z}_3 \times U(1)$.
Let us consider a model with a triplet field $\Phi=(\phi_{\mathbf{1}_{0}}^q, \phi_{\mathbf{1}_{1}}^q, \phi_{\mathbf{1}_{2}}^q)^T$,
where $q \neq 0 $ is a $U(1)$ charge.
A group action of $\mathbb{Z}_3 \times U(1)$ is given as
\begin{align}
\rho_{\Phi}(a,\theta) =
\begin{pmatrix}
e^{iq\theta} & 0 & 0\\
0 & \omega e^{iq\theta} & 0\\
0 & 0 & \omega^2 e^{iq\theta} \\
  \end{pmatrix},
\end{align}
where $\theta $ is a parameter of $U(1)$.
Obviously the same CP-like transformation as in Eq.~\eqref{eq:Z3}
satisfies the consistency condition.
Thus the corresponding automorphism $u_{\textrm{CP-like}}$ is given by a direct product of two specific automorphisms as
\begin{align}
&u_{\textrm{CP-like}}^{\mathbb{Z}_3\times U(1)} = u^{\mathbb{Z}_3}_\textrm{CP-like} \otimes u^{U(1)}_{CP}, \\
&u^{\mathbb{Z}_3}_{\textrm{CP-like}}: a \mapsto a, \quad \quad
u^{U(1)}_{\textrm{CP}}: \theta \mapsto -\theta.
\label{eq:U1CP}
\end{align}
On the contrary to the previous case,
both the two exchanged fields of $\phi_{\mathbf{1}_1}$ and $\phi_{\mathbf{1}_2}^*$
have the same $\mathbb{Z}_3$ charge, but opposite $U(1)$ charges.
Therefore these two particles should be distinguished under the CP-like transformation.

We should stress here that a CP-like transformation $(a, \theta) \mapsto (a, -\theta)$
is only possible for models with specific field content. 
Since the existence of two related fields of $\phi_{\mathbf{1}_1}^{q}$ and $\phi_{\mathbf{1}_1}^{-q}$ 
is not automatically ensured, 
while for the physical CP transformation 
a field $\phi_{\mathbf{1}_1}^{q}$ and their conjugate field $\phi_{\mathbf{1}_2}^{-q}$ 
should be ensured by the reality of the relativistic quantum field theory.
Thus the CP-like symmetric model generally requires the existence 
of the CP-like partner in addition to their complex conjugate field (anti-particle). 

\section{General CP Symmetries and CP Asymmetry}\label{sec:CP-like}

We study the general CP symmetric models with internal symmetry.
In the following, we first show constructions of various models with general CP symmetries.
We use a discrete symmetry group for illustration purposes,
where a special attention is paid to the CP-like symmetry.
To explore the impact of the CP-like symmetry in comparison with the proper CP symmetry,
we study physical scattering amplitudes of the CP asymmetry.
Although the physical CP symmetry is in general violated in CP-like symmetric models,
it is important to notice that different CP violating amplitudes can be related to each other
through the CP-like transformations.
We will find that such correspondences can be classified by representations of initial and final states in the scattering amplitudes.
We then discuss a possibility of particle number generations through the CP violating processes
along with these correspondences.
Subsequently we consider extending CP-like transformation to general continuous symmetry groups.
An outer automorphism of order two is used to define a CP-like transformation.
We then propose to construct a new class of CP-like symmetric model with continuous symmetry groups
which include $U(1)$ and general non-Abelian Lie groups.

\subsection{General CP Transformation in $\Delta(27)$ Model}\label{sec:Delta27}

We study an explicit model with $\Delta(27)$ symmetry for illustration purposes.
$\Delta(27)$ is isomorphic to $(\mathbb{Z}_3 \times \mathbb{Z}_3) \rtimes \mathbb{Z}_3$, and
all elements of $\Delta(27)$ is represented by $a^i a'^j b^k$, where $i,j,k = 0,1,2$,
where $a,a',$ and $b$ satisfy $b a' = a b, ba = a^2 a'^2 b, a a' = a' a$ and $a^3 = a'^3 = b^3 =e$.
There are 11 conjugacy classes in $\Delta(27)$:
\begin{align} \notag
C_{1a}:& \{ e \}, & C_{3a}:& \{ a, a', a^2 a'^2\},
  \\ \notag
C_{3b}:& \{ a^2, a'^2, a a' \}, & C_{3c}:& \{ b, a a'^2 b, a^2a'b \},
  \\ \notag
C_{3d}:& \{ b^2, a^2 a' b^2, a a'^2 b^2 \}, & C_{3e}:& \{ aa'b, a^2b, a'^2 b\},
  \\ \notag
C_{3f}:& \{ a'b^2, a^2 a'^2 b^2, ab^2 \}, & C_{3g}:& \{ ab, a' b, a^2a'^2 b\},
  \\ \notag
C_{3h}:& \{ a'^2 b^2, a^2 b^2, a a' b^2 \}, & C_{3i}:& \{ a a'^2 \},
  \\
C_{3j}:& \{ a^2 a' \},
\end{align}
and hence there are nine singlets $\mathbf{1}_i$ and two triplets, $\mathbf{3}$ and $\mathbf{3}^*$.
$\mathbf{3}^*$ is the complex conjugate representation of $\mathbf{3}$.
The complex conjugate representations of the singlets are also given by
\begin{align}
  \mathbf{1}^*_1 &= \mathbf{1}_2,
  &\mathbf{1}^*_3 &= \mathbf{1}_6,
  &\mathbf{1}^*_4 &= \mathbf{1}_8,
  &\mathbf{1}^*_5 &= \mathbf{1}_7,
\end{align}
and vice versa.
$\mathbf{1}_0$ is the trivial singlet.
Character table of $\Delta(27)$ is summarized in Tab.~\ref{tab:character_D27}.
As is known, the group $\Delta(27)$ has no CIA~\cite{Chen:2014tpa}.
Therefore there is no proper CP transformation in models with $\Delta(27)$ symmetry.
\begin{table}[thbp]
	\centering
	\begin{tabular}{lrrrrrrrrrrr}
		& $C_{1a}$ &$C_{3a}$ & $C_{3b}$ & $C_{3c}$ & $C_{3d}$ & $C_{3e}$ & $C_{3f}$ & $C_{3g}$ & $C_{3h}$ & $C_{3i}$ & $C_{3j}$ \\
		& $1$ &$3$ & $3$ & $3$ & $3$ & $3$ & $3$ & $3$ & $3$ & $1$ & $1$ \\
	 $\Delta(27)$ & $e$ &$a$ & $a^2$ & $b$ & $b^2$ & $a^2b$ & $ab^2$ & $ab$ & $a^2 b^2$ & $aa'^2$ & $a^2a'$ \\
	 \hline
	 $\mathbf{1}_0$ & $1$ &$1$ & $1$ & $1$ & $1$ & $1$ & $1$ & $1$ & $1$ & $1$ & $1$ \\
	 $\mathbf{1}_1$ & $1$ &$1$ & $1$ & $\omega^2$ & $\omega$ & $\omega^2$ & $\omega$ & $\omega^2$ & $\omega$ & $1$ & $1$ \\
	 $\mathbf{1}_2$ & $1$ &$1$ & $1$ & $\omega$ & $\omega^2$ & $\omega$ & $\omega^2$ & $\omega$ & $\omega^2$ & $1$ & $1$ \\
	 $\mathbf{1}_3$ & $1$ &$\omega^2$ & $\omega$ & $1$ & $1$ & $\omega$ & $\omega^2$ & $\omega^2$ & $\omega$ & $1$ & $1$ \\
	 $\mathbf{1}_4$ & $1$ &$\omega^2$ & $\omega$ & $\omega^2$ & $\omega$ & $1$ & $1$ & $\omega$ & $\omega^2$ & $1$ & $1$ \\
	 $\mathbf{1}_5$ & $1$ &$\omega^2$ & $\omega$ & $\omega$ & $\omega^2$ & $\omega^2$ & $\omega$ & $1$ & $1$ & $1$ & $1$ \\
	 $\mathbf{1}_6$ & $1$ &$\omega$ & $\omega^2$ & $1$ & $1$ & $\omega^2$ & $\omega$ & $\omega$ & $\omega^2$ & $1$ & $1$ \\
	 $\mathbf{1}_7$ & $1$ &$\omega$ & $\omega^2$ & $\omega^2$ & $\omega$ & $\omega$ & $\omega^2$ & $1$ & $1$ & $1$ & $1$ \\
	 $\mathbf{1}_8$ & $1$ &$\omega$ & $\omega^2$ & $\omega$ & $\omega^2$ & $1$ & $1$ & $\omega^2$ & $\omega$ & $1$ & $1$ \\
	 $\mathbf{3}$ & $3$ &$0$ & $0$ & $0$ & $0$ & $0$ & $0$ & $0$ & $0$ & $3\omega$ & $3\omega^2$ \\
	 $\mathbf{3}^*$ & $3$ &$0$ & $0$ & $0$ & $0$ & $0$ & $0$ & $0$ & $0$ & $3\omega^2$ & $3\omega$ \\
	 \hline
	\end{tabular}
	\caption{Character table of $\Delta(27)$.}
	\label{tab:character_D27}
\end{table}

\subsubsection{CP-like transformation for $\Delta(27)$}

There are several CP-like transformations for $\Delta(27)$ model.
One example of CP-like transformations we examine here is
\begin{align}
  \mathbf{3} &\to U_s \mathbf{3}^*,
  \label{eq:CPL_D27_3}
\end{align}
where $U_s$ is given by {\cite{Trautner:2016ezn}}
\begin{align}
  U_s &=
  \begin{pmatrix}
    \omega^2 & 0 & 0\\
    0 & 0 & 1\\
    0 & 1 & 0
  \end{pmatrix}.
\end{align}
Since $\mathbf{3}$ is a faithful representation of $\Delta(27)$,
the outer automorphism introducing this CP-like transformation is specified without ambiguity.
It is given by
\begin{align}
  u:(a, a', b) \mapsto (a^2 a'^2, a', a^2b^2),
\end{align}
and this automorphism transforms the conjugacy classes as
\begin{align}
  C_{3c} &\leftrightarrow C_{3h},
  &C_{3d} &\leftrightarrow C_{3g},
  &C_{3e} &\leftrightarrow C_{3f},
  &C_{3i} &\leftrightarrow C_{3j}.
  \label{eq:CP_d27}
\end{align}
$C_{3a}$ and $C_{3b}$ are fixed under the map $u$.
To satisfy this relation, the CP-like transformation for the singlets are determined
as follows,
\begin{align}
  \mathbf{1}_0 &\to \mathbf{1}_0^*,
  &\mathbf{1}_1 &\to \mathbf{1}_1^*,
  &\mathbf{1}_2 &\to \mathbf{1}_2^*,
  &\mathbf{1}_3 &\to \mathbf{1}_7^*,
  &\mathbf{1}_4 &\to \mathbf{1}_8^*,
  \notag
  \\
  \mathbf{1}_5 &\to \mathbf{1}_6^*,
  &\mathbf{1}_6 &\to \mathbf{1}_5^*,
  &\mathbf{1}_7 &\to \mathbf{1}_3^*,
  &\mathbf{1}_8 &\to \mathbf{1}_4^*.
  \label{eq:CPL_D27_1}
\end{align}
This CP-like transformation is represented by a unitary matrix $A$,
\begin{align}
  \mathbf{1}_i &\to A_{ij} \mathbf{1}_j^*,~~~~~\mathrm{with}~~~~~
  A = \begin{pmatrix}
    1 & 0 & 0 & 0 & 0 & 0 & 0 & 0 & 0\\
    0 & 1 & 0 & 0 & 0 & 0 & 0 & 0 & 0\\
    0 & 0 & 1 & 0 & 0 & 0 & 0 & 0 & 0\\
    0 & 0 & 0 & 0 & 0 & 0 & 0 & 1 & 0\\
    0 & 0 & 0 & 0 & 0 & 0 & 0 & 0 & 1\\
    0 & 0 & 0 & 0 & 0 & 0 & 1 & 0 & 0\\
    0 & 0 & 0 & 0 & 0 & 1 & 0 & 0 & 0\\
    0 & 0 & 0 & 1 & 0 & 0 & 0 & 0 & 0\\
    0 & 0 & 0 & 0 & 1 & 0 & 0 & 0 & 0\\
  \end{pmatrix}.
\end{align}
As shown above, this transformation does not transform all singlets to their own conjugate.
While this is not a physical CP transformation,
$\mathbf{1}_{0,1,2}$ are properly transformed to their complex conjugates under the CP-like transformation.
Thus this CP-like transformation can be identified as a physical CP transformation for a model without $\mathbf{1}_{3,..., 8}$.
Generally speaking, if a model has only triplets (or up to a couple of non-trivial singlets), 
there exists a certain CP-like transformation
which can be regarded as a physical CP transformation \cite{Hagedorn:2014wha}.

\subsubsection{Inconsistent CP transformation for $\Delta(27)$}\label{sec:inconsistentCP}

As explained above, if a model has a sufficient number of irreducible representations,
the CP-like transformation ceases to be a physical CP transformation.
Thus the following physical CP transformation,
\begin{align}
\textrm{CP}_{\textrm{inc}}:&
\begin{cases}
\mathbf{1}_i \to \mathbf{1}_i^*, \ \ \ (i=0, \cdots, 8) \\
\mathbf{3} \to U_s \mathbf{3}^*,
\end{cases}
  \label{eq:CP_inc}
\end{align}
should be an inconsistent CP transformation.
This is because there is no element $h \in G$ satisfying $U_s \rho^*_{\mathbf{3}}(g) U_s^\dagger = \rho_{\mathbf{3}}(h)$,
and $\rho^*_{{\mathbf 1}_i}(g) = \rho_{{\mathbf 1}_i} (h)$ for any $g$ and $i$.
Thus we cannot realize physical CP invariance in models with the internal symmetry of $\Delta(27)$.

To see differences of the CP-like and the inconsistent physical CP transformations,
we consider a simple model.
We introduce nine singlet scalars $\phi_{i}, \ (i = 0,...,8)$ and one triplet Dirac fermion $\Psi = (\psi_1, \psi_2, \psi_3)^T$,
where $\phi_i$ and $\Psi$ are in representations of $\mathbf{1}_i$, and $\mathbf{3}$ under $\Delta(27)$, respectively.
The $\Delta(27)$-invariant Yukawa interactions are given as
\begin{align}
  \mathcal{L}_{yukawa}
  =& \sum_{\substack{i = 0 , ..., 8}} y_i
  [\phi_i \otimes (\bar \Psi \otimes \Psi)_{\mathbf{1}^*_i}] + (h.c.)
  \notag
  \\
  =&
  \sum_{\substack{i = 0 , ..., 8\\ j,k = 1,2,3}} y_i (M_{i^*})_{jk}
  \phi_i \bar \psi_j \psi_k +
  \sum_{\substack{i = 0 , ..., 8\\ j,k = 1,2,3}} y_i^* (M_{i^*}^\dagger)_{jk}
  \phi_i^\dagger \bar \psi_j \psi_k
  \label{eq:Yukawa_D27}
\end{align}
where $y_i$ is a complex constant and $(M_{i^*})_{jk}$ is the CG coefficients of $\Delta(27)$, which are summarized in App.~\ref{app:CG_D27}.
Imposing the CP-like invariance in the model,
the coupling constants should satisfy the following conditions,
\begin{align}
  y_0 &= y_0^*, &y_1& = y_1^*, &y_2& = y_2^*,\notag\\
  \omega y_7 &= y_3^*, & \omega y_8 & = y_4^*, & \omega y_6 & = y_5^*,
  \label{eq:D27_1}
  \\
  \omega y_5 &= y_6^*, & \omega y_3 & = y_7^*, & \omega y_4 & = y_8^*.\notag
\end{align}

On the other hand, imposing inconsistent physical CP invariance given in Eq.~\eqref{eq:CP_inc},
we obtain vanishing Yukawa couplings.
This is because some Yukawa terms are transformed to prohibited Yukawa terms under the inconsistent CP transformation.
For example, $\phi_3 \otimes (\bar \Psi \otimes \Psi)_{\mathbf{1}^*_3} = \phi_3 \otimes (\bar \Psi \otimes \Psi)_{{\mathbf{1}}_6}$ is transformed to $\phi_3^\dagger \otimes (\omega^2 \bar{\Psi}_j \otimes \Psi_k)_{{\mathbf{1}}_7}$
under the inconsistent CP transformation.
This term is not $\Delta(27)$ invariant.
Thus allowed Yukawa couplings are restricted to real values as follows,
\begin{align}
  y_0 &= y_0^*, &y_1& = y_1^*, &y_2& = y_2^*,
  \label{eq:D27_2}
\end{align}
and the other couplings of $y_{3,\cdots,8}$ are zero.
Obviously the inconsistent CP symmetry is different from the CP-like symmetry.
We notice, however, that the conditions for $y_{0,1,2}$ in Eq.~\eqref{eq:D27_2}
are the same as the one obtained in the CP-like symmetry in Eq.~\eqref{eq:D27_1}.
Thus CP-like transformation is identified as the physical one if the model
does not have $\phi_{3,..., 8}$.

\subsection{Physical Implications of CP-like Symmetry}

We study physical implications of CP-like symmetry in detail.
Let us first review the CP transformation of the scattering amplitude
of multi-particle states $\{\phi_{\mathbf{r}_1}, \phi_{\mathbf{r}_2},... \} \to \{ \phi_{\mathbf{r}'_1}, \phi_{\mathbf{r}_2'},...\}$,
where
${\mathbf{r}_i}$ is an irreducible representation of the internal symmetry group $G$.
For concreteness, $G$ is assumed to be a discrete group.
In the following analysis we restrict the asymptotic states
to a direct product of single particle states made of elementary fields
for simplicity.\footnote{It may be interesting to consider extending
to more general states including non-product representations such as composite and entangled states.
For a recent study of the CP transformation of composite state, see~\cite{Bischer:2022rvf}.}
Neglecting the internal (wave function) structure,
the asymptotic states can be simply decomposed into those in the irreducible representations.
Thus we only consider that the initial and final states of $i$ and $f$
are in the irreducible representation of $\mathbf{r}_i$ and $\mathbf{r}_f$ of $G$,
where we note that $\mathbf{r}_i = \mathbf{r}_f$ thanks to the symmetry.

The cross section of this process is given by the Lorentz invariant $S$-matrix elements $i \mathcal{M}$
given as
\begin{align} \notag
  &i (2\pi)^4 \mathcal{M}_{\{\phi_{\mathbf{r}_1}(p_1), \phi_{\mathbf{r}_2}(p_2),... \} \to \{\phi_{\mathbf{r}'_1}(p'_1), \phi_{\mathbf{r}_2'}(p'_2),...\}}
  \delta^{(4)}(p_{\textrm{init}} - p_{\textrm{final}})
  \\
  &=
  \lim_{T \to \infty} \bra{p'_1, p'_2,...}
  e^{-2i\mathcal{H}T}
  \ket{p_1, p_2,... },
\end{align}
where $p_i$ and $p'_i$ represent the momenta of the initial and final particles in representation of ${\mathbf{r}_i}$ and ${\mathbf{r}'_i}$, respectively.
$p_{\textrm{init}}$ and $p_{\textrm{final}}$ denote the total momenta of the initial and final states.
The physical CP transformation relates a scattering amplitude and
the identical process but with antiparticles of momenta $\mathbf{k}_i=-\mathbf{p}_i, \mathbf{k}_i'=-\mathbf{p}_i'$ as follows,
\begin{align} \notag
  &\lim_{T \to \infty}
  \bra{p'_1, p'_2,...}
  \mathcal{CP}^{-1}
  \mathcal{CP}
  e^{-2i\mathcal{H}T}
  \mathcal{CP}^{-1}
  \mathcal{CP}
  \ket{p_1, p_2,... }
  \\
  &=
  \lim_{T \to \infty}
  \bra{k'_1, k'_2,...}
  e^{-2i\mathcal{CP} \mathcal{H} \mathcal{CP}^{-1} T}
  \ket{k_1, k_2,... },
\end{align}
where $\mathcal{CP}$ is the physical CP transformation operator.
If the theory is invariant under the physical CP transformation,
the operator $\mathcal{CP}$ satisfies $[\mathcal{H}, \mathcal{CP}]=0$, and hence we obtain
\begin{align}
  \mathcal{M}_{
  \{\phi^{\mathrm{CP}}_{\mathbf{r}_1}(k_1), \phi^{\mathrm{CP}}_{\mathbf{r}_2}(k_2),... \} \to \{ \phi^{\mathrm{CP}}_{\mathbf{r}'_1}(k'_1), \phi^{\mathrm{CP}}_{\mathbf{r}_2'}(k'_2),...\}}
  = \mathcal{M}_
  {\{\phi_{\mathbf{r}_1}(p_1), \phi_{\mathbf{r}_2}(p_2),... \} \to \{ \phi_{\mathbf{r}'_1}(p'_1), \phi_{\mathbf{r}_2'}(p'_2),...\}},
  \label{eq:S-matrix_CP}
\end{align}
where $\phi_{\mathbf{r}_i}^{\textrm{CP}}(k_i)$ is the antiparticle of $\phi_{\mathbf{r}_i}(p_i)$ with inverse momentum.
Therefore the scattering amplitude is invariant under exchanging particles and antiparticles with inverse momenta.
It is worth noting that the physical CP transformation flips the sign of the particle numbers,
so that the physical CP invariance implies that both processes of particle generations and particle annihilation can occur with equal probability.
Thus the CP violation is necessary for number generation mechanisms such as baryogenesis.

Next, we consider the CP-like transformation of the scattering amplitudes.
If a model is invariant under a CP-like transformation, we obtain
\begin{align}
  \mathcal{M}_{\{\phi^{\textrm{CP-like}}_{\mathbf{r}_1}(k_1), \phi^{\textrm{CP-like}}_{\mathbf{r}_2}(k_2),... \}
  \to
  \{ \phi^{\textrm{CP-like}}_{\mathbf{r}'_1}(k'_1), \phi^{\textrm{CP-like}}_{\mathbf{r}_2'}(k'_2),...\}}
  = \mathcal{M}_
  {\{\phi_{\mathbf{r}_1}(p_1), \phi_{\mathbf{r}_2}(p_2),... \} \to \{ \phi_{\mathbf{r}'_1}(p'_1), \phi_{\mathbf{r}_2'}(p'_2),...\}},
  \label{eq:S-matrix_CP-like}
\end{align}
where $\phi^{\textrm{CP-like}}_{\mathbf{r}_i}(k_i)$ denotes the CP-like transformed particle of $\phi_{\mathbf{r}_i}(p_i)$ with inverse momentum.
By definition, we have at least one particle $\phi_{\mathbf{r}_i}$
which is transformed to ${\phi}^*_{\mathbf{r}_{j \neq i}}$ under the CP-like transformation.
Therefore Eq.~\eqref{eq:S-matrix_CP-like} does not guarantee the vanishing CP asymmetry.

From the perspective of the CP-like transformation,
it is convenient to classify the representations for the asymptotic states of the physical amplitudes
into the following three classes:
\begin{enumerate}
\renewcommand{\labelenumi}{\Alph{enumi}.}
  \item Representation which is properly transformed to its complex conjugate representation under the CP-like transformation.
  For instance $\mathbf{3}, \mathbf{3}^*$ and $\mathbf{1}_{0,1,2}$ in the previous $\Delta(27)$ model
  are properly transformed to their complex conjugate representations under Eq.~\eqref{eq:CP_d27}.
  \item Representation which is transformed to the same representation under the CP-like transformation, i.e. the CP-like eigenstate.
  For instance, $\mathbf{1}_4$ in the $\Delta(27)$ model is transformed $\mathbf{1}_4 \to \mathbf{1}^*_8 = \mathbf{1}_4$.
  \item Representation which is transformed to an irrelevant representation under the CP-like transformation
  different from neither its complex conjugate representation nor the original representation.
  $\mathbf{1}_{3,5,6,7}$ in the $\Delta(27)$ model are classified into this class.
\end{enumerate}

In the case of class A, the CP-like transformation of the initial state $i$ is
$i \to i^{\textrm{CP-like}} = \bar{i}$ (up to phase),
where $\bar{i}$ is the complex conjugate state of $i$ with inverse momentum.
Thus the final state $f$ is also transformed to its complex conjugate state as
$f \to f^{\textrm{CP-like}} = \bar{f}$ (up to phase).
The CP-like symmetry relates two different matrix elements as follows,
\begin{align}\label{eq:classA}
  |\mathcal{M}_{\{i\} \to \{f\}}|
  =
  |\mathcal{M}_{\{i^\textrm{CP-like}\} \to \{f^\textrm{CP-like} \}}|
  = |\mathcal{M}_{\{\bar{i}\} \to \{\bar{f}\}}|.
\end{align}
The physical CP violation cannot be observed from this process.

As for the case of class B,
the field $\phi_{\mathbf{r}_i}$ is transformed as $\phi_{\mathbf{r}_i} \to \phi_{\mathbf{r}_i^*}^*$.
Since $\phi_{\mathbf{r}_i}$ and $\phi_{\mathbf{r}_i^*}^*$ have the same representation under $G$,
both two fields could be kinematically mixed.
In fact, from the symmetry point of view the following mass terms should be generally allowed,
\begin{align}
  m^2 \left( \phi_{\mathbf{r}_i}^*\phi_{\mathbf{r}_i}^{\phantom *} + \phi_{\mathbf{r}^*_i}^* \phi_{\mathbf{r}^*_i}^{\phantom *} \right)
  + \mu^2 \phi_{\mathbf{r}_i^*} \phi_{\mathbf{r}_i} + (c.c.)
  =
(\phi_{\mathbf{r}_i}^*, \phi_{\mathbf{r}^*_i}^{\phantom *})
\begin{pmatrix}
m^2 & \mu^{2*} \\
\mu^{2} & m^2
\end{pmatrix}
\begin{pmatrix}
  \phi_{\mathbf{r}_i}^{\phantom *}\\
  \phi_{{\mathbf{r}}^*_i}^*
\end{pmatrix}.
\end{align}
This mass matrix is diagonalized by the CP-like eigenstates $\phi_{\mathbf{r}_i}^{\pm} = \frac1{\sqrt 2} (\phi_{\mathbf{r}_i} \pm \phi_{{\mathbf{r}}^*_i}^*)$,
and the particle (energy) eigenstates should be also given by the CP-like eigenstates $\phi_{\mathbf{r}_i}^{\pm}$ rather than $\phi_{\mathbf{r}_i}$.
If the states $i$ and $f$ are CP-like eigenstates,
the CP-like symmetry implies that
\begin{align}\label{eq:classB}
|\mathcal{M}_{\{i\} \to \{f\}}| = |\mathcal{M}_{\{i^\textrm{CP-like}\} \to \{f^\textrm{CP-like} \}}|  = |\mathcal{M}_{\{i\} \to \{f\}}| \ ({\rm with \ inverse \ momenta}).
\end{align}
There is no relationship between $\mathcal{M}_{\{i\} \to \{f\}}$ and $\mathcal{M}_{\{\bar{i}\} \to \{\bar{f}\}}$.
Therefore in principle we can observe the physical CP violation from this process.

In the last case of class C,
if the initial and final states transform as $i \to \bar{i'}$ and $f \to \bar{f}'$ by the CP-like
transformation, where $\mathbf{r}_{i'} = \mathbf{r}_{f'} \neq \mathbf{r}_{i} = \mathbf{r}_{f}$.
We obtain
\begin{align}\label{eq:classC}
  \mathcal{M}_{\{i\} \to \{f\}}
  =
  \mathcal{M}_{\{\bar{i'}\} \to \{\bar{f'}\}}.
\end{align}
From this process the physical CP violation can also be observed.
In addition, this relation means that the two CP-violating amplitudes are equivalent.

\begin{figure}[thbp]
  \centering
  \begin{tikzpicture}[baseline=(o.base)]
    \begin{feynhand}
          \vertex (a) at (-1,0) {$\phi_{\mathbf{r}_1}$};
          \vertex (c) at (1,1) {${\psi}_{\mathbf{r}_2}$};
          \vertex (b) at (1,-1) {$\bar{\psi}_{\mathbf{r}_3}$};
          \vertex [grayblob] (o) at (0,0) {};
          \propag [sca] (a) to (o);
          \propag [fermion] (o) to (c);
          \propag [anti fermion] (o) to (b);
       \end{feynhand}
  \end{tikzpicture}
  \caption{3-point decay.}
  \label{Fig:3-point_decay}
\end{figure}
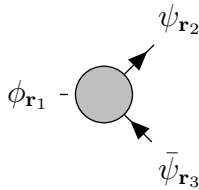

The differences of the above three amplitudes are clearly obtained when we consider a three-point decay process
as depicted in Fig.~\ref{Fig:3-point_decay},
which will be studied in models with $\Delta(27)$ in the next subsection.

\subsubsection{CP-like symmetry in $\Delta(27)$ model: class B}
\label{sec:class_B}

To illustrate the physical implications of the CP-like symmetry,
we calculate a decay amplitude whose initial state belongs to class B
in a model with $\Delta(27)$ symmetry.
We introduce one complex scalar singlet $\phi_4^+$, a triplet scalar fields $\Phi_{\mathbf{3}}$,
two singlet fermions $\psi_{5},$  $\psi_{6}$, and a triplet fermion $\Psi_{\mathbf{3}}$.
The field contents are summarized in Tab.~\ref{tab:CPmodel_D27}.
The CP-like transformation of the fields is given as
\begin{align}\notag
  \textrm{CP-like}:
  ~
  &\phi_4^+ \to \phi_4^+,
  &\Phi_{\mathbf{3}}& \to U_s \Phi_{\mathbf{3}}^*,
  \\ 
  &\psi_5 \to C \psi_6^*,
  &\psi_6& \to C \psi_5^*,
  &\Psi_{\mathbf{3}}& \to U_s C \Psi_{\mathbf{3}}^*,
\end{align}
where $C=-i \gamma_2 \gamma_0$ is the standard charge conjugation matrix for spinors.
$\phi_4^+$ is the CP-like eigenstate.
The Yukawa terms are given by
\begin{align} \notag
  \mathcal{L}_{\textrm{Yukawa}} =
  &y_1 \phi_4^+ \bar{\psi}_6 \psi_5
  + y_2 \phi_4^+ \otimes (\bar{\Psi}_{\mathbf{3}} \otimes \Psi_{\mathbf{3}})_{\mathbf{1}_8}
  + y_3 \bar{\psi}_5 \otimes (\Phi_{\mathbf{3}}^* \otimes \Psi_{\mathbf{3}})_{\mathbf{1}_5}
  \\
  &+ y_4 \bar{\psi}_6 \otimes (\Phi_{\mathbf{3}}^* \otimes \Psi_{\mathbf{3}})_{\mathbf{1}_6}
  + (h.c.).
\end{align}
The CP-like invariance constrains $y_4 = y_3^*$,
while $y_1$ and $y_2$ are the free parameters.
The scalar potential is irrelevant to discussion in this section, and we omit it here.
Assuming that $\phi_4^+$ is much heavier than $\psi_5$ and $\bar{\psi}_6$,
we calculate the decay amplitude of $\phi_4^+ \to \psi_6 \bar{\psi}_5$.
Since the product state of ${\mathbf{1}_6} \otimes \mathbf{1}^*_5 = {\mathbf{1}_4}$,
the initial and final states are in the same representation under $\Delta(27)$ as expected.
As explained in Sec.~\ref{sec:inconsistentCP}, the Yukawa term is not invariant under the physical CP transformation in Eq.~\eqref{eq:CP_inc},
we expect that the nonzero value of a CP asymmetry
$\epsilon_{\phi_4^+ \to \psi_6 \bar{\psi}_5} \equiv |\mathcal{M}_{\phi_4^+ \to \psi_6 \bar{\psi}_5}|^2- |\bar{\mathcal{M}}_{\bar{\phi}_4^+ \to \bar{\psi}_6 {\psi}_5}|^2$
can be observed.
At the one-loop order the matrix element  $\mathcal{M}_{ \phi_4^+ \to \psi_6 \bar{\psi}_5 }$ is given by the following diagrams as
\begin{align} \notag
   \begin{tikzpicture}[baseline=(o.base)]
      \begin{feynhand}
         \vertex (a) at (-1,0) {$\phi_4^+$};
         \vertex (b) at (1,-1) {$\bar{\psi}_5$};
         \vertex (c) at (1,1) {$\psi_6$};
         \vertex [dot] (o) at (0,0);
         \propag [sca] (a) to (o);
         \propag [fermion] (o) to (c);
         \propag [anti fermion] (o) to (b);
      \end{feynhand}
   \end{tikzpicture}
   +
   \begin{tikzpicture}[baseline=(o.base)]
      \begin{feynhand}
         \vertex (a) at (-1,0) {$\phi_4^+$};
         \vertex (b) at (1,-1) ;
         \vertex (c) at (1,1) ;
         \vertex (d) at (2, -1) {$\bar{\psi}_5$};
         \vertex (e) at (2, 1) {$\psi_6$};
         \vertex [dot] (o) at (0,0);
         \propag [sca] (a) to (o);
         \propag [fermion] (b) to [edge label = $\Psi_{\mathbf{3}}$](o);
         \propag [anti fermion] (c) to [edge label' = $\Psi_{\mathbf{3}}$] (o);
         \propag [anti fermion] (b) to (d);
         \propag [fermion] (c) to (e);
         \propag [sca] (b) to [edge label'= $\Phi_{\mathbf{3}}$] (c);
      \end{feynhand}
   \end{tikzpicture}
   =& -i \left\{ y_1  - y_2 y_3^* y_4 \textrm{tr}\left( M_6 M_8 M_5^\dagger \right)I \right\}\bar u^s(p)v^{s'}(p')
   \\
   =& -i \left( y_1  - 3 \omega^2 y_2 y_3^* y_4 I \right)\bar u^s(p)v^{s'}(p'),
\end{align}
where $I$ is the one-loop integral, and $p, p'$ are the momenta of the final state fermions.
The explicit form of $I$ is given in App.~\ref{App:1-loop}.
$M_i$ are the CG coefficients of $\Delta(27)$ given in App.~\ref{app:CG_D27}.
The physical CP asymmetry is observed as
\begin{align}
  \epsilon_{\phi_4^+ \to \psi_6 \bar{\psi}_5} \propto \mathrm{Im}\left(
  3\omega^2  y_1^* y_2 y_3^* y_4 \right)\mathrm{Im}\, I \neq 0.
  \label{eq:e465}
\end{align}
Thus the decay process of the class B state is CP asymmetric.

\begin{table}[htbp]
  \centering
  \begin{tabular}{| c | c cccc |}
    \hline
    & $\phi_4^+$ & $\Phi_{\mathbf{3}}$ & $\psi_{5}$ & $\psi_{6}$ & $\Psi_{\mathbf{3}}$ \\
    \hline
    $\Delta(27)$ & $\mathbf{1}_4$  & $\mathbf{3}$  & $\mathbf{1}_5$  & $\mathbf{1}_6$ & $\mathbf{3}$ \\
    mass & $m_4$  & $m_s$  & $m_5$  & $m_6$ & $m_f$ \\
    \hline
  \end{tabular}
  \caption{Field contents of $\Delta(27)$ model.}
  \label{tab:CPmodel_D27}
\end{table}

We also find that this decay process can generate the particle numbers of $\psi_5$ and $\psi_6$,
since $\psi_5$ and $\psi_6$ are in different representations and are different particles.
We note, however, that the sum of the particle numbers of $\psi_{5,6}$ is conserved, i.e.
\begin{align}
  N_{\psi_5} + N_{\psi_6} = 0,
  \label{eq:N56}
\end{align}
where $N_{\psi_{5,6}} \neq 0 $ are the particle numbers of $\psi_{5,6}$.
The relation in Eq.~\eqref{eq:N56} is analogous to other particle number generation processes
such as the GUT baryogenesis \cite{Yoshimura:1978ex} and the sphaleron process \cite{Kuzmin:1985mm},
in which the number of $B-L$ is conserved thanks to the internal (gauge) symmetry.
In this model the conservation of $N_{\psi_5} + N_{\psi_6}$ is
guaranteed by the CP-like symmetry.
The decay of the CP-like eigenstate could be the origin of matter generations.

\subsubsection{CP-like Symmetry in $\Delta(27)$ Model: Class C}\label{sec:classC}

We consider a model with nine singlet scalars $\phi_i$ and $N$ triplet fermions $\Psi^{I}_{\mathbf{3}}$,
\begin{align}
  \Psi^I_{\mathbf{3}} &= (\psi_1^{I},\psi_2^{I},\psi_3^{I})^T,
  &I& = 1,...,N.
\end{align}
The field contents are summarized in Tab.~\ref{tab:MC_D27}.
The mass terms of $\phi_i$ and $\Psi_{\mathbf{3}}^I$ are given as
\begin{align}
  \mathcal{L}_{\textrm{mass}} = m_i^2 \phi_i^* \phi_i + m_I \bar{\Psi}_{\mathbf{3}}^I \Psi_{\mathbf{3}}^I,
\end{align}
where we omit the mixing mass terms,
since they are irrelevant to the following analysis.
The Yukawa terms are almost the same as those in Eq.~\eqref{eq:Yukawa_D27}, but Yukawa couplings are generalized to $y_{i}^{IJ}$.
At the one-loop order, the matrix element $\mathcal{M}_{ \phi_i \to \psi^I_j \bar{\psi}^J_k }$ \ $(i \neq 0, 1, 2, 4, 8)$
is given by the following diagrams as
\begin{align} \notag
   &\begin{tikzpicture}[baseline=(o.base)]
      \begin{feynhand}
         \vertex (a) at (-1,0) {$\phi_i$};
         \vertex (b) at (1,-1) {$\bar{\psi}_k^J$};
         \vertex (c) at (1,1) {$\psi_j^I$};
         \vertex [dot] (o) at (0,0);
         \propag [sca] (a) to (o);
         \propag [fermion] (o) to (c);
         \propag [anti fermion] (o) to (b);
      \end{feynhand}
   \end{tikzpicture}
   +
   \begin{tikzpicture}[baseline=(o.base)]
      \begin{feynhand}
         \vertex (a) at (-1,0) {$\phi_i$};
         \vertex (b) at (1,-1) ;
         \vertex (c) at (1,1) ;
         \vertex (d) at (2, -1) {$\bar{\psi}_k^J$};
         \vertex (e) at (2, 1) {$\psi_j^I$};
         \vertex [dot] (o) at (0,0);
         \propag [sca] (a) to (o);
         \propag [fermion] (b) to [edge label = $\bar{\psi}_n^L$](o);
         \propag [anti fermion] (c) to [edge label' = $\psi_m^K$] (o);
         \propag [anti fermion] (b) to (d);
         \propag [fermion] (c) to (e);
         \propag [sca] (b) to [edge label'= $\phi_l$] (c);
      \end{feynhand}
   \end{tikzpicture}
   \\
   &= -i \left(
   (Y_{i}^{IJ})_{jk}
   - \sum_{\substack{l,K,L}}
   \left[
    Y_{l}^{IK\dagger}Y^{KL}_{i}Y^{LJ}_{l} + Y_{l}^{IK} Y^{KL}_{i}Y^{LJ\dagger}_{l}
   \right]_{jk} I^{KL}_{l}
   \right)\bar u^s(p)v^{s'}(p'),
\end{align}
where $I_{l}^{KL}$ is the loop integral with $\phi_l, \psi_{\mathbf{3}}^K, \psi_{\mathbf{3}}^L$.
$Y_i^{IJ}$ is the Yukawa matrix given by $Y_i^{IJ} = y^{IJ}_{i} M_{i^*}$.
As a result, we calculate a CP asymmetry,
\begin{align}
  \epsilon_{\phi_i\to \Psi_{\mathbf{3}}^I \bar{\Psi}_{\mathbf{3}}^J}
  \equiv& \sum_{\substack{j,k=1}}^3 \left( |\mathcal{M}_{\phi_i \to \psi^I_j \bar{\psi}^J_k}|^2 - |\bar{\mathcal{M}}_{\bar{\phi}_i \to \bar{\psi}^I_i {\psi}^J_k}|^2 \right)
  \nonumber
  \\
  =& -4 \sum_{\substack{l, K,L}} \mathrm{Im}\left(
  \textrm{tr}\,
  Y_{i}^{IJ\dagger} \left[Y_{l}^{IK\dagger}Y^{KL}_{i}Y^{LJ}_{l} + Y_{l}^{IK} Y^{KL}_{i}Y^{LJ\dagger}_{l}\right]
  \right)
  \mathrm{Im}\, I^{KL}_{l}
  \nonumber
  \\
  =& - 8 \sum_{\substack{l, K,L}} \mathrm{Re}\left(y_{l}^{IK *} y_{l}^{LJ}
  \textrm{tr}\,M_{{l}^*}^\dagger M_{i^*} M_{{l}^*} M_{{i}^*}^\dagger
  \right)
  \mathrm{Im}\, y_{i}^{IJ*} y_{i}^{KL}
  \mathrm{Im}\, I^{KL}_{l}, \label{eq:CPV-formula}
\end{align}
where we sum over the indices of $j, k$ from 1 to 3 for the final states $\mathbf{3}$,
since those three components are equivalent under the $\Delta(27)$ symmetry.
This summation is important to prove that there is no general CP transformation that cancels the CP asymmetry
for a specific channel.\footnote{See App.~\ref{sec:trace}.}
In terms of $\Delta(27)$ symmetry this summation should correspond to extracting
$\mathbf{1}_i$ from the tensor products of $\mathbf{3} \otimes {\mathbf{3}}^*$.
From the conditions for CP-like symmetry in Eq.~\eqref{eq:D27_1}, $y_{i}^{IJ}$ is complex.
Hence when $N \geq 2$, even though the theory is CP-like invariant,
$y_i^{IJ*} y_i^{KL}$ is complex, and we observe the physical CP violation of $\epsilon_{\phi_i\to \Psi_{\mathbf{3}}^I \bar{\Psi}_{\mathbf{3}}^J} \neq 0$ 
at the one-loop level.

\begin{table}[htbp]
  \centering
  \begin{tabular}{| c | c c |}
    \hline
    & $\phi_i$ & $\Psi^{I}_{\mathbf{3}}$ \\
    \hline
    $\Delta(27)$ & $\mathbf{1}_i$ & $\mathbf{3}$ \\
    \hline
  \end{tabular}
  \caption{Matter contents of $\Delta(27)$ model. $i = 0, ..., 8,$ and $I = 1,...,N$.}
  \label{tab:MC_D27}
\end{table}

In addition, it follows from Eq.~\eqref{eq:classC}
that the CP-like transformation relates $\epsilon_{\phi_i \to \Psi^I \bar{\Psi}^J}$ to the other CP asymmetry
of $\epsilon_{\phi_{i'} \to \Psi^I \bar{\Psi}^J}$, where $i'$ is the index of the CP-like conjugated state of $\phi_i$ given in Eq.~\eqref{eq:CPL_D27_1}.
For instance, $\phi_3$ is related to $\phi_7^*$.
Using the following relations
\begin{align}
  \mathrm{tr}\,M_{l^*}^\dagger M_{3^*} M_{l^*} M_{3^*}^\dagger =
  (\mathrm{tr}\,M_{l^{\prime *}}^\dagger M_{7^*} M_{l^{\prime*}} M_{7^*}^\dagger)^*, \ \ \
    y_l^{IK *} y_l^{LJ} = (y_{l'}^{IK *} y_{l'}^{LJ})^*,
\end{align}
we obtain
\begin{align} \notag
  \epsilon_{\phi_3 \to \Psi^I_{\mathbf{3}} \bar{\Psi}_{\mathbf{3}}^J}
  =& -8
  \sum_{l', K,L}
  \mathrm{Re}\, \left(
  (y_{l'}^{IK *} y_{l'}^{LJ})^*
  (\mathrm{tr}\,M_{l^{\prime*}}^\dagger M_{7^*} M_{{l}^{\prime*}} M_{7^*}^\dagger)^*
  \right)
  \mathrm{Im}\, (y_7^{IJ*} y_7^{KL})^*
  \mathrm{Im}\, I^{KL}_{l'}
  \\
  =& - \epsilon_{\phi_7 \to \Psi^I_{\mathbf{3}} \bar{\Psi}^J_{\mathbf{3}}},
\end{align}
where we use $I_l^{KL} = I_{l'}^{KL}$, because the masses of $\phi_l$ and $\phi_{l'}$ are the same
due to the CP-like symmetry,
and the minus sign in the last line comes from $\mathrm{Im}\, (y_7^{IJ*} y_7^{KL})^* = -\mathrm{Im}\, y_7^{IJ*} y_7^{KL}$.
Thus we obtain
\begin{align}
  \epsilon_{\phi_3 \to \Psi^I_{\mathbf{3}} \bar{\Psi}_{\mathbf{3}}^J}+ \epsilon_{\phi_7 \to \Psi_{\mathbf{3}}^I \bar{\Psi}_{\mathbf{3}}^J} = 0.
\end{align}
The CP asymmetries of the two different amplitudes are related by the CP-like transformation as expected.

In both class B and C cases, the physical CP violation is observable.
In addition, since two particles in different representations are exchanged by the CP-like transformation,
two different particle numbers can be generated through the corresponding CP-violating scattering process,
although the sum of the two exchanged particle numbers is conserved by the CP-like transformation.
It is noted that our results can also be applied to the pair annihilation process of two fermions to one complex scalar field.
Thus, the CP asymmetry also implies the violation of particle number conservation of the complex scalar fields such as $\phi_4^+$ and $\phi_{3,7}$ in the models presented here.
The only difference between class B and C is that the CP-eigenstate has no CP-like partner,
and its particle number is not related to other antiparticles nor conserved, i.e., $N_{\phi_4^+} \neq 0$.
In any case, it is crucial to consider a state of either class B or C for number generation.
This mechanism might be useful for phenomenological purposes, e.g., for baryogenesis and asymmetric DM.

\subsection{CP-like Symmetry for Continuous Groups} \label{sec:ContinuousCP-like}

We discuss a possible extension of CP-like transformations
to general groups.
As is known, since most of popular simple Lie groups
such as $SU(N) \ (N \geq 3)$ and Abelian group $U(1)$
have the unique outer automorphism group of $\mathbb{Z}_2$
which is identified with the proper CP transformation \cite{Grimus:1995zi},
the CP-like transformation is not well defined for such simple groups.
Let us provide a specific explanation by using a $U(1)$ symmetry.
For a multiplet field $\Phi = (\phi_p, \phi_q)^T$,
where $\phi_{p,q}$ have $U(1)$ charges of $p, q \neq 0$,
a $U(1)$ transformation is represented as
\begin{align}
\Phi \to \Phi' = \rho_\Phi(\theta) \Phi,
\quad \quad
\rho_\Phi(\theta)
=
\begin{pmatrix}
e^{ip\theta} & 0 \\
0 & e^{iq\theta}
\end{pmatrix},
\ \ \theta \in \mathbb{R}.
\end{align}
A candidate CP-like transformation is given as
\begin{align}
\Phi(x) \to U \Phi^*(\tilde{x}),  \ \
U =
\begin{pmatrix}
0 & 1 \\
1 & 0
\end{pmatrix}.
\end{align}
The consistency condition of
$U \rho_\Phi^*(\theta) U = \rho_\Phi(\theta')$
leads to the following relation as
\begin{align}
\begin{pmatrix}
e^{-iq\theta} & 0 \\
0 & e^{-ip\theta}
\end{pmatrix}
=
\begin{pmatrix}
e^{ip\theta'} & 0 \\
0 & e^{iq\theta'}
\end{pmatrix}
\ \Rightarrow \
p=\pm q.
\end{align}
As a result, two automorphisms are allowed.
In the case of $p=q$,
this is a proper CP transformation,
which is based on the unique outer automorphism $u^{U(1)}_{\textrm{CP}}$ as given in Eq.~\eqref{eq:U1CP}.
On the other hand in the case of $p=-q$
it corresponds to an identity map
\begin{align}\label{eq:U1CP-like}
u^{U(1)}_{\textrm{CP-like}}: \theta \mapsto \theta,
\end{align}
where two exchanged fields of $\phi_p$ and $\phi_{-p}^*$
are in the same irreducible representation (class B).
These two are kinematically mixed to be CP-like basis
\begin{align}
\phi_\pm = \phi_p + \phi_{-p}^*.
\end{align}
Under this automorphism there is no representation belongs to class A or C except for the trivial singlet,
so that this is not a genuine CP-like transformation.

For the reasons above, we shall extend slightly further
to consider non-simple groups of $G = G_1 \times G_2$,
where we assume that both $G_1$ and $G_2$ have a complex conjugation outer automorphism.
We provide different ways to construct models with CP-like transformations as follows.

One trivial way to construct a CP-like symmetric model
is to use the following direct product of two automorphisms as
\begin{align}
  u_{\textrm{CP-like}}^{G_1\times G_2} = u^{G_1}_{\textrm{CP-like}} \times u^{G_2}_{\textrm{CP}},
\end{align}
where $u^{G_1}_{\textrm{CP-like}}$ is the identity map of the group $G_1$,
while $u^{G_2}_{\textrm{CP}}$ corresponds to the $\mathbb{Z}_2$ outer automorphism of the group $G_2$.
Thus the CP-like transformation for a direct product representation of $(\mathbf{r}_1, \mathbf{r}_2)$ is given as
\begin{align}
  \textrm{CP-like} : (\mathbf{r}_1, \mathbf{r}_2) \to (\mathbf{r}_1, \mathbf{r}^*_2),
\end{align}
where $\mathbf{r}_{1}$ and $\mathbf{r}_{2}$ are some non-trivial irreducible representations of $G_{1}$ and $G_{2}$, respectively,
and $\mathbf{r}^*_2$ is the complex conjugate representation of $\mathbf{r}_2$.
Accordingly the direct product representations are categorized into the aforementioned three classes
in the following manner,
\begin{align}
\textrm{Class A}: (\mathbf{1}_1, \mathbf{r}_2), \quad \quad
\textrm{Class B}: (\mathbf{r}_1, \mathbf{1}_2), \quad \quad
\textrm{Class C}: (\mathbf{r}_1, \mathbf{r}_2),
\end{align}
where
$\mathbf{1}_{1}$ and $\mathbf{1}_{2}$ are the trivial singlet representations of $G_{1}$ and $G_{2}$, respectively.

An example for this type of CP-like transformation is easily constructed in a model with $U(1)_1 \times U(1)_2$ symmetry.
Let us introduce a multiplet field $\Phi=(\phi_p, \phi_{q})^T$,
where $\phi_{p}$ and $\phi_{q}$ have two $U(1)$ charges $p_{1,2}$, and $q_{1,2}$ under $U(1)_1 \times U(1)_2$.
A group action on $\Phi$ is represented as
\begin{align}
\Phi \to \Phi' = \rho_\Phi(\theta_1, \theta_2) \Phi,
\quad \quad
\rho_\Phi(\theta_1, \theta_2)
=
\begin{pmatrix}
e^{i(p_1\theta_1+p_2\theta_2)} & 0 \\
0 & e^{i(q_1\theta_1+q_2\theta_2)}
\end{pmatrix},
\ \ \theta_{1,2} \in \mathbb{R}.
\end{align}
For example, when $q_1=-p_1$ and $q_2=p_2$ the following CP-like transformation
\begin{align}
\Phi(x) \to U_{\textrm{CP-like}}^{U(1)_1\times U(1)_2} \Phi^*(\tilde{x}),  \ \
U_{\textrm{CP-like}}^{U(1)_1\times U(1)_2} =
\begin{pmatrix}
0 & 1 \\
1 & 0
\end{pmatrix},
\end{align}
satisfies the consistency conditions.
The corresponding automorphism $u_{\textrm{CP-like}}^{U(1)_1 \times U(1)_2}$ is
\begin{align}\label{eq:U1xU1}
u_{\textrm{CP-like}}^{U(1)_1 \times U(1)_2}& : (\theta_1, \theta_2) \mapsto (\theta_1, -\theta_2).
\end{align}
Thus the CP-like partners of $\phi_p$ and $\phi_q$ have the opposite $U(1)_1$ charges but
the same $U(1)_2$ charge.
If the model has this CP-like symmetry,
it requires the existence of a CP-like partner in addition to their complex conjugate fields.
In the same way, it is possible to construct a CP-like symmetric model with internal symmetry group $G_1 \times G_2$.

In addition, in the case of $U(1)$ two independent $U(1)$ transformations can mix,
so that it is possible to find a non-trivial CP-like transformation for generic values of $U(1)$ charges.
When $p_1 q_2 - p_2 q_1 \neq 0$,
a general solution to the consistency condition is as follows,
\begin{align}
\notag
&  \begin{pmatrix}
     0 & 1\\
     1 & 0
  \end{pmatrix}
  \rho_\Phi(\theta_1, \theta_2)^*
  \begin{pmatrix}
     0 & 1\\
     1 & 0
  \end{pmatrix}^\dagger
=
\rho_\Phi(\theta_1', \theta_2')
\\
\quad \quad  \Rightarrow \quad \quad
&  \begin{pmatrix}
    \theta_1'\\
    \theta_2'
  \end{pmatrix}
  = A
  \begin{pmatrix}
    \theta_1\\
    \theta_2
  \end{pmatrix},
\quad \quad
A =
  \frac1{p_1 q_2 - p_2 q_1}
  \begin{pmatrix}
    p_1p_2-q_1q_2 & p_2^2-q_2^2 \\
    q_1^2-p_1^2 & q_1q_2-p_1p_2
  \end{pmatrix}.
  \label{eq:auto_U1}
\end{align}
This is a CP-like transformation since it exchanges the states of linearly independent charges.
To see it more clearly
we diagonalize the automorphism corresponding to the CP-like transformation in Eq.~\eqref{eq:auto_U1} as,
\begin{align}
  \begin{pmatrix}
    \tilde{\theta}_1\\
    \tilde{\theta}_2
  \end{pmatrix}
  = M
\begin{pmatrix}
\theta_1\\
\theta_2
\end{pmatrix}
\end{align}
where $M$ is a diagonalization matrix of $A$, and $\tilde{\theta}_{1}$ and $\tilde{\theta}_{2}$ are
transformation parameters of $\tilde{U}(1)_1 \times \tilde{U}(1)_2 \simeq U(1)_1 \times U(1)_2$
(linear combinations of original two $U(1)$ groups).
From the facts that $\det{A}=-1$, in the new basis
the automorphism is given as
\begin{align}
  \begin{pmatrix}
    \tilde{\theta}_1'\\
    \tilde{\theta}_2'
  \end{pmatrix}
  =
  \begin{pmatrix}
    1 & 0\\
    0 & -1
  \end{pmatrix}
  \begin{pmatrix}
    \tilde{\theta}_1\\
    \tilde{\theta}_2
  \end{pmatrix}.
\end{align}
The $\tilde{U}(1)_2$ charge reverses,
while the $\tilde{U}(1)_1$ charge is unchanged.
Thus we obtain a CP-like transformation for $U(1)_1 \times U(1)_2$ symmetry.

We should stress here again that a CP-like transformation $(\mathbf{r}_1, \mathbf{r}_2) \mapsto (\mathbf{r}_1, \mathbf{r}_2^*)$
is only possible for models with specific field content. 
Since the existence of two related fields of $\phi_{\mathbf{r}_1, \mathbf{r}_2}$ and $\phi_{\mathbf{r}_1, \mathbf{r}_2^*}$ 
is not automatically ensured, 
while for the physical CP transformation 
a field $\phi_{\mathbf{r}_1, \mathbf{r}_2}$ and their conjugate field $\phi_{\mathbf{r}_1^*, \mathbf{r}_2^*}$ 
should be ensured by the reality of the relativistic quantum theory.
Thus the CP-like symmetric model generally requires the existence 
of the CP-like partner in addition to their complex conjugate field (anti-particle).

\section{CP/CP-like Symmetry and Spontaneous Symmetry Breaking}\label{sec:SSB}

In this section we first discuss the origin of the CP-like symmetry in the CP-like symmetric model.
In the previous sections, we have shown that
models with CP-like symmetry for type I groups (i.e., the absence of CIA) 
exhibit a physical CP violation.
It is important to notice that such type I groups 
can be embedded into simpler groups like $SU(N) \ (N \geq 3)$, 
in which a proper CP transformation can be defined.
This observation implies a potential scenario in certain dynamical models:
when a vacuum expectation value (VEV) gives rise to a symmetry breaking from $G \to H$,
a physical CP is simultaneously violated due to a spontaneous change of the proper CP $\to$ CP-like,
even if the VEV does not break the CP symmetry, that is, the vacuum remains invariant under the original CP transformation.
The spontaneous CP violation based on the mechanism was first proposed
in a model with $SU(3) \to T_7$~\cite{Ratz:2016scn}, where $T_7$ is a type I group.
Thus, it might be expected that
a limited class of groups, such as type I, could manifest such spontaneous CP violation.
We note, however, that as we will illustrate later, 
a wider class of CP-like symmetric models can be derived from a proper CP symmetric model,
even if the subgroup $H$ possesses CIA.

If $H$ does not have any CIA, the physical CP transformation in the broken phase
should be an inconsistent CP transformation, and CP asymmetry should be observed.
There are two possibilities for spontaneous CP violation in the broken phase.
If the inconsistent physical CP transformation in the broken phase is given by
a product of the original CP transformation and a broken element of $G\backslash H$,
the CP violation is directly caused by the SSB of the internal symmetry,
and hence the CP asymmetry is proportional to the VEV giving rise to the SSB.
On the other hand, if this is not the case,
the inconsistent physical CP transformation in the broken phase should remain inconsistent even in the symmetric phase,
although the theory is invariant under the same CP(-like) transformation in both phases.
Therefore the CP asymmetry may not necessarily be proportional to the VEV.
We will show that the latter type of CP asymmetry really occurs through SSB.
In this case, the CP violation genuinely comes from the group structure.
We will carefully study the CP asymmetry of several physical observables in the broken phases,
by which we elucidate the relations between the physical CP violation in the broken phase and
the CP invariance in symmetric phase in detail.

Next, we discuss the fate of the CP-like symmetry.
Let us assume that $H$ is a type I group.
As mentioned above, if a model is invariant under $H$ and CP-like transformations,
the physical CP is violated in general.
Here one question arises immediately:
is there a possibility that the CP-like transformation would become a proper CP transformation after further symmetry breaking of $H \to H'$?
If the VEV does not break the CP-like symmetry, it appears to be possible.
A trivial example is when $H$ is completely broken down to $H' = I$,
resulting in all fields belonging to the trivial singlet representation ({\it multiplet merging}),
making it impossible to distinguish between CP-like partners. 
Accordingly the CP-like transformation will spontaneously change to a proper CP transformation.
In a more general scenario where $H' \neq I$,
if two mutually exchanged representations, $\mathbf{r}_i$ and $\mathbf{r}_j$, belong to the same representation after SSB,
the transformation matrix in Eq.~\eqref{eq:U_CPlike} should change to that of a proper CP transformation in Eq.~\eqref{eq:properCP}.
On the contrary, the spontaneous CP violation may occur
when two mutually exchanged fields under a proper CP transformation split into two distinct representations ({\it multiplet splitting}).
The details of such examples will be shown in what follows.

\subsection{$\Delta(54)\times U(1)$ Model}

In this subsection, we study  symmetry breaking pattern of model with $\Delta(54) \times U(1)$.
The discrete group $\Delta(54)$ has 10 irreducible representations of
two real singlets $\mathbf{1}_{0,1}$, four pseudoreal doublets $\mathbf{2}_{1,2,3,4}$ and four complex triplets $\mathbf{3}_{1,2}, \mathbf{3}^*_{1,2}$.
See App.~\ref{sec:A1} for the details of $\Delta(54)$.
Since there is no CIA of $\Delta(54)$,
we consider the following CP-like transformation,
\begin{align}
  \mathbf{3}_i &\to U_s \mathbf{3}_i^*,
  &\mathbf{2}_1& \to \mathbf{2}_1^*,
  &\mathbf{2}_2& \to \mathbf{2}_4^*,
  &\mathbf{2}_{3}& \to S_2 \mathbf{2}_{3}^* ,
  &\mathbf{2}_4& \to \mathbf{2}_2^*,
  &\mathbf{1}_{0,1} \to \mathbf{1}_{0,1}^*,
  \label{eq:CP-like}
\end{align}
where $U_s$ and $S_2$ are unitary matrices given by
\begin{align}
  U_s &=
  \begin{pmatrix}
    \omega^2 & 0 & 0\\
    0 & 0 & 1\\
    0 & 1 & 0
  \end{pmatrix}, \
  &S_2  &=
  \begin{pmatrix}
    0 & 1 \\
    1 & 0
  \end{pmatrix}.
  \label{eq:S2}
\end{align}
This CP-like transformation corresponds to an automorphism
\begin{align}\label{eq:autoDelta54}
  u^{\Delta(54)}_{\textrm{CP-like}}~:~(a, a', b, c) \mapsto (a^2 a'^2, a', a^2b^2,c),
\end{align}
on $\Delta(54)$,\footnote{It is straightforward to check that 
$U_s \rho_{\mathbf{3}_i}(g)^* U_s^\dagger = \rho_{\mathbf{3}_i}(u_{\textrm{CP-like}}(g))$, and $S_2 \rho_{\mathbf{2}_3}(g)^* S_2^\dagger = \rho_{\mathbf{2}_3}(u_{\textrm{CP-like}}(g))$.}
and $e^{i \theta} \to e^{-i \theta}$ on $U(1)$.

\begin{table}[thbp]
  \centering
  \begin{tabular}{|c|ccccccc|}
    \hline
    & $\Psi^1$ & $\Psi^2$ & $\Psi^3$ & $\Psi^4$ & $S$ & $T$ & $R$
    \\
    \hline
    $\Delta(54)$ & $\mathbf{3}_1$  & $\mathbf{3}_1$ & $\mathbf{3}_1$  & $\mathbf{3}_1$ & $\mathbf{1}_1$ & $\mathbf{2}_3$  & $\mathbf{2}_1$
    \\
    $U(1)$  &  $1$ & $2$ & $3$ & $4$  & $0$ & $-1$ & $-2$\\
    \hline
  \end{tabular}
  \label{tab:model}
  \caption{Irreducible representations and $U(1)$ charges of the fields.
  $\Psi^I$ are Dirac fermions, $T, R$ are complex scalars, $S$ is a real scalar.}
\end{table}

We consider a Lagrangian with
\begin{align}
  \mathcal{L}
  = \bar \Psi^{I}(i \slashed{\partial} - m_\Psi^{I} )\Psi^{I}
  + \frac12 \partial_\mu S \partial^\mu S
  + \partial_\mu T^\dagger \partial^\mu T
  + \partial_\mu R^\dagger \partial^\mu R
  - \mathcal{L}_{\textrm{Yukawa}}
  - \mathcal{V}(S, T, R),
\end{align}
where we introduce two complex doublet scalar fields $T, R$, one real singlet scalar $S$,
and have four triplet Dirac fermions $\Psi^{1,2,3,4}$.
Their charges and representations are summarized in Tab.~\ref{tab:model}.
The interaction term $\mathcal{L}_{\textrm{Yukawa}}$ is given by
\begin{align} \notag
  \mathcal{L}_{\textrm{Yukawa}}
  =
  &
  t_1 [T \otimes (\bar{\Psi}^1 \otimes \Psi^2)_{\mathbf{2}_3}]_{\mathbf{1}_0}
  + t_2 [T \otimes (\bar{\Psi}^3 \otimes \Psi^4)_{\mathbf{2}_3}]_{\mathbf{1}_0}
  \\
  &
  + r_1 [R \otimes (\bar{\Psi}^1 \otimes \Psi^3)_{\mathbf{2}_1}]_{\mathbf{1}_0}
  + r_2 [R \otimes (\bar{\Psi}^2 \otimes \Psi^4)_{\mathbf{2}_1}]_{\mathbf{1}_0}
  + (h.c.),
\end{align}
where $t_i$ and $r_j$ are Yukawa couplings.
Since $\mathbf{3}^*_1 \otimes \mathbf{3}_1$ does not include $\mathbf{1}_1$, there are no Yukawa couplings between $S$ and the triplet fermions.
The tensor products are written as
\begin{align} \notag
  [T \otimes (\bar{\Psi}^1 \otimes \Psi^2)_{\mathbf{2}_3}]_{\mathbf{1}_0}
  =& M'_{ijk} T_i \bar{\Psi}_j^1 \Psi_k^2
  \\
  [R \otimes (\bar{\Psi}^1 \otimes \Psi^3)_{\mathbf{2}_1}]_{\mathbf{1}_0}
  =& N'_{ijk} R_i \bar{\Psi}_j^1 \Psi_k^3,
\end{align}
where
\begin{align} \notag
  M'_{1jk} &=
  \begin{pmatrix}
    0 & 0 & 1\\
    \omega & 0 & 0\\
    0 & \omega^2 & 0
  \end{pmatrix},~~
  &M'_{2jk} &=
  \begin{pmatrix}
    0 & 1 & 0\\
    0 & 0 & \omega^2\\
    \omega & 0 & 0
  \end{pmatrix},\\
  N'_{1jk} &=
  \begin{pmatrix}
    1 & 0 & 0 \\
    0 & \omega & 0\\
    0 & 0 & \omega^2
  \end{pmatrix},~~
  &N'_{2jk} & =
  \begin{pmatrix}
    1 & 0 & 0\\
    0 & \omega^2 & 0\\
    0 & 0 & \omega
  \end{pmatrix}.
\end{align}
We note that the above CG coefficients are the same as those of $\Delta(27)$:
$M'_{1} = M_4, M'_{2} = M_8, N'_{1} = M_1$ and $N'_{2} = M_2$ in App.~\ref{app:CG_D27}.
The scalar potential is given by
\begin{align}
  \mathcal{V} =\,
  &
  \frac12 m_S^2 S^2 +
  m_T^2 (T^\dagger \otimes T)_{\mathbf{1}_0}
  + m_R^2 (R^\dagger \otimes R)_{\mathbf{1}_0}
  + \kappa_1 S (T^\dagger \otimes T)_{\mathbf{1}_1}
  + \kappa_2 S (R^\dagger \otimes R)_{\mathbf{1}_1}
  \notag
  \\
  &
  + \frac{\lambda_1}4 (T^\dagger \otimes T)_{\mathbf{1}_0}^2
  + \frac{\lambda_2}4 (T^\dagger \otimes T)_{\mathbf{1}_1}^2
  + \frac{\lambda_3}4 (R^\dagger \otimes R)_{\mathbf{1}_0}^2
  + \frac{\lambda_4}4 (R^\dagger \otimes R)_{\mathbf{1}_1}^2
  \notag
  \\
  &+ \lambda_5 (T^\dagger \otimes T)_{\mathbf{1}_0} (R^\dagger \otimes R)_{\mathbf{1}_0}
  + \lambda_6 (T^\dagger \otimes T)_{\mathbf{1}_1}(R^\dagger \otimes R)_{\mathbf{1}_1}
  \notag
  \\
  & + \lambda_7 S^2 (T^\dagger \otimes T)_{\mathbf{1}_0}
  + \lambda_8 S^2 (R^\dagger \otimes R)_{\mathbf{1}_0}
  + \lambda_{9} S^4
  + (h.c.),
  \label{eq:potential_D54}
\end{align}
where $\kappa_i$ and $\lambda_j$ are complex parameters.
The tensor products are given by
\begin{align} \notag
  (T^\dagger \otimes T)_{\mathbf{1}_0} &= |T_1|^2 + |T_2|^2,
  &(R^\dagger \otimes R)_{\mathbf{1}_0} &= |R_1|^2 + |R_2|^2,
  \notag
  \\
  (T^\dagger \otimes T)_{\mathbf{1}_1} &= |T_1|^2 - |T_2|^2,
  &(R^\dagger \otimes R)_{\mathbf{1}_1} &= |R_1|^2 - |R_2|^2.
\end{align}

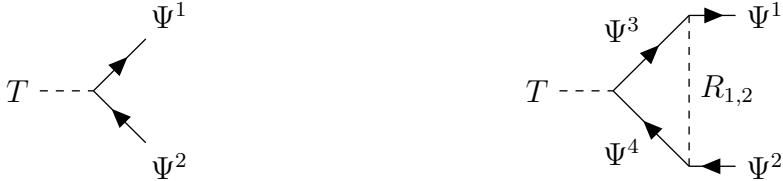
\begin{figure}[thbp]
  %\centering
  \begin{minipage}[b]{0.45\linewidth}
    \centering
    %\tikzset{external/force remake}
    \begin{tikzpicture}[baseline=(o.base)]
       \begin{feynhand}
          \vertex (a) at (-1,0) {$T$};
          \vertex (b) at (1,-1) {$\Psi^2$};
          \vertex (c) at (1,1) {$\Psi^1$};
          \vertex [dot] (o) at (0,0);
          \propag [sca] (a) to (o);
          \propag [fermion] (o) to (c);
          \propag [anti fermion] (o) to (b);
       \end{feynhand}
    \end{tikzpicture}
  \end{minipage}
  \begin{minipage}[b]{0.45\linewidth}
    \centering
    \begin{tikzpicture}[baseline=(o.base)]
      \begin{feynhand}
        \vertex (a) at (-1,0) {$T$};
        \vertex (b) at (1,-1) ;
        \vertex (c) at (1,1) ;
        \vertex (d) at (2, -1) {$\Psi^2$};
        \vertex (e) at (2, 1) {$\Psi^1$};
        \vertex [dot] (o) at (0,0);
        \propag [sca] (a) to (o);
        \propag [fermion] (b) to [edge label = $\Psi^4$](o);
        \propag [anti fermion] (c) to [edge label' = $\Psi^3$] (o);
        \propag [anti fermion] (b) to (d);
        \propag [fermion] (c) to (e);
        \propag [sca] (b) to [edge label'= $R_{1,2}$] (c);
      \end{feynhand}
    \end{tikzpicture}
  \end{minipage}
  \caption{Decay diagram of $T \to \Psi^1 \bar{\Psi}^2$.}
  \label{Fig:CP_D54}
\end{figure}

All the fields in the model are properly transformed to their complex conjugate fields under the CP-like transformation (See Eq.~\eqref{eq:CP-like}),
and hence this is regarded as a physical CP transformation.\footnote{Since $S$ is real, $S$ transforms to $S$ itself under the CP-like transformation.}
By imposing the CP-like invariance in this model
we obtain the following constrains on the couplings as
\begin{align}
\kappa_1 =& 0,
&\lambda_6 =& 0,
&  t_i =& |t_i| \omega,
& r_{1,2}, \kappa_2, \lambda_i (i \neq 6) \in \mathbb{R},
\end{align}
where we notice that $(T^\dagger \otimes T)_{\mathbf{1}_1}$ is CP(-like) odd and $(R^\dagger \otimes R)_{\mathbf{1}_1}$ is CP(-like) even.
It should be noted here that this model has complex couplings, while it is invariant under the physical CP transformation by construction.
In fact, for example, the CP asymmetry of the partial decay width of $T \to \Psi^1 \bar{\Psi}^2$ at the one-loop level is given by the interference of diagrams written in Fig.~\ref{Fig:CP_D54}.
It is given as
\begin{align} \notag
  \epsilon_{T \to \Psi^1 \bar{\Psi}^2}
  \propto&
  \sum_{\substack{a, b = 1,2}} \mathrm{Im}\left(
  t_1^* r_1 t_2 r_2^*
  \mathrm{tr} M^{\prime \dagger}_{a} N'_{b} M'_{a} N^{\prime \dagger}_{b}
  \right)
  \\
  =& \mathrm{Im}\left(
  -6 t_1^* r_1 t_2 r_2^*
  \right)
  = 0.
\end{align}
In the last line, we use the fact that $r_{1,2}$ are real and the phases of $t_{1,2}$ are the same.

\subsubsection{Spontaneous CP Violation: $\Delta(54) \times U(1) \to \Delta(27) \times U(1)$ }

We consider a spontaneous symmetry breaking of $\Delta(54) \times U(1) \to \Delta(27) \times U(1)$.
Since $\Delta(27)$ is inconsistent with a proper CP transformation,
we expect that the physical CP transformation for $\Delta(54)$ turns into CP-like transformation in the broken phase.
The irreducible representations of $\Delta(54)$ are decomposed to those of $\Delta(27)$ as follows,
\begin{align} \notag
  \mathbf{2}_1 &= \mathbf{1}_{2} \oplus \mathbf{1}_{1},
  &\mathbf{1}_{i} &= \mathbf{1}_0,
  \\ \notag
  \mathbf{2}_2 &= \mathbf{1}_5 \oplus \mathbf{1}_7,
  &\mathbf{3}_{i} &= \mathbf{3},
  \\ \notag
  \mathbf{2}_3 &= \mathbf{1}_8 \oplus \mathbf{1}_4,
  &\mathbf{3}^*_{i} &= \mathbf{3}^*,
  \\
  \mathbf{2}_4 &= \mathbf{1}_6 \oplus \mathbf{1}_3,
\end{align}
where the left-hand side corresponds to the irreducible representations of $\Delta(54)$, and the right-hand side $\Delta(27)$.
Thus $\Delta(54) \times U(1)$ is broken down to $\Delta(27) \times U(1)$ by a nonzero VEV for the non-trivial singlet $S$,
\begin{align}
  \braket{S} &= v \in \mathbb{R}.
\end{align}
This vacuum does not violate the original CP symmetry,
since $\braket{S} \in \mathbb{R}$ is invariant under the CP transformation defined in Eq.~\eqref{eq:CP-like}.
This vacuum is stable if $m_S^2$ is negative and $\lambda_{9}$ is positive.
The doublets of $T,R$ are decomposed into the singlets of $\Delta(27)$.
The degenerate masses of $R_1$ and $R_2$ split by $\braket{S}$.
On the other hand, the masses of $T_1$ and $T_2$ remain degenerate even after SSB,
because they are related by the CP-like transformation.
We summarize the representations and mass eigenstates in Tab.~\ref{tab:model_SSB}.
\begin{table}[thbp]
  \centering
  \begin{tabular}{|c|cccccc|}
    \hline
    & $\Psi^I$ & $S$ & $T_1$ & $T_2$ & $R_1$ & $R_2$
    \\
    \hline
    $\Delta(27)$ & $\mathbf{3}$  & $\mathbf{1}_0$ & $\mathbf{1}_8$ & $\mathbf{1}_4$ & $\mathbf{1}_2$ & $\mathbf{1}_1$
    \\
    $U(1)$  &  $q^I$ & $0$ & $-1$ & $-1$ & $-2$ & $-2$
    \\
    $m$  &  $m_{\Psi}^I$ & $\sqrt{-\lambda_{9} m_S^2}$ & $m_T$ & $m_T$ & $\sqrt{m_R^2 + \kappa_2 v}$ & $\sqrt{m_R^2 - \kappa_2 v}$
    \\
    \hline
  \end{tabular}
  \label{tab:model_SSB}
  \caption{Irreducible representations and $U(1)$ charges of the fields, and mass eigenvalues.}
\end{table}

This model is inconsistent with the physical CP transformation
because there are more than three fields in different singlet and triplet representations.
Thus physical CP transformation turns into a CP-like transformation in the broken phase while the vacuum is invariant
under the original CP transformation.
The CP-like transformation for $\Delta(27)\times U(1)$ model are exactly the same as
those in Eqs.~\eqref{eq:CPL_D27_3}, \eqref{eq:CPL_D27_1}, namely
\begin{align}
  \mathbf{3} &\to U_s \mathbf{3}^*,
  &\mathbf{1}_{0,1,2} \to& \mathbf{1}_{0,1,2}^*,
  &\mathbf{1}_4 \to& \mathbf{1}_8^*,
\end{align}
The field contents in the broken phase are similar to those of the model studied in Sec.~\ref{sec:classC} (up to $U(1)$ charges).
The CP asymmetry of the partial decay width of the complex scalar fields in the broken phase is similarly calculable.
For instance, the CP asymmetry of the $T_1 \to \Psi^1 \bar{\Psi}^2$ is given by
\begin{align} \notag
\epsilon_{T_1 \to \Psi^1 \bar{\Psi}^2}
  =&
  \sum_{a = 1,2}
  |t_1||t_2| r_1  r_2
  \mathrm{Im}\left(
  \mathrm{tr}
  M_{1}^{\prime \dagger}
  N'_{a}
  M'_{1}
  N_{a}^{\prime \dagger}
  \right)
  \mathrm{Im}\, I_{a}^{34}
  \\
  = &
  \frac{3}{2} |t_1||t_2| r_1  r_2
  \left(
  - \mathrm{Im}\, I_{1}^{34} + \mathrm{Im}\, I_{2}^{34}
  \right)
\end{align}
where $I_{a}^{34} \ (a=1,2)$ corresponds to the one-loop integral with $R_a$ and $\Psi^{3,4}$ (See Fig.~\ref{Fig:CP_D54}).
Since the masses of $R_1$ and $R_2$ are not degenerate in the broken phase, and hence,
$- \mathrm{Im}\, I_{1}^{34} + \mathrm{Im}\, I_{2}^{34} \neq 0$,
we observe the physical CP violation $\epsilon_{T_1 \to \Psi^1 \bar{\Psi}^2} \neq 0$.
On the other hand, $\epsilon_{T_2 \to \Psi^1 \bar{\Psi}^2}$ is given by
\begin{align} \notag
  \epsilon_{T_2 \to \Psi^1 \bar{\Psi}^2}
  =&
  \sum_{a = 1,2}
  |t_1||t_2| r_1  r_2
  \mathrm{Im}\left(
  \mathrm{tr}
  M_{2}^{\prime \dagger}
  N'_{a}
  M'_{2}
  N_{a}^{\prime \dagger}
  \right)
  \mathrm{Im}\, I_{a}^{34}
  \\
  = &
  \frac{3}{2} |t_1||t_2| r_1  r_2
  \left(
  \mathrm{Im}\, I_{1}^{34} - \mathrm{Im}\, I_{2}^{34}
  \right) \neq 0.
\end{align}
Thus we obtain
\begin{align}
  \epsilon_{T_1 \to \Psi^1 \bar{\Psi}^2} + \epsilon_{T_2 \to \Psi^1 \bar{\Psi}^2} =0.
\end{align}
The magnitudes of both CP asymmetries for $T_1$ and $T_2$ are the same, while their signs are opposite
due to the CP-like symmetry, as shown previously in Sec.~\ref{sec:classC}.

Here we should comment on the role of the VEV in the physical CP asymmetry.
As shown above, the CP asymmetry $\epsilon_{T_i \to \Psi^1 \bar{\Psi}^2}$ is proportional to
the mass difference of $R_1$ and $R_2$ in the one-loop integrals,
thus $\epsilon_{T_i \to \Psi^1 \bar{\Psi}^2}$ should be proportional to
$\braket{S}$.
This result might imply that the physical CP is violated by the VEV.
In other words all the physical CP asymmetries would vanish when $\braket{S} \to 0$.
We note, however, that this is not always the case,
since as stated above the vacuum does not violate
the original CP symmetry.\footnote{A VEV dependent CP violation was first studied in a CP-like $T_7$ model in \cite{Ratz:2016scn},
where the decay asymmetry for a particle in CP-like eigenstate (class B) was calculated.}
In fact,
as it will turn out that physical CP transformation for $\Delta(27) \times U(1)$ is not given by a product of the CP-like transformation and a broken element,
physical CP is violated even in the symmetric phase,
and hence there is no reason to make the CP asymmetry proportional to $\braket{S}$.

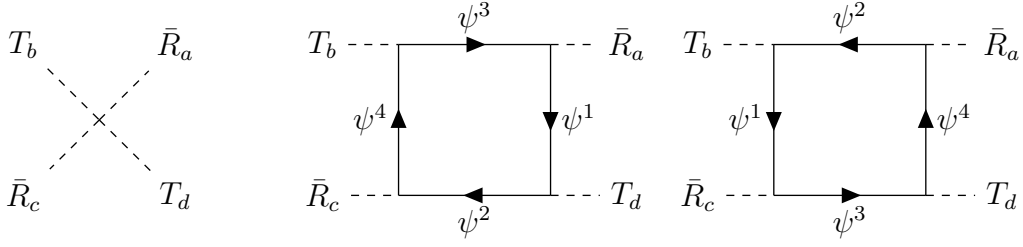
\begin{figure}[thbp]
  \begin{minipage}[b]{0.3\linewidth}
    \centering
    \begin{tikzpicture}[baseline=(o.base)]
       \begin{feynhand}
          \vertex (a) at (-1,1) {$T_b$};
          \vertex (b) at (-1,-1) {$\bar{R}_c$};
          \vertex (c) at (1,1) {$\bar{R}_a$};
          \vertex (d) at (1,-1) {$T_d$};
          \vertex [dot] (o) at (0,0);
          \propag [sca] (a) to (o);
          \propag [sca] (o) to (c);
          \propag [sca] (o) to (b);
          \propag [sca] (o) to (d);
       \end{feynhand}
    \end{tikzpicture}
  \end{minipage}
  \begin{minipage}[b]{0.3\linewidth}
    \centering
    \begin{tikzpicture}[baseline=0cm]
       \begin{feynhand}
          \vertex (a) at (-1,1) {$T_b$};
          \vertex (b) at (-1,-1) {$\bar{R}_c$};
          \vertex (c) at (3,1) {$\bar{R}_a$};
          \vertex (d) at (3,-1) {$T_d$};
          \vertex [dot] (x) at (0,1);
          \vertex [dot] (y) at (2,1);
          \vertex [dot] (z) at (2,-1);
          \vertex [dot] (o) at (0,-1);
          \propag [sca] (a) to (x);
          \propag [sca] (b) to (o);
          \propag [fermion] (o) to [edge label = $\psi^4$](x);
          \propag [fermion] (x) to [edge label = $\psi^3$](y);
          \propag [fermion] (y) to [edge label = $\psi^1$](z);
          \propag [fermion] (z) to [edge label = $\psi^2$](o);
          \propag [sca] (y) to (c);
          \propag [sca] (z) to (d);
       \end{feynhand}
    \end{tikzpicture}
  \end{minipage}
  \begin{minipage}[b]{0.3\linewidth}
    \centering
    \begin{tikzpicture}[baseline=0cm]
       \begin{feynhand}
          \vertex (a) at (-1,1) {$T_b$};
          \vertex (b) at (-1,-1) {$\bar{R}_c$};
          \vertex (c) at (3,1) {$\bar{R}_a$};
          \vertex (d) at (3,-1) {$T_d$};
          \vertex [dot] (x) at (0,1);
          \vertex [dot] (y) at (2,1);
          \vertex [dot] (z) at (2,-1);
          \vertex [dot] (o) at (0,-1);
          \propag [sca] (a) to (x);
          \propag [sca] (b) to (o);
          \propag [anti fermion] (o) to [edge label = $\psi^1$](x);
          \propag [anti fermion] (x) to [edge label = $\psi^2$](y);
          \propag [anti fermion] (y) to [edge label = $\psi^4$](z);
          \propag [anti fermion] (z) to [edge label = $\psi^3$](o);
          \propag [sca] (y) to (c);
          \propag [sca] (z) to (d);
       \end{feynhand}
    \end{tikzpicture}
  \end{minipage}
  \caption{4-point scattering of $T_b \bar{R}_c \to T_d \bar{R}_a$.}
  \label{Fig:CP_4point_D54}
\end{figure}

To show this point clearly, we also calculate the CP asymmetry of $2 \to 2$ scattering amplitudes of four complex scalars
as depicted in Fig.~\ref{Fig:CP_4point_D54}.\footnote{There are other one-loop diagrams contributing to the 4-point scattering,
but the other diagrams do not contribute to the CP asymmetry
because there are no complex couplings without Yukawa couplings in this model.
Therefore the CP asymmetry should be obtained from the diagrams including fermions at the one-loop level.}
The tree level and one-loop diagrams are calculated as
\begin{align} \notag
  \mathcal{M}^{\textrm{tree}}_{b\bar{c} \to d\bar{a}} =&
    -i \lambda_5  \delta_{ac} \delta_{bd}
  \\ \notag
  \mathcal{M}^{\textrm{1-loop}}_{b\bar{c} \to d\bar{a}} =& (-i)^4
  (
  r_2 t_2 r_1^* t_1^* \textrm{tr} N_a' M_b' N_c^{\prime \dagger} M_d^{\prime \dagger}
  +
  r_1 t_1 r_2^* t_2^* \textrm{tr} N_a' M_b' N_c^{\prime \dagger} M_d^{\prime \dagger}
  ) I_{\textrm{1-loop}}
  \\ \notag
  =& (-i)^4
  2 r_2 |t_2| r_1 |t_1|
  \textrm{tr} N_a' M_b' N_c^{\prime \dagger} M_d^{\prime \dagger}
  I_{\textrm{1-loop}}
  \\
  =&
  \begin{cases}
    6 r_1 r_2 |t_1 t_2| \omega^2 \delta_{ac}\delta_{bd} I_{\textrm{1-loop}}
    &(a = b)
    \\
    6 r_1 r_2 |t_1 t_2| \omega \delta_{ac}\delta_{bd} I_{\textrm{1-loop}}
    &(a\neq b)
  \end{cases},
\end{align}
where we use that $r_1, r_2$ are real, and the phases of $t_1$ and $t_2$ are the same.
As a result, the CP asymmetry at the one-loop level is given by
\begin{align}
  \epsilon_{T_b \bar{R}_c \to T_d \bar{R}_a} =
  \begin{cases}
    3 \sqrt{3} \lambda_5 r_1 r_2 |t_1t_2| \delta_{ac}\delta_{bd}
    \im I_{\textrm{1-loop}} &(a=b)
    \\
    -3\sqrt{3} \lambda_5 r_1 r_2 |t_1t_2|  \delta_{ac}\delta_{bd}
    \im I_{\textrm{1-loop}}
    &(a\neq b)
  \end{cases}.
  \label{eq:CP_asym_broken}
\end{align}
Thus the CP violation is observed in the $2 \to 2$ scattering processes.
We should emphasize that the CP asymmetry $\epsilon_{T_b \bar{R}_c \to T_d \bar{R}_a}$
does not depend on the mass splitting of the doublets, or scalar interaction parameters that include $\braket{S}$.
Thus $\epsilon_{T_b \bar{R}_c \to T_d \bar{R}_a}$ is not proportional to $\braket{S}$ at the one-loop level.
The spontaneous CP violation is caused by the
splitting of the multiplicity due to the VEV of the scalar field.
Actually this specific CP asymmetry is non-zero even in the symmetric phase.
Nevertheless, this process does not violate the CP symmetry in the symmetric phase.
This is because $T_{1,2}$ and $R_{1,2}$ are in the same multiplet in $\Delta(54)$,
and hence we can sum over the flavor indices $a, b$ of $T_a$ and $\bar{R}_b$ in external lines
in the scattering amplitudes.
Thus the cross section should be given by
\begin{align}
  \left(\frac{d \sigma}{d \Omega}\right)_{T\bar{R} \to T \bar{R}} \propto \sum_{a,b,c,d}
  |\mathcal{M}_{T_b\bar{R}_c \to T_d \bar{R}_a}|^2.
\end{align}
This value is CP symmetric because
\begin{align} \notag
  \left(\frac{d \sigma}{d \Omega}\right)_{T\bar{R}\to T\bar{R}}
  -
  \left(\frac{d \bar{\sigma}}{d \Omega}\right)_{\bar{T}{R}
  \to \bar{T} {R}}
  &\propto \sum_{a,b,c,d} \left(
  |\mathcal{M}_{T_b\bar{R}_c \to T_d \bar{R}_a}|^2
  -|\mathcal{M}_{\bar{T}_b {R}_c \to \bar{T}_d {R}_a}|^2 \right)
  \\ \notag
  &=
  \epsilon_{T_1 \bar{R}_1 \to T_1 \bar{R}_1} + \epsilon_{T_1 \bar{R}_2 \to T_1 \bar{R}_2} + \epsilon_{T_2 \bar{R}_1 \to T_2 \bar{R}_1} + \epsilon_{T_2 \bar{R}_2 \to T_2 \bar{R}_2}
  \\ \notag
  &=
  2\left(\epsilon_{T_1 \bar{R}_1 \to T_1 \bar{R}_1} +  \epsilon_{T_2 \bar{R}_1 \to T_2 \bar{R}_1}\right)
  \quad \quad (\because \Delta(54))
  \\
  &=0,
\end{align}
where $\epsilon_{T_1 \bar{R}_1 \to T_1 \bar{R}_1}$ and
$\epsilon_{T_2 \bar{R}_1 \to T_2 \bar{R}_1}$ cancel out each other (See Eq.~\eqref{eq:CP_asym_broken}).
This cancellation is also consistent with the fact that there exists a general CP transformation
associated with a unitary matrix $S_2$ in Eq.~\eqref{eq:S2}.
As shown here, the summation over the internal indices
is important to prove the existence of the CP violation (see App.~\ref{sec:trace} for a general proof).

This result is also understood from a calculation of the matrix elements in a diagonal basis for the CP-like transformations.
We take the following CP basis
\begin{align}
\tilde{T}_{\pm} &= \frac{1}{\sqrt 2}(T_1 \pm T_2), \quad
\tilde{R}_{1,2} =R_{1,2}
\end{align}
on which the CP-like transformation in Eq.~\eqref{eq:CP-like} diagonally acts as
\begin{align}\label{eq:CP-like_t}
\textrm{CP-like}:
\begin{cases}
\tilde{T}_{\pm} \to \tilde{T}_{\pm}^{\textrm{CP-like}} = \pm \tilde{T}^*_{\pm}, \\
R_{1,2} \to R_{1,2}^{\textrm{CP-like}} = R^*_{1,2}.
\end{cases}
\end{align}
From the properties in Eq.~\eqref{eq:CP-like_t}
we see that the CP-like symmetry makes a connection between two amplitudes as
\begin{align}\label{eq:M4pt}
\mathcal{M}_{\tilde{T}_x \bar{R}_c \to \tilde{T}_y \bar{R}_a}
=
\mathcal{M}_{\tilde{T}_x^{\textrm{CP-like}} \bar{R}_c^{\textrm{CP-like}} \to \tilde{T}_y^{\textrm{CP-like}} \bar{R}_a^{\textrm{CP-like}}}
=
xy \mathcal{M}_{\bar{\tilde{T}}_x R_c \to \bar{\tilde{T}}_y R_a}
\end{align}
where $x,y = \pm$.
Thus
the matrix element in this CP-basis is invariant under the CP-like transformation.
In the broken phase, however, we have to distinguish two fields $T_1$ and $T_2$ due to a multiplet splitting.
Therefore we specify either $T_1$ or $T_2$ as the initial and final states.
For example, a scattering amplitude of $T_1 \bar{R}_a \to T_1 \bar{R}_c$
can be obtained as a sum of the CP invariant matrix elements as
\begin{align}
\mathcal{M}_{T_1 \bar{R}_a \to T_1 \bar{R}_c}
&= \frac12\left(
\mathcal{M}_{\tilde{T}_+ \bar{R}_a \to \tilde{T}_+ \bar{R}_c}
+
\mathcal{M}_{\tilde{T}_+ \bar{R}_a \to \tilde{T}_- \bar{R}_c}
+
\mathcal{M}_{\tilde{T}_- \bar{R}_a \to \tilde{T}_+ \bar{R}_c}
+
\mathcal{M}_{\tilde{T}_- \bar{R}_a \to \tilde{T}_- \bar{R}_c}
\right).
\end{align}
We note that while each term is CP invariant,
there are interferences between them,
and hence it is not guaranteed that $\mathcal{M}_{T_1 \bar{R}_a \to T_1 \bar{R}_c}$ is also CP invariant.
In fact, from Eq.~\eqref{eq:M4pt} we easily see that there is a
relationship between two different matrix elements for different states as
\begin{align}
\mathcal{M}_{T_1 \bar{R}_a \to T_1 \bar{R}_c}
=&
\mathcal{M}_{\bar{T}_2 R_a \to \bar{T}_2 R_c},
\end{align}
but there is no relation between $\mathcal{M}_{T_1 \bar{R}_a \to T_1 \bar{R}_c}$
and $\mathcal{M}_{\bar{T}_1 R_a \to \bar{T}_1 R_c}$.
This is also consistent with the general result obtained in Eq.~\eqref{eq:classC}.

As shown above,
we explicitly confirm that
when the vacuum does not violate the original CP(-like) symmetry,
it remains after the SSB.
Nevertheless, the spontaneous CP violation occurs due to multiplet splitting.
To be specific, $S_2$ can be regarded as a physical CP transformation matrix at the level of $\Delta(54)$.
However, it becomes to a CP-like transformation after the SSB,
since $S_2$ exchanges $T_1$ and $T_2$ in different irreducible representations at the level of $\Delta(27)$.
Here we note that
a physical CP transformation in the broken phase
is an inconsistent CP transformation,
since it should be defined as
\begin{align}
  \mathbf{3}_i &\to U \mathbf{3}_i^*,
  &\mathbf{2}_1& \to \mathbf{2}_1^*,
  &\mathbf{2}_2& \to \mathbf{2}_4^*,
  &\mathbf{2}_{3}& \to \mathbf{2}_{3}^* ,
  &\mathbf{2}_4& \to \mathbf{2}_2^*,
  &\mathbf{1}_{0,1} \to \mathbf{1}_{0,1}^*.
\label{eq:physicalCP}
\end{align}
Obviously this transformation is inconsistent with any automorphisms of $\Delta(54)$ and $\Delta(27)$,
and it is not given by any product of original CP transformation and a broken element of $\Delta(54)\backslash \Delta(27)$.
We also note that this model is not invariant under the physical CP transformation (Eq.~\eqref{eq:physicalCP}) from the beginning.
Therefore for some scattering amplitudes we could observe a ``CP violation'' in both symmetric and broken phases,
e.g. $\epsilon_{T_b \bar{R}_c \to T_d \bar{R}_a}$,
where we recall that the complex conjugate states are defined by using the physical (but inconsistent) CP transformation defined in Eq.~\eqref{eq:physicalCP}.
In fact we explicitly confirm that this CP asymmetry is not proportional to the VEV at leading order.
Remarkably enough, in the broken phase this CP asymmetry
proves evidence of the physical CP violation,
while in the symmetric phase this does not mean the physical CP violation
due to the existence of a general CP transformation.
It should be noted that the results presented here do not depend on basis for the transformation matrix $(S_2)$
as explicitly shown in the calculations with a different (CP-like) basis.

The aforementioned result can be directly derived through group-theoretical analysis. 
Using the properties of $\Delta(27)$ as a type I group (no proper CP) and as a normal subgroup of $\Delta(54)$, 
we begin with the assumption that 
a physical CP transformation $CP_{\rm phys}$ at the level of $\Delta(27) \times U(1)$ 
can be represented as $CP_{\rm phys} = \rho(\tilde{g}) \circ CP$, 
where $CP$ is the original CP transformation at the level of $\Delta(54) \times U(1)$, 
and $\tilde{g} \in \Delta(54)\backslash \Delta(27)$, a broken element. 
Consequently, for any $h \in \Delta(27)$, there exists $h' \in \Delta(27)$ such that 
$\tilde{g}^{-1} h \tilde{g}= h'$, thus ensuring that $CP_{\rm phys}$ 
satisfies the consistency condition, 
\begin{align} \notag
CP_{\rm phys}^{-1} \circ \rho(h) \circ CP_{\rm phys} =& CP^{-1} \circ \rho(\tilde{g}^{-1}h\tilde{g}) \circ CP 
\\ \notag
=& CP^{-1} \circ \rho(h') \circ CP 
\\
=& \rho(u^{\Delta(54)}_{\textrm{CP-like}}(h')) \in H. 
\end{align}
However, this assumption leads to a contradiction, as type I groups do not have any proper CP transformations, 
implying that $CP_{\rm phys}$ should not satisfy the consistency condition. 
Therefore, $CP_{\rm phys}$ cannot be represented as a product of $CP$ and any broken element. 
This conclusion can be extended as follows: In scenarios involving the spontaneous change 
of a physical CP to a CP-like via SSB from $G \to H$, 
if the unbroken symmetry $H$ is a type I group and a normal subgroup of $G$, 
then any physical CP transformation defined in the broken phase ceases to exhibit symmetry, even in the symmetric phase.

In the next section we will further investigate the spontaneous breaking
of the internal symmetry group of $\Delta(27) \times U(1)$
in light of the relationship between the CP-like and
physical CP transformations.
We will see that when $\Delta(27)\times U(1)$ symmetry is broken down further by the scalar fields VEV of $T, R$,
the role of the original CP transformation can change depending on the symmetry breaking pattern.
We study the following three CP-like invariant vacua: $\braket{R_i}\in \mathbb{R}$, $\braket{T_1} = \braket{T_2}^*$, and their combination.
The first one breaks $U(1)$ and the internal symmetry is broken down to $\Delta(27)$.
Since there is no CIA in $\Delta(27)$, it is still CP-like symmetric,
and the physical CP symmetry remains violated as well.
On the other hand, in other two cases we will see some interesting phenomena will occur
such as the spontaneous CP restorations due to a multiplet merging.

\subsubsection{Spontaneous CP restoration: $\Delta(27) \times U(1) \to \mathbb{Z}_3^{a^2a'} \times \mathbb{Z}^{a^2 b}_3$}
We consider the VEVs of the scalar fields
\begin{align}
  \braket{T_1} = \braket{T_2}^* = \frac{v_T}{\sqrt{2}} e^{i\theta},
\end{align}
and $\braket{R_i} = 0$.
This vacuum does not violate the CP-like symmetry.
$v_T$ is given by
\begin{align}
  v_T = \pm \sqrt{\frac{-2m_T^2}{\lambda_1}},
\end{align}
where $m_T$ and $\lambda_1$ is the real parameter of the scalar potential Eq.~\eqref{eq:potential_D54}.
This vacuum is stable if $\lambda_2>0$ and $m_R^2 - \frac{4\lambda_5}{\lambda_1} m_T^2  \pm 2 \kappa_2 \braket{S} > 0$.
The vacuum is invariant under $a^2 a'$ and $a^2 b$, which are $\mathbb{Z}_3$ generators of $\Delta(27)$, and the $U(1)$ is completely broken down.
Thus symmetry is broken to $\mathbb{Z}_3^{a^2a'} \times \mathbb{Z}_3^{a^2b}$.
The irreducible decompositions are given as
\begin{align}
  \mathbf{3}_1 =& \mathbf{1}_{20}\oplus \mathbf{1}_{21}\oplus \mathbf{1}_{22},
  &\mathbf{1}_{4,8} =& \mathbf{1}_{00},
  &\mathbf{1}_{1} =& \mathbf{1}_{01},
  &\mathbf{1}_{2} =& \mathbf{1}_{02},
\end{align}
where the left-hand side corresponds to the irreducible representations of $\Delta(27)$,
and the right-hand side $\mathbb{Z}_3^{a^2a'}\times \mathbb{Z}_3^{a^2b}$.
The lower indices of $\mathbf{1}_{ij}$ denote the charges of $\mathbb{Z}_3^{a^2a'}\times \mathbb{Z}_3^{a^2b}$,
\begin{align} \notag
  a^2a':& \mathbf{1}_{ij} \to \omega^i \mathbf{1}_{ij},
  \\
  a^2b:& \mathbf{1}_{ij} \to \omega^j \mathbf{1}_{ij}.
\end{align}
The representations are summarized in Tab.~\ref{tab:model_Z3*Z3}.
Since the vacuum is invariant under the CP-like transformation at $\Delta(27)$,
it is consistent with an automorphism of $\mathbb{Z}_3 \times \mathbb{Z}_3$,
\begin{align}
u^{\mathbb{Z}_3^{a^2a'}\times \mathbb{Z}_3^{a^2b}} :
(a^2 a', a^2 b) \mapsto (a a'^2, a' b^2) = ((a^2 a')^{-1}, (a^2 b)^{-1}).
\end{align}
Thus all group elements in $\mathbb{Z}_3^{a^2a'}\times \mathbb{Z}_3^{a^2b}$
are transformed to their own inverses,
this is the proper CP transformation (CIA) at the level of $\mathbb{Z}_3^{a^2a'}\times \mathbb{Z}_3^{a^2b}$.
This result is also consistent with the fact that the two exchanged fields $T_1$ and $T_2^*$ under the CP-like transformation
belong to the same representation $\mathbf{1}_{00}$ as shown in Tab.~\ref{tab:model_Z3*Z3}.
Due to the multiplet merging, the CP-like transformation given by $S_2$ spontaneously
changes to the proper CP transformation.

\begin{table}[thbp]
\centering
\begin{tabular}{|c|cccc|}
  \hline
  & $\Psi^i$ & $S, T_1, T_2$ & $R_1$ & $R_2$
  \\
  \hline
  $\mathbb{Z}_3^{a^2 a'}\times\mathbb{Z}_3^{a^2 b}$
  & $\mathbf{1}_{20}\oplus\mathbf{1}_{21}\oplus\mathbf{1}_{22}$
  & $\mathbf{1}_{00}$
  & $\mathbf{1}_{01}$
  & $\mathbf{1}_{02}$
  \\
  \hline
\end{tabular}
\caption{The representations of the matter fields.}   \label{tab:model_Z3*Z3}
\end{table}

\subsubsection{CP-like symmetry for group with CIA: $\Delta(27) \times U(1) \to \mathbb{Z}_3^{a^2 a'} \times \mathbb{Z}^{a'}_3$}
If $\braket{R_1}$ and $\braket{R_2}$ develop non-zero real VEVs, the unbroken symmetry can change.
We consider the following VEVs
\begin{align}
\braket{R_1} = v_1, \ \ \braket{R_2} = v_2, \ \ \braket{T} = 0,
\ \ v_i \in \mathbb{R}.
\end{align}
This vacuum is realized if $m_R^2 <0$, $\lambda_3 >0$, and $v_1^2 - v_2^2 = -\frac{2\kappa_2}{\lambda_4} \braket{S}$.
This vacuum has two flat directions, but is free from tachyon.
The symmetry group of $\Delta(27) \times U(1)$ is broken to $\mathbb{Z}_3^{a^2 a'} \times \mathbb{Z}_3^{a'}$,
The irreducible decompositions are given as
\begin{align}
\mathbf{3} =& \mathbf{1}_{20}\oplus \mathbf{1}_{21}\oplus \mathbf{1}_{22},
&\mathbf{1}_{4} =& \mathbf{1}_{02},
&\mathbf{1}_{8} =& \mathbf{1}_{01},
&\mathbf{1}_{1} =& \mathbf{1}_{00},
&\mathbf{1}_{2} =& \mathbf{1}_{00},
\end{align}
where the left-hand side corresponds to the irreducible representations of $\Delta(27)$,
and the right-hand side $\mathbb{Z}_3^{a^2 a'} \times \mathbb{Z}_3^{a'}$.
The lower index of $\mathbf{1}_{ij}$ denotes the charges of $\mathbb{Z}_3^{a^2 a'} \times \mathbb{Z}_3^{a'}$,
\begin{align} \notag
a^2a':& \mathbf{1}_{ij} \to \omega^i \mathbf{1}_{ij},
\\
a':& \mathbf{1}_{ij} \to \omega^j \mathbf{1}_{ij}.
\end{align}
From Eq.~\eqref{eq:autoDelta54},
the automorphism of $\mathbb{Z}_3^{a^2 a'} \times \mathbb{Z}_3^{a'}$ is given as
\begin{align}
u^{\mathbb{Z}_3^{a^2 a'} \times \mathbb{Z}_3^{a'}} :
(a^2 a', a') \mapsto (aa'^2, a').
\end{align}
Thus it is CP-like transformation rather than physical (proper) CP,
although $\mathbb{Z}_3^{a^2 a'} \times \mathbb{Z}_3^{a'}$ has a CIA.

If $\braket{R_i}$ and $\braket{T_i}$ develops VEVs at the same time, the internal symmetry is broken further.
Suppose the vacuum is given by
\begin{align}
\braket{R_1} &\in \mathbb{R},
&\braket{R_2}& = 0,
&\braket{T_1}& = \braket{T_2}^*\neq 0.
\end{align}
This vacuum does not violate the CP-like symmetry, but break $\Delta(27) \times U(1)$ to $\mathbb{Z}_3^{a^2a'}$, which is the center of $\Delta(27)$.
The consistent automorphism is $a^2 a'$ to $a a^{\prime 2} = (a^2 a')^{-1}$.
Hence it becomes to a physical (proper) CP transformation.

\subsubsection{Other breaking patterns of $\Delta(27) \times U(1)$}
If there is an additional triplet scalar $U$ and it develops the VEV given by
\begin{align}
  \braket{U} \propto (\omega, 0, 0)^T,
\end{align}
$\Delta(27)$ is broken to $\mathbb{Z}_3$ generated by $a'$.
This vacuum is CP(-like) invariant.
Since $a'$ is fixed under $u_{\textrm{CP}}$, CP-like symmetry is still CP-like at the level of $\mathbb{Z}_3^{a'}$, and physical CP is violated.
Thus we can realize $\mathbb{Z}_3$ symmetric model with CP-like symmetry.

If the internal symmetry is completely broken down
the CP-like transformation can be regarded as a physical CP transformation due to a multiplet merging,
since all the fields should belong to the same representation of the trivial singlet.

\begin{figure}[htbp]
  \centering
  \includegraphics[width=0.9\textwidth]{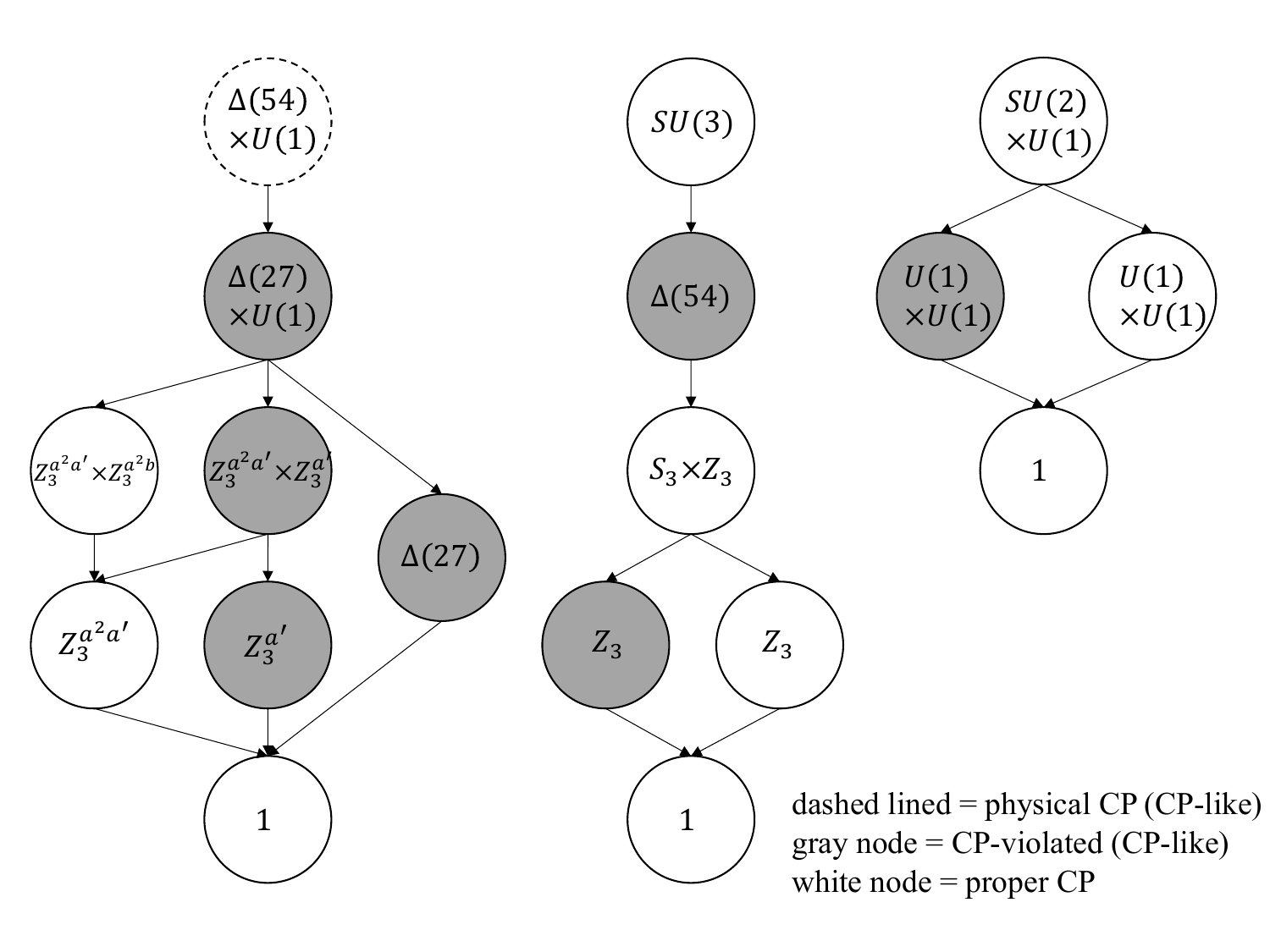}
  \caption{The relationship between symmetry breaking of the internal symmetry and
  the resultant property of the CP transformation.
  We only consider CP(-like) invariant vacuum.
  The gray node represents a CP violated (with CP-like symmetric) vacuum.
  The white node represents a physical CP symmetric vacuum,
  where
  the CP transformation is based on a CIA on the internal symmetry group.
  The broken lined node represents a physical CP (CP-like) symmetric vacuum.
  We only show a part of the breaking patterns for $\Delta(54) \times U(1)$, and
  there are other possible breaking patterns.
  We also show the breaking pattern of $SU(3)$ and $SU(2)\times U(1)$ models which are studied in the following subsections.}
  \label{fig:chart}
\end{figure}

We summarize the symmetry breaking patterns of $\Delta(54)\times U(1)$ model obtained aforementioned processes in Fig.~\ref{fig:chart}.
We find that various non-standard CP violation/restoration scenarios appear.
It is shown that both the proper CP and CP-like transformations can be mutually converted
due to the mechanisms of the multiplet splitting and merging,
which depends on which subgroup remains after the SSB.
We will study its cosmological and phenomenological implications.

\subsection{$SU(3)$ Model}

Here we show another example in a model with the proper CP symmetry for continuous symmetry group $SU(3)$.
Since the methods for calculations of CP asymmetry and analyses of the scalar potentials
are the same as in the previous subsection,
we only discuss a possible breaking pattern and their symmetries based on the group structures for simplicity.

We start with a model with symmetry group $G=SU(3)$ and a proper CP invariance.
The unique $\mathbb{Z}_2$ outer automorphism of $SU(3)$ 
is defined as 
\begin{align}
\label{eq:autoSU3}
u_{\textrm{CP}}: g \to h \in G, \textrm{\ such \ that \ } 
\rho_{\mathbf{3}}(g)^* = \rho_{\mathbf{3}}(h), 
\end{align}
where $\mathbf{3}$ is the fundamental representation of $SU(3)$, 
and $\rho_{\mathbf{3}}(g)$ is a $3\times 3$ special unitary matrix. 
Based on $u_\textrm{CP}$, 
we define a proper CP transformation as 
\begin{align}
\textrm{proper CP}: \mathbf{3} \to \mathbf{3}^*. 
\end{align}

\subsubsection{Spontaneous CP violation : $SU(3) \to \Delta(54)$}
If $SU(3)$ is broken down to $\Delta(54)$,
the proper CP transformation spontaneously changes to a CP-like transformation.
Given an explicit form of the group elements of $\Delta(54)$ as shown in Eqs. \eqref{eq:Delta54triplet1} and \eqref{eq:Delta54triplet2}, 
one can easily see that the fundamental representation $\mathbf{3}$ of $SU(3)$ 
is identified with $\mathbf{3}_2$ of $\Delta(54)$.
Thus the automorphism of $\Delta(54)$ is also determined from Eq.~\eqref{eq:autoSU3} as
\begin{align}
\label{eq:autoD54}
  u^{\Delta(54)}~:~(a, a', b, c) \mapsto (a^2, a'^2, b,c).
\end{align}
From the matrix representations for four doublet fields $\phi_{\mathbf{2}_{1,2,3,4}}$ (see App.~\ref{sec:A1}),
we immediately see that the corresponding CP transformation matrix
is given a CP-like form
for the multiplet $\Phi_{\mathbf{2}} = (\phi_{\mathbf{2}_{1}}, \phi_{\mathbf{2}_{2}}, \phi_{\mathbf{2}_{3}}, \phi_{\mathbf{2}_{4}})^T$,
\begin{align}
\textrm{CP-like}: \Phi_{\mathbf{2}} \to U_{\Phi_\mathbf{2}} \Phi_{\mathbf{2}}^*,
\quad \quad
  U_{\Phi_\mathbf{2}} = \begin{pmatrix}
    \mathbf{1} & \mathbf{0} & \mathbf{0} & \mathbf{0} \\
    \mathbf{0} & \mathbf{0} & \mathbf{1} & \mathbf{0} \\
    \mathbf{0} & \mathbf{1} & \mathbf{0} & \mathbf{0} \\
    \mathbf{0} & \mathbf{0} & \mathbf{0} & \mathbf{1} \\
  \end{pmatrix},
\end{align}
While the triplets $\mathbf{3}_{1,2}$, and two singlets $\mathbf{1}_{0,1}$
properly transform into their complex conjugate representations under the CP transformation.
From the fact that irreducible decompositions of the adjoint representation $\mathbf{8}$ of $SU(3)$ are
given as $\mathbf{8} = \mathbf{2}_1 \oplus \mathbf{2}_2 \oplus \mathbf{2}_3 \oplus \mathbf{2}_4$ (see Eq.~\eqref{eq:33}),
we also see that
the original CP transformation
exchanges two different representations of $\phi_{\mathbf{2}_{2}}, \phi_{\mathbf{2}_{3}}$ in $\Delta(54)$.
Therefore the CP transformation spontaneously changes a CP-like transformation due to a multiplet splitting.

\subsubsection{Spontaneous CP restoration : $\Delta(54) \to S_3 \times \mathbb{Z}_3$}
Let us consider the spontaneous breaking of $\Delta(54) \to S_3 \times \mathbb{Z}_3$.
All group elements of $S_3$ and $\mathbb{Z}_3$ are given as $\{e,b,c,b^2,bc,b^2c \}$,
and $\{e, aa'^2, a^2a' \}$, respectively.
There are three conjugacy classes in $S_3$:
\begin{align}
C_{1}:& \{ e \}, & C_{2}:& \{ b, b^2 \}, & C_{3}:& \{ c, bc, b^2c \},
\end{align}
and there are two singlets, and one doublet.
We denote all the irreducible representations
of $S_3 \times \mathbb{Z}_3$ as
the three doublets $\mathbf{2}_i$ and six singlets $\mathbf{1}_i$, $\mathbf{1}'_i$,
where $i=0,1,2$ represents a $\mathbb{Z}_3$ charge.
From Eq.~\eqref{eq:autoD54},
we see that the original CP transformation becomes to a CIA of $S_3 \times \mathbb{Z}_3$.
Let us also see the irreducible decompositions;
\begin{align} \notag
\mathbf{1}_{0} &= \mathbf{1}_0, & \mathbf{1}_{1} &= \mathbf{1}'_0,
  \\ \notag
\mathbf{2}_1 &= \mathbf{2}_{0}, & \mathbf{2}_2 &= \mathbf{2}_0,
  \\ \notag
\mathbf{2}_3 &= \mathbf{2}_0, & \mathbf{2}_4 &= \mathbf{1}_0 \oplus \mathbf{1}_0',
  \\ \notag
\mathbf{3}_{1} &= \mathbf{1}_1 \oplus \mathbf{2}_1, & \mathbf{3}_{2} &= \mathbf{1}'_1 \oplus \mathbf{2}_1,
\\
\mathbf{3}_{1}^* &= \mathbf{1}_2 \oplus \mathbf{2}_2, & \mathbf{3}_{2}^* &= \mathbf{1}'_2 \oplus \mathbf{2}_2,
\end{align}
where the left-hand side corresponds to the irreducible representations of $\Delta(54)$, and the right-hand side $S_3 \times \mathbb{Z}_3$.
This result is also consistent with the proper CP transformation.
Since two mutually exchanged fields $\phi_{\mathbf{2}_2}$ and $\phi_{\mathbf{2}_3}$ by the CP-like transformation in $\Delta(54)$
eventually belong to the same representation of $\mathbf{2}_0$ in $S_3 \times \mathbb{Z}_3$.
Thus the physical CP symmetry can spontaneously emerge from the CP-like symmetry.

\subsubsection{Spontaneous CP violation : $S_3 \times \mathbb{Z}_3 \to \mathbb{Z}'_3$}
Next, we consider the breaking of $S_3 \times \mathbb{Z}_3 \to \mathbb{Z}_3$.
$S_3 \times \mathbb{Z}_3$ has two different subgroups
of $\mathbb{Z}_3: \{e, aa'^2, a^2a' \}$, and $\mathbb{Z}'_3: \{e, b, b^2 \}$.
It is important to notice that while these two subgroups are both represented by $\mathbb{Z}_3$,
the properties of the corresponding automorphism vary depending on the generators of each subgroup. 

As for $\mathbb{Z}_3: \{e, aa'^2, a^2a' \}$, the automorphism in Eq.~\eqref{eq:autoD54} correspond
to the (class-inverting) outer automorphism, so that the corresponding CP transformation is the proper CP transformation.

On the other hand, in the case of $\mathbb{Z}'_3: \{e, b, b^2 \}$,
the automorphism in Eq.~\eqref{eq:autoD54} correspond
to the identity map (non-CIA),
so that the corresponding CP transformation is the CP-like transformation.
Let us also see the irreducible decompositions of $S_3 \times \mathbb{Z}_3 \to \mathbb{Z}'_3$;
\begin{align}
\mathbf{1}_{i} &= \mathbf{1}_0, & \mathbf{1}'_{i} &= \mathbf{1}_0, & \mathbf{2}_i &= \mathbf{1}_1 \oplus \mathbf{1}_2,
\ \ \ (i=0,1,2)
\end{align}
where the left-hand side corresponds to the irreducible representations of $S_3 \times \mathbb{Z}_3$, and the right-hand side $\mathbb{Z}'_3$.
Thus two components of a doublet, which are connected by the CP transformation,
split into two fields of $\phi_{\mathbf{1}_1}$ and $\phi_{\mathbf{1}_2}$.
It is interesting to investigate a relation between the physical CP transformation ($CP_\textrm{phys}$) at the level of $\mathbb{Z}'_3$ and 
the proper CP transformation at the level of $S_3 \times \mathbb{Z}_3$.
Since $\mathbb{Z}'_3$ has a CIA, $CP_\textrm{phys}$ is given by a proper CP transformation 
which should correspond to the outer automorphism of 
\begin{align} 
\label{eq:physCP_Z3}
u^\textrm{phys}~:~b \mapsto b^2. 
\end{align}
This automorphism $u^\textrm{phys}$ is obviously different from $u^{\Delta(54)}$ in Eq.~\eqref{eq:autoD54}. 
We notice, however, that a CP transformation of $CP_\textrm{phys} = \rho(c) \circ CP$, 
given by a product of the original CP transformation ($CP$) and a broken element of $c \in S_3 \times \mathbb{Z}_3\backslash \mathbb{Z}'_3$, 
satisfies the following relation,  
\begin{align} \notag
CP_\textrm{phys}^{-1} \circ \rho(b) \circ CP_\textrm{phys} =& 
CP^{-1} \circ \rho(c^{-1} \, b \, c) \circ CP 
\\ \notag
=& CP^{-1} \circ \rho(b^2) \circ CP 
\\ \notag
=& \rho(u^{\Delta(54)}(b^2)) 
\\
=& \rho(b^2).
\end{align}
This CP transformation is consistent with the CIA, $u^\textrm{phys}$ in Eq.~\eqref{eq:physCP_Z3}.
Thus in this scenario of the spontaneous change of a proper CP to a CP-like transformation 
via SSB of $S_3 \times \mathbb{Z}_3 \to \mathbb{Z}'_3$
the physical (proper) CP transformation $CP_\textrm{phys}$ is actually 
given by a product of a CP-like transformation and a broken element, 
so that the physical CP symmetry is spontaneously broken by the broken element and 
the CP asymmetry should be explicitly proportional to the VEV, 
while there exists CP-like symmetry in the broken phase.  

The symmetry breaking patterns of $SU(3)$ and CP are summarized in Fig.~\ref{fig:chart}.

\subsection{$SU(2) \times U(1)$}

We show yet another simple example in which
a CP-like symmetric model with continuous internal symmetry group can be
derived from a proper CP symmetric model.
As an example, we consider $SU(2)\times U(1)$ symmetric model,
and introduce a scalar field $S = (S_1, S_2, S_3)^T$, and a Dirac fermion field $\Psi^q = (\Psi^q_1,\Psi^q_2,\Psi^q_3)^T$ in the triplet representation of $SU(2)$.
$q$ denotes $U(1)$ charge and $S$ is neutral under $U(1)$.
A group action on a triplet field $\Psi^q$ (and $S$) is represented as
\begin{align}
\Psi^q \to \Psi'^q =& \rho_{\Psi^q}(\vec{c}, \theta) \Psi^q,
\quad \quad
\rho_{\Psi^q}(\vec{c}, \theta)
= e^{i \vec{\sigma} \cdot \vec{c}} e^{iq \theta},
\\ \notag
\vec{c}=&(c_1,c_2,c_3),  \quad \quad \theta, c_i \in \mathbb{R}.
\end{align}
where $\vec{\sigma}=(\sigma_1, \sigma_2, \sigma_3)$
is the Pauli vector in a three-dimensional representation,
which is given in terms of the $\mathfrak{su}(2)$ generators,
\begin{align}
\sigma_1 &= \frac 1{\sqrt{2}} \begin{pmatrix}
0 & 1 & 0\\
1 & 0 & 1\\
0 & 1 & 0\\
\end{pmatrix},
&
\sigma_2& = \frac 1{\sqrt{2}}
\begin{pmatrix}
0 & -i & 0\\
i & 0 & -i\\
0 & i & 0\\
\end{pmatrix},
&
\sigma_3& =
\begin{pmatrix}
1 & 0 & 0\\
0 & 0 & 0\\
0 & 0 & -1\\
\end{pmatrix}.
\end{align}
A proper CP transformation of $\Psi^q$ is given as
\begin{align}
\Psi^q(x) \to - U C \Psi^{q*}(\tilde{x}),
\label{eq:SU(2)CP}
\end{align}
where $U$ is an unitary matrix which acts on $SU(2)$ indices, and $C$ denotes the charge conjugation matrix.
The complex conjugation automorphism of $SU(2)\times U(1)$ is represented as a direct product of the two automorphisms as
$u_{\textrm{CP}}^{SU(2)\times U(1)} = u_{\textrm{CP}}^{SU(2)} \times u_{\textrm{CP}}^{U(1)}$,
and the consistency condition for $SU(2)$ is given as, 
\begin{align}
\label{eq:SU2consitency}
U \left( e^{i \vec{\sigma} \cdot \vec{c}} \right)^* U^\dagger = e^{i \vec{\sigma} \cdot \vec{c^\prime}}
= e^{i \vec{\sigma} \cdot R \cdot \vec{c}},
\end{align}
where $R$ is an $O(3)$ matrix~\cite{Grimus:1995zi}.
For $U={\bf 1}_{3\times3}$ we obtain $R= \textrm{diag} (-1,1,-1)$. 
Then the corresponding automorphism $u_{\textrm{CP}}^{SU(2)}$ is given as 
\begin{align}
\label{eq:autoSU2}
u_{\textrm{CP}}^{SU(2)} : 
\sigma_1c_1 \mapsto \sigma_1c_1'= -\sigma_1c_1,  \quad \ 
\sigma_2c_2 \mapsto \sigma_2c_2'= \sigma_2c_2, \quad \ 
\sigma_3c_3 \mapsto \sigma_3c_3'= -\sigma_3c_3.
\end{align}

The mass and interaction terms for $\Psi^q$ for a $SU(2)\times U(1)$ symmetric model 
can be written as 
\begin{align}
\mathcal{L}_{\Psi_q} = -m \bar{\Psi}^q \Psi^q 
    + y \bar{\Psi}^q (S \otimes \Psi^q)_{\mathbf{3}}
    + \frac{\lambda }{\Lambda} (S\otimes S)_{\mathbf{5}} (\bar{\Psi}^q \otimes \Psi^q)_{\mathbf{5}}
    + (h.c.). 
    \label{eq:yuk_su2}
\end{align}
where we add non-renormalizable terms for later convenience.

\subsubsection{CP-like Symmetry: $SU(2) \times U(1) \to U(1) \times U(1)$}
 
The group $SU(2)$ is spontaneously broken to $U(1)$ by a nonzero VEV for a triplet scalar $S$.
Let us assume that the VEV is given as
\begin{align}\label{eq:VEV1}
\braket{S} = (v, 0, v)^T \ \ (v \in \mathbb{R}),
\end{align}
the vacuum is invariant under the CP transformation with $U = \mathbf{1}_{3\times 3}$.
The unbroken $U(1)$ is generated by $\sigma_2$, and we refer to it as $U(1)_{\sigma_2}$.
When $c_1=c_3=0$, from the consistency condition in Eq.~\eqref{eq:SU2consitency}
we obtain 
\begin{align}
CP^{-1} \circ (e^{i \vec{\sigma} \cdot \vec{c}})^* \circ CP = e^{i \vec{\sigma} \cdot R \cdot \vec{c} }
\quad \quad \Rightarrow \quad \quad CP^{-1} \circ (e^{i \sigma_2c_2})^* \circ CP = e^{i \sigma_2 c_2}.
\end{align}
Thus we see that the proper CP transformation at the level of $SU(2)$ becomes
to a CP-like transformation at the level of $U(1)_{\sigma_2} \in SU(2)$.  
The triplet fields are decomposed into the eigenstates for the remaining symmetry of $U(1)_{\sigma_2} \times U(1)$.
The eigenstates are given by
\begin{align}
\Psi_{+}^q &= -\frac{i}2 \Psi_1^q + \frac{1}{\sqrt{2}}\Psi_2^q + \frac{i}{2} \Psi_3^q,
\nonumber \\
\Psi_-^q &= \frac{i}2 \Psi_1^q + \frac{1}{\sqrt{2}}\Psi_2^q - \frac{i}{2} \Psi_3^q,
\nonumber
\\
\Psi_0^q &= \frac{\Psi_1^q + \Psi_3^q}{\sqrt{2}},
\label{eq:eigenstates_sigma2}
\end{align}
as well as $S$.
The lower index indicates the $U(1)_{\sigma_2}$ charge.
For a multiplet field of these eigenstates $\tilde{\Psi}^q = (\Psi_{+}^q, \Psi_{-}^q, \Psi_{0}^q)^T$,
the original proper CP transformation (Eq.~\eqref{eq:SU(2)CP}) can be represented as
a CP-like form,
\begin{align}
\textrm{CP-like} : \tilde{\Psi} \to - U_{\tilde{\Psi}} C \tilde{\Psi}^*,
\quad \quad
U_{\tilde{\Psi}} = \begin{pmatrix}
0 & 1 &0 \\
1 & 0 & 0 \\
0 & 0 & 1
\end{pmatrix},
\end{align}
which indicates that the corresponding automorphism of the generators for $U(1)_{\sigma_2} \times U(1)$ is give as
\begin{align}
u^{U(1)_{\sigma_2} \oplus U(1)} :
(c_2, \theta) \mapsto (c_2, -\theta).
\end{align}
It should be noted that this automorphism is consistent with the CP-like transformation defined in Eq.~\eqref{eq:U1xU1}.

As for the vacuum stability of the model,
one can investigate the following general renormalizable scalar potential,
\begin{align}
\mathcal{V}(S)
=& \frac{m^2}2 S^{\dagger} \cdot S + \frac{\mu^2}{2} (S \otimes S)_0
\notag \\
&
+ \frac{\lambda_1}4 \left[ (S\otimes S)_0 \right]^2
+ \frac{\lambda_2}4 (S^\dagger \cdot S) (S \otimes S)_0
\notag \\
&
+ \frac{\lambda_3}4 (S^\dagger\cdot S)^2
+ \frac{\lambda_4}4 (S^* \otimes S^*)_0 (S \otimes S)_0
+ (h.c.),
\label{eq:potential}
\end{align}
where $X^\dagger \cdot Y$ and $(X\otimes Y)_0$ are the two $SU(2)$ invariant tensor products
\begin{align}
X^\dagger \cdot Y = x_1^* y_1 + x_2^* y_2 + x_3^* y_3,
\quad \quad \quad
(X\otimes Y)_0 = x_1 y_3 - x_2 y_2 + x_3 y_1,
\end{align}
for two triplets of $X = (x_1, x_2, x_3)^T$ and $Y = (y_1,y_2,y_3)^T$.
As a result, there exists a parameter region where the CP-like symmetric vacuum given in Eq.~\eqref{eq:VEV1} is stable.
Therefore we show that a CP-like symmetric model with continuous internal symmetry can be naturally derived from a proper CP symmetric model.

While we use a basis where the unbroken group is represented by $U(1)_{\sigma_2}$,
all $U(1)$ subgroups of $SU(2)$ are completely equal,
as their matrix representations can be transformed through a basis change.
Hence breaking to one or the other subgroup should have the same phenomenological consequence.
Given that the CP transformation with $U = \mathbf{1}_{3\times 3}$ properly maps $U(1)_{\sigma_3}$ to its complex conjugate,
it might be inferred that the model can exhibit a physical CP symmetry for $U(1)_{\sigma_3}$ invariant vacuum 
after a basis change, suggesting that a CP-like symmetric vacuum for continuous groups might not exist.
This apparent puzzle can be resolved by considering the basis change of the CP transformation matrix $U$ simultaneously. 

A basis change for $\Psi$ is generally represented by $3 \times 3$ unitary matrix $V$ as
\begin{align}
\Psi' =& V \Psi, \\ 
\rho_{\Psi'}(g) =& V \rho_\Psi(g) V^\dagger, \quad \quad g \in SU(2).  
\end{align}
We can choose $V U(1)_{\sigma_2}V^\dagger = U(1)_{\sigma_3}$, where $\Psi'$ corresponds to a $U(1)_{\sigma_3}$ eigenstate. 
We refer to the original basis for $\Psi$ as the $\sigma_2$ basis, and the new basis for $\Psi'$ as the $\sigma_3$ basis, respectively.
Then the CP transformation is also changed to
\begin{align}\label{eq:U2}
\textrm{CP}: \Psi' \to - V U C \Psi^* = - V U V^T C \Psi'^* \equiv - U' C\Psi'^*,  \quad \quad 
U' = 
\begin{pmatrix} 0 & 0 & 1 \\ 0 & 1 & 0 \\ 1 & 0 & 0 
\end{pmatrix}.
\end{align}
The CP transformation in the $\sigma_3$ basis is accompanied by a unitary matrix $U' = VUV^T= VV^T$. 
Therefore, from the consistency condition the automorphism of unbroken $U(1)_{\sigma_3}$ associated with $U'$ is obtained as
\begin{align}\label{eq:CP-likeU'}
U' U(1)_{\sigma_3}^* U'^\dagger =& V V^T (V U(1)_{\sigma_2}V^\dagger)^* (V V^T )^\dagger
    \notag
    \\
    =& V U(1)_{\sigma_2} V^\dagger  
    \notag
    \\
    =& U(1)_{\sigma_3} \neq U(1)_{\sigma_3}^*.
\end{align}
Thus the CP transformation in the $\sigma_3$ basis is also CP-like for $U(1)_{\sigma_3}$, and not the physical one.
The automorphism group associated with the CP transformation remains invariant under this basis change.
Here we note that the VEV for a triplet field $S' \equiv V S$ (in the $\sigma_3$ basis) 
should be given as 
\begin{align}\label{eq:VEV2}
\braket{S'} = (0, \sqrt{2}v, 0)^T \ \ (v \in \mathbb{R}),
\end{align}
to be invariant under $U(1)_{\sigma_3}$ transformations, 
which is apparently different from the form of $\braket{S}=(v, 0, v)^T$ in the original basis,
although the physical results do not depend on the basis.

To clearly observe the phenomenological consequences in the CP-like symmetric vacuum, 
let us examine the mass spectra for the model given in Eq.~\eqref{eq:yuk_su2}. 
For $\braket{S}=(v,0,v)^T$, we obtain the mass matrix $M$ for $\Psi^q_i$, 
\begin{align}
    M 
    &= 
    \begin{pmatrix}
        m - \frac{\lambda v^2}{3 \Lambda} - \frac{\lambda^* v^{*2}}{3 \Lambda} & -yv + y^* v^* & -\frac{\lambda v^2}{\Lambda}-\frac{\lambda^* v^{*2}}{\Lambda}\\
        yv - y^* v^* & m + \frac{2\lambda v^2}{3 \Lambda} + \frac{2\lambda^* v^{*2}}{3 \Lambda} & -yv+y^* v^*\\
        -\frac{\lambda v^2}{\Lambda}-\frac{\lambda^* v^{*2}}{\Lambda} & yv-y^* v^* & m- \frac{\lambda v^2}{3 \Lambda} - \frac{\lambda^* v^{*2}}{3 \Lambda}\\
    \end{pmatrix}
    \notag
    \\
    &=
    \begin{pmatrix}
      m - \frac{2\lambda v^2}{3 \Lambda} & 0 & -\frac{2\lambda v^2}{\Lambda}\\
      0 & m + \frac{4\lambda v^2}{3 \Lambda} & 0\\
      -\frac{2\lambda v^2}{\Lambda} & 0 & m- \frac{2\lambda v^2}{3 \Lambda}\\
  \end{pmatrix}.
\end{align}
The relative sign of $yv$ and $y^*v^*$ comes from anti-symmetric nature of $(\mathbf{3}\otimes \mathbf{3})_{\mathbf{3}}$.
Since the model is assumed to be invariant under $\textrm{CP}$ with $U={\bf 1}_{3\times3}$,
the coupling constants $y$ and $\lambda$ as well as $v$ are real.
Using the mass eigenstates in Eq.~\eqref{eq:eigenstates_sigma2}, 
we obtain the mass eigenvalues,
\begin{align}
m_{\Psi_0^q} = m - \frac{8\lambda v^2}{3 \Lambda}, \ \ m_{\Psi_\pm^q} = m + \frac{4\lambda v^2}{3 \Lambda}.
\end{align}
As shown above, two charged states $\Psi_\pm^q$ are degenerate due to the CP-like symmetry,
since the CP transformation with $U={\bf 1}_{3\times3}$ exchanges ${\Psi_+^q}$ and $({\Psi_-^q})^*$ (See Eq.~\eqref{eq:eigenstates_sigma2}).

Since there is no other symmetry, the theory should exhibit physical CP violation.
To explicitly show that this model indeed violates physical CP symmetry, 
we can examine the complex phases of the Yukawa couplings.
In the broken phase, the effective Yukawa interactions in Eq.~\eqref{eq:yuk_su2} can be expressed in terms of 
$U(1)_{\sigma_2}$ eigenstates in Eq.~\eqref{eq:eigenstates_sigma2} as 
\begin{align}
  \mathcal{L}_{yukawa} =& i y \{ S_0 (-\bar{\Psi}_+^q \Psi_+^q + \bar{\Psi}_-^q \Psi_-^q) 
  + S_- (\bar{\Psi}_0^q \Psi_+^q - \bar{\Psi}_-^q \Psi_0^q)
  + S_+ (\bar{\Psi}_+^q \Psi_0^q - \bar{\Psi}_0^q \Psi_-^q)
  \}
  \notag
  \\
  &+ \frac{\sqrt{2} \lambda v}{\Lambda} \{ S_- (\bar{\Psi}_0^q \Psi_+^q + \bar{\Psi}_-^q \Psi_0^q) + S_+ (\bar{\Psi}_+^q \Psi_0^q  + \bar{\Psi}_0^q \Psi_-^q)
  \}
  \notag\\
  &- \frac{2\sqrt{2}}3 \frac{\lambda v}{\Lambda} S_0
  (\bar{\Psi}_+^q \Psi_+^q - 2\bar{\Psi}_0^q \Psi_0^q
  +\bar{\Psi}_-^q \Psi_-^q)
  +(h.c.).
\end{align}
It is obvious that a part of complex phases of the Yukawa couplings must remain by field redefinition.
For instance, it is impossible to simultaneously rotate away the following two complex phases in the Yukawa couplings,  
\begin{align}
\mathcal{L}_{yukawa}  \ni
\left( -i y- \frac{2\sqrt{2}}3 \frac{\lambda v}{\Lambda} \right) S_0 \bar{\Psi}_+^q \Psi_+^q
+ \left( i y - \frac{2\sqrt{2}}3 \frac{\lambda v}{\Lambda} \right) S_0 \bar{\Psi}_-^q \Psi_-^q.
\end{align} 
Therefore, a vacuum that is invariant under the CP-like transformation necessarily violates physical CP symmetry.
It is remarkable that the imaginary part of the Yukawa couplings can be dynamically generated 
through the CG coefficients even in the context of continuous symmetry groups, 
while all the couplings and the VEV can be real parameters in terms of the triplet fields in the symmetric phase.

\subsubsection{Proper CP Symmetry: $SU(2) \times U(1) \to U(1) \times U(1)$}

If the vacuum is given by $\braket{S} = (0, \sqrt{2}v, 0)^T$, 
$U(1)_{\sigma_3}$ symmetry is realized.
The mass matrix $M$ for $\Psi_i^q$ takes a diagonal form as, 
\begin{align}
    M = 
    \begin{pmatrix}
        m - 2\sqrt{2}  y v +\frac{4\lambda}{3\Lambda} v^2& 0 & 0\\
        0 & m -\frac{8\lambda}{3\Lambda} v^2 & 0\\
        0 & 0 & m + 2\sqrt{2} y v +\frac{4\lambda}{3\Lambda} v^2\\
    \end{pmatrix}.
    \label{eq:mass_split}
\end{align}
We obtain mass splitting because the CP transformation with $U={\bf 1}_{3\times3}$
does not relate among $\Psi_{i}^q$ with different $U(1)$ charges.
Since $v$ is assumed to be real for the CP invariance, 
there is no complex phase in any of the couplings in the $\Psi_{i}^q$ basis, 
and the physical CP symmetry is preserved in the broken phase.
Thus these two vacua of $\braket{S} = (v, 0, v)^T$ and $\braket{S} = (0, \sqrt{2}v, 0)^T$ are physically distinct, 
even though the representations for both VEVs can be transformed through a basis change. 

Let us also investigate the basis dependence of the CP transformation matrix.
As suggested by the basis change in Eq.~\eqref{eq:U2}, 
we consider $U=U'$ as an example of a different definition of a charge conjugation CP transformation for $SU(2)$, 
which we refer to as CP$'$ transformation.
Since CP$'$ properly maps $U(1)_{\sigma_2}$ to its complex conjugate, acting as a physical CP transformation, 
we assume $\braket{S} = (v, 0, v)^T$ with $v \in \mathbb{R}$.
The $\textrm{CP}'$ invariance requires $y$ to be purely imaginary because of anti-symmetric nature of the tensor product of $(\mathbf{3}\otimes \mathbf{3})_{\mathbf{3}}$.
Under this symmetry, 
the three mass eigenvalues are provided by
\begin{align}
    m - \frac{8\lambda v^2}{3 \Lambda}, m + \frac{4\lambda v^2}{3 \Lambda} \pm \sqrt{2}i (y-y^*)v.
\end{align}
This results in a splitting of  the mass degeneracy. 
By replacing $\textrm{Im}\, y$ by $y$, we obtain the same mass eigenvalues as those given by Eq. \eqref{eq:mass_split}.
Thus, we find that this model with CP$'$ and $\braket{S} = (v, 0, v)^T$ is equivalent to the model with CP and $\braket{S} = (0, \sqrt{2}v, 0)^T$,
as expected from the basis change from $\sigma_2$ to $\sigma_3$. 

The symmetry breaking patterns of $SU(2)\times U(1)$ and CP are also summarized in Fig.~\ref{fig:chart}.

\section{Conclusion}
\label{sec:Conclusion}

In this paper we have extensively discussed the general CP transformations in quantum field theory with internal global symmetry including continuous groups.
A special attention has been paid to the CP-like transformation.
Since the CP-like transformation exchanges a particle and a different (anti)particle,
the physical CP symmetry is in general violated even if the theory has a CP-like symmetry~\cite{Chen:2014tpa}.
Despite its interesting property,
the phenomenological aspects of the CP-like symmetric models have not been fully investigated so far.
To address physical consequences of the CP-like symmetry
we have computed the various scattering amplitudes.
As a result it has been found that
physical scattering processes that exhibit a physical CP asymmetry
can also generate the matter/antimatter asymmetry.
We have classified these CP-violating scattering amplitudes
based on the representations of the asymptotic states under the CP-like transformation,
by which a required condition for generating particle number asymmetry can be easily obtained.
One striking feature of the CP-like symmetric model is that,
while there exist physical processes that can exhibit particle number generations,
the resultant particle number asymmetries are constrained by the CP-like symmetry.
Assuming that the two different particles of $\psi_i$ and $\bar{\psi}_j$ are related to each other via the CP-like transformation,
we have explicitly calculated particle number generations of both particles $N_{\psi_i}$ and $N_{\psi_j}$ from
a CP-violating scattering process, where it has been shown that the following relation about
the number violations should hold in general,
\begin{align}
N_{\psi_i} + N_{\psi_j} = 0.
\end{align}
This relation is analogous to other particle number generation processes
such as the GUT baryogenesis \cite{Yoshimura:1978ex} and the sphaleron process \cite{Kuzmin:1985mm}.
We also have found CP-like eigenstate which has no CP-like partner, and hence, its particle number is not related to other antiparticles nor conserved.
Existence of these two classes of states is crucial for CP violation in CP-like symmetric models.
The mechanism for generating particle number asymmetry might be useful for phenomenological
purposes, e.g. for baryogenesis and asymmetric DM,
which we will study in the future work.

We also have discussed a dynamical origin of the CP-like symmetry and the fate of the CP-like symmetry.
It has turned out that a proper CP and a CP-like transformation can be mutually converted
through the SSB,
where it is found that the mechanisms of the multiplet splitting/merging play an important role.
As we have pointed out that the emergence of CP-like symmetry is not directly related to the absence of the CIA,
CP-like symmetric models with a wider class of internal symmetry groups can be
obtained from proper CP symmetric models through the SSB.
Various examples of spontaneous CP violation/restoration have been explicitly shown,
in which we have provided a simple way to construct a CP-like symmetric model
with continuous symmetries from an underlying proper CP symmetric theory via the SSB.
Our results suggest that some quantum charge under a continuous symmetry
may be defined to be invariant under the CP-like transformation.
The CP-like symmetric model may shed light on the origin of the physical CP violation
and simultaneously provides us with novel frameworks for new physics.

Our analyses can be extended to other symmetries such as local (gauge) symmetry and the modular symmetry.
The modular symmetry includes CP transformation in its nature,
which also acts on the flavor symmetry as an automorphism.
It would be also possible to generalize the non-CP automorphisms.
The above generalizations and applications to phenomenology may be interesting and will be studied elsewhere.

\section*{Acknowledgements}

H. O. is supported in part by JSPS KAKENHI Grants No. 21K03554 and No. 22H00138.
S. U is supported in part by JSPS KAKENHI Grant No. 22K14039.
The authors thank the Yukawa Institute for Theoretical Physics at Kyoto University.
Discussions during the YITP workshop YITP-W-23-10 on "Progress in Particle Physics 2023"
were useful to complete this work.

\appendix

\section{Group Property of $\Delta(54)$}\label{sec:A1}

The discrete group $\Delta(54)$
is isomorphic to $\Delta(27) \rtimes \mathbb{Z}_2 \cong ((\mathbb{Z}_3 \times \mathbb{Z}_3)\rtimes \mathbb{Z}_3) \rtimes \mathbb{Z}_2$.
It means all elements of $\Delta(54)$ can be represented by $a^i a'^j b^k c^l$ ($i,j,k = 0,1,2$ and $l = 0,1$),
where $a, a', b$ corresponds to the elements of $\Delta(27)$,
and $c$ corresponds to the element of $\mathbb{Z}_2$.
The commutation relations including $c$ are given as $ca = a a' c$, $ca' = a'^2 c$, and $cb = b^2 c$,
and those for $\Delta(27)$ are given in Sec.~\ref{sec:Delta27}.
There are 10 conjugacy classes in $\Delta(54)$:
\begin{align}
\notag
  C_{1a} :& \{e\},~~C_{3e} : \{a a'^2 \},~~~~C_{3f} : \{a^2 a' \},
  \\ \notag
  C_{3a} :& \{b, b^2, a^2 a' b, a a'^2 b, a a'^2 b^2, a^2 a' b^2 \},
  \\ \notag
  C_{3b} :& \{a, a', a^2, aa', a'^2, a^2 a'^2 \},
  \\ \notag
  C_{3c} :& \{a^2 b, a'^2b , a a' b, a b^2, a'b^2 , a^2 a'^2 b^2\},
  \\ \notag
  C_{3d} :& \{ab, a'b, a^2 a'^2 b, a^2 b^2, a a' b^2, a'^2 b^2 \},
  \\ \notag
  C_{2a} :& \{c, a'c, bc, a'^2c, b^2c, ab^2c, a a'bc, a^2b^2c, a^2 a'^2 b c \},
  \\ \notag
  C_{6a} :& \{a c, a a' c, a a'^2 c, a'bc, a^2 b c , a'^2 b^2c, a a'^2 b c, a a'^2 b^2 c, a^2 a'^2 b^2 c \},
  \\
  C_{6b} :& \{a^2c, a b c, a^2 a' c , a'b^2c, a'^2 b c, a^2 a'^2 c, a a' b^2 c, a^2 a'b c,  a^2 a' b^2 c\}.
\end{align}
Therefore $\Delta(54)$ has 10 irreducible representations of
two singlets $\mathbf{1}_{0,1}$, four doublets $\mathbf{2}_{1,2,3,4}$ and four triplets $\mathbf{3}_{1,2}, \mathbf{3}^*_{1,2}$.
The character table of $\Delta(54)$ is summarized in Tab.~\ref{tab:character_D54}.
\begin{table}[thbp]
	\centering
	\begin{tabular}{lrrrrrrrrrr}
		& $C_{1a}$ &$C_{3a}$ & $C_{3b}$ & $C_{3c}$ & $C_{3d}$ & $C_{2a}$ & $C_{6a}$ & $C_{6b}$ & $C_{3e}$ & $C_{3f}$ \\
		degeneracy & $1$ &$6$ & $6$ & $6$ & $6$ & $9$ & $9$ & $9$ & $1$ & $1$ \\
	 $\Delta(54)$ & $e$ &$b$ & $a'$ & $ba'b$ & $ba'$ & $c$ & $ac$ & $a^2c$ & $a a'^2$ & $a^2a'$ \\
	 \hline
	 $\mathbf{1}_0$  & $1$ &$1$ & $1$ & $1$ & $1$ & $1$ & $1$ & $1$ & $1$ & $1$ \\
	 $\mathbf{1}_1$  & $1$ &$1$ & $1$ & $1$ & $1$ & $-1$ & $-1$ & $-1$ & $1$ & $1$ \\
	 $\mathbf{2}_1$  & $2$ &$-1$ & $2$ & $-1$ & $-1$ & $0$ & $0$ & $0$ & $2$ & $2$ \\
	 $\mathbf{2}_2$  & $2$ &$-1$ & $-1$ & $-1$ & $2$ & $0$ & $0$ & $0$ & $2$ & $2$ \\
	 $\mathbf{2}_3$  & $2$ &$-1$ & $-1$ & $2$ & $-1$ & $0$ & $0$ & $0$ & $2$ & $2$ \\
	 $\mathbf{2}_4$  & $2$ &$2$ & $-1$ & $-1$ & $-1$ & $0$ & $0$ & $0$ & $2$ & $2$ \\
	 $\mathbf{3}_1$  & $3$ &$0$ & $0$ & $0$ & $0$ & $1$ & $\omega$ & $\omega^2$ & $3\omega$ & $3\omega^2$ \\
	 $\mathbf{3}^*_1$  & $3$ &$0$ & $0$ & $0$ & $0$ & $1$ & $\omega^2$ & $\omega$ & $3\omega^2$ & $3\omega$ \\
	 $\mathbf{3}_2$  & $3$ &$0$ & $0$ & $0$ & $0$ & $-1$ & $-\omega$ & $-\omega^2$ & $3\omega$ & $3\omega^2$ \\
	 $\mathbf{3}^*_2$  & $3$ &$0$ & $0$ & $0$ & $0$ & $-1$ & $-\omega^2$ & $-\omega$ & $3\omega^2$ & $3\omega$ \\
	 \hline
	\end{tabular}
	\caption{Character table of $\Delta(54)$.}
	\label{tab:character_D54}
\end{table}

The faithful representation is given by the triplets.
A matrix representation of the triplets is given by
\begin{align}\label{eq:Delta54triplet1}
  \rho_{\mathbf{3}_1}(a) &= \rho_{\mathbf{3}_2}(a) =
  \begin{pmatrix}
    \omega & 0 & 0\\
    0 & \omega^2 & 0\\
    0 & 0 & 1\\
  \end{pmatrix},~~&
  \rho_{\mathbf{3}_1}(a') &= \rho_{\mathbf{3}_2}(a') =
  \begin{pmatrix}
    1 & 0 & 0\\
    0 & \omega & 0\\
    0 & 0 & \omega^2\\
  \end{pmatrix},
%  \notag
  \\ \label{eq:Delta54triplet2}
  \rho_{\mathbf{3}_1}(b) &= \rho_{\mathbf{3}_2}(b) =
  \begin{pmatrix}
    0 & 1 & 0\\
    0 & 0 & 1\\
    1 & 0 & 0\\
  \end{pmatrix},~~&
  \rho_{\mathbf{3}_1}(c) &= -\rho_{\mathbf{3}_2}(c) =
  \begin{pmatrix}
    1 & 0 & 0\\
    0 & 0 & 1\\
    0 & 1 & 0\\
  \end{pmatrix}.
%  \notag
\end{align}
$\mathbf{3}^*_i$ is the complex conjugate representation of $\mathbf{3}_i$.
The matrix representations of the doublets are then given by
\begin{align}
  \rho_{\mathbf{2}_1}(a) &= \rho_{\mathbf{2}_1}(a') =
  \begin{pmatrix}
    1 & 0 \\
    0 & 1
  \end{pmatrix},
  &\rho_{\mathbf{2}_1}(b) =&
  \begin{pmatrix}
    \omega & 0 \\
    0 & \omega^2
  \end{pmatrix},
  &\rho_{\mathbf{2}_1}(c) =&
  \begin{pmatrix}
    0 & 1 \\
    1 & 0
  \end{pmatrix},
  \\
  \rho_{\mathbf{2}_2}(a) &= \rho_{\mathbf{2}_2}(a') =
  \begin{pmatrix}
    \omega^2 & 0 \\
    0 & \omega
  \end{pmatrix},
  &\rho_{\mathbf{2}_2}(b) =&
  \begin{pmatrix}
    \omega & 0 \\
    0 & \omega^2
  \end{pmatrix},
  &\rho_{\mathbf{2}_2}(c) =&
  \begin{pmatrix}
    0 & 1 \\
    1 & 0
  \end{pmatrix},
  \\
  \rho_{\mathbf{2}_3}(a) &= \rho_{\mathbf{2}_3}(a') =
  \begin{pmatrix}
    \omega & 0 \\
    0 & \omega^2
  \end{pmatrix},
  &\rho_{\mathbf{2}_3}(b) =&
  \begin{pmatrix}
    \omega & 0 \\
    0 & \omega^2
  \end{pmatrix},
  &\rho_{\mathbf{2}_3}(c) =&
  \begin{pmatrix}
    0 & 1 \\
    1 & 0
  \end{pmatrix},
  \\
  \rho_{\mathbf{2}_4}(a) &= \rho_{\mathbf{2}_4}(a') =
  \begin{pmatrix}
    \omega & 0 \\
    0 & \omega^2
  \end{pmatrix},
  &\rho_{\mathbf{2}_4}(b) =&
  \begin{pmatrix}
    1 & 0 \\
    0 & 1
  \end{pmatrix},
  &\rho_{\mathbf{2}_4}(c) =&
  \begin{pmatrix}
    0 & 1 \\
    1 & 0
  \end{pmatrix}.
\end{align}
Since $\rho_{\mathbf{2}_i}(a^{(\prime)})^* = S_2^\dagger \rho_{\mathbf{2}_i}(a^{(\prime)}) S_2$, and $\rho_{\mathbf{2}_i}(b)^* = S_2^\dagger \rho_{\mathbf{2}_i}(b) S_2$,
where $S_2$ is the matrix given in Eq.~\eqref{eq:S2}.
These doublets are pseudoreal representations.
We also have the trivial singlet $\mathbf{1}_0$ and a nontrivial singlet $\mathbf{1}_1$,
$\rho_{\mathbf{1}_1}(a) = \rho_{\mathbf{1}_1}(a') = \rho_{\mathbf{1}_1}(b) = 1$ and $\rho_{\mathbf{1}_1}(c) = -1$.

The outer automorphism group of $\Delta(54)$ is isomorphic to $S_4$, which is equivalent to the permutation of 4 doublets.
It follows that $\mathbf{2}_i \to \mathbf{2}_i$ corresponds to the identity element of the outer automorphism,
and we cannot introduce consistent CP transformation which is based on $\mathbf{3}_i \to \mathbf{3}_i^*$
and $\mathbf{2}_i \to \mathbf{2}_i$ simultaneously.

\section{Clebsch-Gordon Coefficients of $\Delta(27)$ and $\Delta(54)$}
\label{app:CG}

In this Appendix, we note CG coefficients of $\Delta(27)$ and $\Delta(54)$.
There are various conventions of the irreducible representations of discrete groups.
We note tensor products relative to our calculations rather than complete table of the algebra.
Complete table of tensor products of these representations and its CG coefficients are summarized in \cite{Chen:2014tpa, Kobayashi:2022moq} and reference therein, but the conventions are a little different.
In this paper, we make the irreducible representations $\Delta(27)$ and CG coefficients coincide with that of $\Delta(54)$ as possible.

\subsection{CG Coefficients of $\Delta(27)$}
\label{app:CG_D27}

The tensor product of $\mathbf{3}$ and ${\mathbf{3}}^*$ is decomposed to nine singlets,
\begin{align}
  {\mathbf{3}}^* \otimes \mathbf{3} = \sum_{i = 0 ,...,8} \mathbf{1}_i.
\end{align}
We introduce $x_{\mathbf{3}^*}  = (x_1, x_2, x_3)^T$ and $y_{\mathbf{3}} = (y_1, y_2, y_3)^T$
Their tensor product is decomposed to
\begin{align} \notag
  (x_{\mathbf{3}^*} \otimes y_{\mathbf{3}})_{\mathbf{1}_0} &=
  x_1 y_1 + x_2 y_2 + x_3 y_3,
  \\ \notag
  (x_{\mathbf{3}^*} \otimes y_{\mathbf{3}})_{\mathbf{1}_1} &=
  x_1 y_1 + \omega x_2 y_2 + \omega^2 x_3 y_3,
  \\ \notag
  (x_{\mathbf{3}^*} \otimes y_{\mathbf{3}})_{\mathbf{1}_2} &=
  x_1 y_1 + \omega^2 x_2 y_2 + \omega x_3 y_3,
  \\ \notag
  (x_{\mathbf{3}^*} \otimes y_{\mathbf{3}})_{\mathbf{1}_3} &=
  x_1 y_3 + x_2 y_1 + x_3 y_2,
  \\ \notag
  (x_{\mathbf{3}^*} \otimes y_{\mathbf{3}})_{\mathbf{1}_4} &=
  x_1 y_3 + \omega x_2 y_1 + \omega^2 x_3 y_2,
  \\ \notag
  (x_{\mathbf{3}^*} \otimes y_{\mathbf{3}})_{\mathbf{1}_5} &=
  x_1 y_3 + \omega^2 x_2 y_1 + \omega x_3 y_2,
  \\ \notag
  (x_{\mathbf{3}^*} \otimes y_{\mathbf{3}})_{\mathbf{1}_6} &=
  x_1 y_2 + x_2 y_3 + x_3 y_1,
  \\ \notag
  (x_{\mathbf{3}^*} \otimes y_{\mathbf{3}})_{\mathbf{1}_7} &=
  x_1 y_2 + \omega x_2 y_3 + \omega^2 x_3 y_1,
  \\ 
  (x_{\mathbf{3}^*} \otimes y_{\mathbf{3}})_{\mathbf{1}_8} &=
  x_1 y_2 + \omega^2 x_2 y_3 + \omega x_3 y_1.
\end{align}
The CG coefficients are written by matrix form.
They are given by
\begin{align} \notag
  M_{0} =
  \begin{pmatrix}
    1 & 0 & 0\\
    0 & 1 & 0\\
    0 & 0 & 1\\
  \end{pmatrix},
  M_{1} =
  \begin{pmatrix}
    1 & 0 & 0\\
    0 & \omega & 0\\
    0 & 0 & \omega^2\\
  \end{pmatrix},
  M_{2} =
  \begin{pmatrix}
    1 & 0 & 0\\
    0 & \omega^2 & 0\\
    0 & 0 & \omega\\
  \end{pmatrix},
  \\ \notag
  M_{3} =
  \begin{pmatrix}
    0 & 0 & 1\\
    1 & 0 & 0\\
    0 & 1 & 0\\
  \end{pmatrix},
  M_{4} =
  \begin{pmatrix}
    0 & 0 & 1\\
    \omega & 0 & 0\\
    0 & \omega^2 & 0\\
  \end{pmatrix},
  M_{5} =
  \begin{pmatrix}
    0 & 0 & 1\\
    \omega^2 & 0 & 0\\
    0 & \omega & 0\\
  \end{pmatrix},
  \\
  M_{6} =
  \begin{pmatrix}
    0 & 1 & 0\\
    0 & 0 & 1\\
    1 & 0 & 0\\
  \end{pmatrix},
  M_{7} =
  \begin{pmatrix}
    0 & 1 & 0\\
    0 & 0 & \omega\\
    \omega^2 & 0 & 0\\
  \end{pmatrix},
  M_{8} =
  \begin{pmatrix}
    0 & 1 & 0\\
    0 & 0 & \omega^2\\
    \omega & 0 & 0\\
  \end{pmatrix},
\end{align}
where
\begin{align}
  (x_{\mathbf{3}^*} \otimes y_{\mathbf{3}})_{\mathbf{1}_i} =
  x_{\mathbf{3}^*}^T M_i y_{\mathbf{3}}.
\end{align}
It is also useful to introduce $M_{{i}^*}$, which satisfies
\begin{align}
  (x_{\mathbf{3}^*} \otimes y_{\mathbf{3}})_{\mathbf{1}_i^*} =
  x_{\mathbf{3}^*}^T M_{i^*} y_{\mathbf{3}}.
\end{align}
It is obvious that $M_{i^*}$ is relevant to $M_{j}$.
\begin{align} \notag
  M_{0^*} &= M_0, &M_{{1}^*}& = M_2, &M_{{2}^*}& = M_1,
  \\ \notag
  M_{3^*} &= M_6, &M_{{4}^*}& = M_8, &M_{{5}^*}& = M_7,
  \\
  M_{6^*} &= M_3, &M_{{7}^*}&  = M_5, &M_{{8}^*}& = M_4.
\end{align}

\subsection{CG Coefficients of $\Delta(54)$}
\label{app:CG_D54}

We concentrate on ${\mathbf{3}}^*_i \otimes \mathbf{3}_i$.
It is decomposed to one singlet and four different doublets.
Decomposition rules of the tensor products are given by
\begin{align} \notag
  (x_{\mathbf{3}^*_i} \otimes y_{\mathbf{3}_i})_{\mathbf{1}_0} &=
  x_1 y_1 + x_2 y_2 + x_3 y_3,
  \\ \notag
  (x_{\mathbf{3}^*_i} \otimes y_{\mathbf{3}_i})_{\mathbf{2}_1} &=
  \begin{pmatrix}
    x_1 y_1 + \omega^2 x_2 y_2 + \omega x_3 y_3\\
    \omega x_1 y_1 + \omega^2 x_2 y_2 + x_3 y_3
  \end{pmatrix},
  \\ \notag
  (x_{\mathbf{3}^*_i} \otimes y_{\mathbf{3}_i})_{\mathbf{2}_2} &=
  \begin{pmatrix}
    \omega x_1 y_3 + x_2 y_1 + \omega^2 x_3 y_2\\
    \omega^2 x_1 y_2 + x_2 y_3 + \omega x_3 y_1
  \end{pmatrix},
  \\ \notag
  (x_{\mathbf{3}^*_i} \otimes y_{\mathbf{3}_i})_{\mathbf{2}_3} &=
  \begin{pmatrix}
    x_1 y_2 + \omega^2 x_2 y_3 + \omega x_3 y_1\\
    \omega x_1 y_3 + \omega^2 x_2 y_1 + x_3 y_2
  \end{pmatrix},
  \\
  (x_{\mathbf{3}^*_i} \otimes y_{\mathbf{3}_i})_{\mathbf{2}_4} &=
  \begin{pmatrix}
    x_1 y_2 + x_2 y_3 + x_3 y_1\\
    x_1 y_3 + x_2 y_1 + x_3 y_2
  \end{pmatrix}.
  \label{eq:33}
\end{align}
We also note the CG coefficients of products of doublets,
\begin{align}
  x_{\mathbf{2}_i}
  \otimes
  y_{\mathbf{2}_i}
  =&
  (x_1 y_2 + x_2 y_1)_{\mathbf{1}_0} \oplus (x_1 y_2 - x_2 y_1)_{\mathbf{1}_1}
  \oplus
  \begin{pmatrix}
    x_2 y_2\\
    x_1 y_1
  \end{pmatrix}_{\mathbf{2}_i}
  \\
  x_{\mathbf{2}_i}
  \otimes
  y_{\mathbf{2}_j}
  =&
  \begin{pmatrix}
    x_2 y_2\\
    x_1 y_1
  \end{pmatrix}_{\mathbf{2}_k}
  \oplus
  \begin{pmatrix}
    x_1 y_2\\
    x_2 y_1
  \end{pmatrix}_{\mathbf{2}_l}.
\end{align}
where $i, j, k ,l$ is a permutation of $ \{ 1,2,3, 4 \}$.
For example,
\begin{align}
  x_{\mathbf{2}_1}
  \otimes
  y_{\mathbf{2}_2}
  =&
  \begin{pmatrix}
    x_2 y_2\\
    x_1 y_1
  \end{pmatrix}_{\mathbf{2}_3}
  \oplus
  \begin{pmatrix}
    x_1 y_2\\
    x_2 y_1
  \end{pmatrix}_{\mathbf{2}_4}.
\end{align}

\section{CP-Violating Amplitude}
\label{App:1-loop}

In this Appendix, we show an explicit form of the one-loop amplitude of decay of a scalar field in our $\Delta(27)$ model studied in Section
\ref{sec:class_B},
\begin{align}
  \begin{tikzpicture}[baseline=(o.base)]
     \begin{feynhand}
        \vertex (a) at (-1,0) {$\phi_4^+$};
        \vertex (b) at (1,-1) ;
        \vertex (c) at (1,1) ;
        \vertex (d) at (2, -1) {$\bar{\psi}_5$};
        \vertex (e) at (2, 1) {$\psi_6$};
        \vertex [dot] (o) at (0,0);
        \propag [sca] (a) to (o);
        \propag [fermion] (b) to [edge label = $\Psi_{\mathbf{3}}$](o);
        \propag [anti fermion] (c) to [edge label' = $\Psi_{\mathbf{3}}$] (o);
        \propag [anti fermion] (b) to (d);
        \propag [fermion] (c) to (e);
        \propag [sca] (b) to [edge label'= $\Phi_{\mathbf{3}}$] (c);
     \end{feynhand}
  \end{tikzpicture}
  =& - 3 i \omega^2 y_2 y_3^* y_4 I \bar u^s(p)v^{s'}(p'),
\end{align}
where $I$ is formally calculated as
\begin{align}
  I \bar{u}^s(p) v^{s'}(p')
  =& \int \frac{d^4 q}{(2\pi)^4}\,
  \frac{i}{q^2 - m_s^2}
  \bar{u}^s(p)
  \frac{i((\slashed{p}+\slashed{q}) + m_f)}{(p+q)^2 - m_f^2}
  \frac{i((-\slashed{p'}+\slashed{q}) + m_f)}{(-p'+q)^2 - m_f^2}
  v^{s'}(p').
\end{align}
This integral is evaluated by using the Feynman parameters technique.
The integral is rewritten as
\footnotesize
\begin{align}
  I \bar{u}^s(p) v^{s'}(p')
  =&
  - i \int \frac{d^4 l}{(2\pi)^4} \int_{0}^{1} dx dy dz\,\delta(x+y+z-1)
  \notag
  \\
  &\hspace{-20mm}\frac{\bar{u}^s(p)
  (\slashed{l}^2 +  y^2 m_5^2 + z^2 m_6^2 -2 yz p' \cdot p
  - (m_5 + m_6 + 2 m_f)(y m_6 + z m_5)+(m_6 + m_f)(m_5 + m_s))
  v^{s'}(p')}
  {(l^2 + (-y^2 + y) p^2 + (-z^2 + z) p'^2 + 2yzp\cdot p' - x m_s^2 -(y + z) m_f^2)^3}.
  \label{eq:integral_I}
\end{align}
\normalsize
We omit a linear term of $l$ since it does not contribute to the integral.
$l$ is given by
\begin{align}
  l =& q + (yp -zp').
\end{align}
We use $\slashed{p}u^s(p) = m_6 u^s(p)$ and $\slashed{p'}v^{s'}(p') = -m_5 v^{s'}(p')$.
Then, we obtain
\begin{align}
  I = -i \int \frac{d^4 l}{(2\pi)^4} \int_{0}^{1} dy \int_{0}^{1-y} dz \,\frac{l^2 + \Delta'} {(l^2 -\Delta +i\epsilon)^3},
\end{align}
where
\begin{align}
  \Delta =& y(y-1) p^2 + z(z-1)p'^2 -2yzp\cdot p' + (1-y-z) m_s^2 + (y + z) m_f^2,
  \notag
  \\
  \Delta' =&
   y^2 m_6^2 + z^2 m_5^2 -2 yz p' \cdot p - (m_6 + m_5 + 2 m_f)(y m_6 + z m_5)+(m_6 + m_f)(m_5 + m_f).
\end{align}
We introduce Wick rotation and substitute $l_0 = i l_E^0$.
We obtain
\begin{align}
  I = \int \frac{d^4 l_E}{(2\pi)^4} \int_{0}^{1} dy \int_{0}^{1-y} dz \,\frac{l_E^2 - \Delta'} {(l_E^2 + \Delta)^3}.
\end{align}
It is evaluated by the dimensional regularization method.
We change $d^4l_E$ to $d^{4-\epsilon} l_E$, and we obtain
\begin{align}
  I
  =&
  \int_0^1 dy \int_0^{1-y} dz\,
  \left\{
  \frac{1}{(4\pi)^{2-\epsilon/2}}\left(2-\frac \epsilon 2 \right) \frac{\Gamma\left(\frac \epsilon 2\right)}{\Gamma(3)}
  \left(\frac{1}{\Delta} \right)^{\frac \epsilon 2}
  - \frac{\Delta'}{(4\pi)^{2-\epsilon/2}}\frac{\Gamma\left(1 - \frac \epsilon 2\right)}{\Gamma(3)}
  \left(\frac{1}{\Delta} \right)^{1-\frac \epsilon 2} \right\}
  \nonumber
  \\
  = &
  \int_0^1 dy \int_0^{1-y} dz\,
  \frac{2}{(4\pi)^2 \sqrt{\pi}}
  \left\{
  \left(
  \frac 4 \epsilon - 2 \log \Delta -2\gamma -1 +2 \log 4\pi\right)
  - \frac{ \sqrt{\pi} \Delta'}{\Delta} + \mathcal{O}(\epsilon) \right\}.
\end{align}
The divergent parts are cancelled by the counter terms.
For instance, we adopt the $\overline{\textrm{MS}}$ scheme.
In this scheme the divergent terms and constant terms given rise from the regularization technique are subtracted by the counter terms.
We obtain
\begin{align}
  I =&
  \int_0^1 dy \int_0^{1-y} dz\,
  \frac{2}{(4\pi)^2 \sqrt{\pi}}
  \left\{
  \left(
  - 2 \log \frac{\Delta}{\mu^2}  \right)
  - \frac{ \sqrt{\pi} \Delta'}{\Delta} \right\}.
\end{align}
where $\mu$ is the renormalization scale, and thus we obtain the finite invariant matrix of particle decay at the one-loop level:
\begin{align}
  i\mathcal{M}_{\phi_4^+ \to \psi_6 \bar{\psi}_5}^{ss'} =
  -i \left( y_1  - 3 \omega^2 y_2 y_3^* y_4 I \right)\bar u^s(p)v^{s'}(p').
\end{align}

On the other hand, the one-loop decay amplitude of antiparticle with inverse momentum is similarly given by
\begin{align}
  \begin{tikzpicture}[baseline=(o.base)]
     \begin{feynhand}
        \vertex (a) at (-1,0) {$\bar{\phi}_4^{+}$};
        \vertex (b) at (1,-1) ;
        \vertex (c) at (1,1) ;
        \vertex (d) at (2, -1) {$\psi_{5}$};
        \vertex (e) at (2, 1) {$\bar{\psi}_{6}$};
        \vertex [dot] (o) at (0,0);
        \propag [sca] (a) to (o);
        \propag [anti fermion] (b) to [edge label = $\psi_{\mathbf{3}}$](o);
        \propag [fermion] (c) to [edge label' = $\psi_{\mathbf{3}}$] (o);
        \propag [fermion] (b) to (d);
        \propag [anti fermion] (c) to (e);
        \propag [sca] (b) to [edge label'= $\phi_{\mathbf{3}}$] (c);
     \end{feynhand}
  \end{tikzpicture}
  =&
  i 3 \omega y_2^* y_3 y_4^* J \bar u^{s'}(-p')v^{s}(-p),
\end{align}
where $J$ is given by
\begin{align}
  J \bar{u}^{s'}(-p') v^{s}(-p) =&
  \int \frac{d^4 q}{(2\pi)^4}\,
  \frac{i}{q^2 - m_s^2}
  \bar{u}^{s'}(-p')
  \frac{i((-\slashed{p'}+\slashed{q}) + m_f)}{(-p'+q)^2 - m_f^2}
  \frac{i((\slashed{p}+\slashed{q}) + m_f)}{(p+q)^2 - m_f^2}
  v^{s}(-p).
  \label{eq:integral_J}
\end{align}
It is obvious that the integrands of Eqs.~\eqref{eq:integral_J} and \eqref{eq:integral_I} are the same except for the spinors.
Thus we find that $I = J$.
We finally obtain the CP asymmetry of $S$-matrix at the one-loop level by
\begin{align}
  \sum_{s,s'}| {\mathcal{M}}_{{\phi}_4^+ \to {\psi}_6 \bar{\psi}_5}^{ss'}|^2
  - |\bar{\mathcal{M}}_{\bar{\phi}_4^+ \to \bar{\psi}_6 {\psi}_5}^{ss'}|^2
  =& - 12
  \mathrm{Im}\, \omega^2 y_1^* y_2 y_3^* y_4 \,
  \mathrm{Im}\, I \times (p - m_6)\cdot(p'+ m_5).
\end{align}
The imaginary part of $I$ comes from $\ln \Delta$.
On the rest frame of $\phi_4^+$, we obtain
\begin{align}
  \Delta &= y(y+z-1)m_6^2 + z(y+z-1)m_5^2 +(1-y-z) m_s^2 + (y+z) m_f^2 -yz m_4^2,
\end{align}
and it takes a negative value if $m_4$ is heavy enough.

\section{Trace of $S$-Matrix and Existence of Generalized CP symmetry}
\label{sec:trace}

A scattering amplitude is CP conserved if there is a physical CP transformation which cancels CP asymmetry.
Nonexistence of physical CP transformation for a certain process is not trivial
since there are many candidates of physical CP transformations in general.
In this appendix, we consider a sufficient condition for physical CP violation.

\begin{figure}[thbp]
  \centering
  %\tikzset{external/force remake}
  \begin{tikzpicture}
    \begin{feynhand}
      \setlength{\feynhandblobsize}{15mm}
      \filldraw (0,-0.1) circle [radius=0.1mm];
      \filldraw (0,0) circle [radius=0.1mm];
      \filldraw (0,0.1) circle [radius=0.1mm];
      \filldraw (4,-0.1) circle [radius=0.1mm];
      \filldraw (4,0) circle [radius=0.1mm];
      \filldraw (4,0.1) circle [radius=0.1mm];
      \vertex (a1) at (0,-1.5) {$\mathbf{r}_1$};
      \vertex (a3) at (0,-0.5) {$\mathbf{r}_2$};
      \vertex (a4) at (0,0.5) {$\mathbf{r}_{n-1}$};
      \vertex (a6) at (0,1.5) {$\mathbf{r}_n$} ;
      \vertex [grayblob] (b) at (2,0) {};
      \vertex (c1) at (4,-1.5) {$\mathbf{r}'_{1}$};
      \vertex (c3) at (4,-0.5) {$\mathbf{r}'_{2}$};
      \vertex (c4) at (4,0.5) {$\mathbf{r}'_{m-1}$};
      \vertex (c6) at (4,1.5) {$\mathbf{r}'_{m}$} ;
      \propag [plain] (a1) to (b);
      \propag [plain] (a3) to (b);
      \propag [plain] (a4) to (b);
      \propag [plain] (a6) to (b);
      \propag [plain] (b) to (c1);
      \propag [plain] (b) to (c3);
      \propag [plain] (b) to (c4);
      \propag [plain] (b) to (c6);
    \end{feynhand}
  \end{tikzpicture}
  \caption{A scattering process.}
  \label{Fig:_diagram1}
\end{figure}
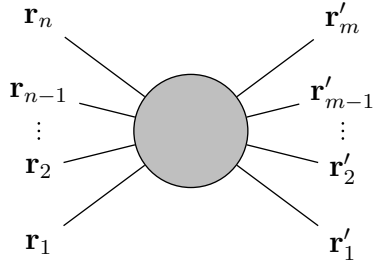

We consider a model with internal symmetry $G$, which contains fields denoted by $\phi_{\mathbf{r}_i} = (\phi_{\mathbf{r}_i}^1, \phi_{\mathbf{r}_i}^2,...,\phi_{\mathbf{r}_i}^{\textrm{dim }\mathbf{r}_i})^T$,
which transforms as an irreducible representation $\mathbf{r}_i$ of $G$.
Let us consider a scattering process depicted by Fig.~\ref{Fig:_diagram1}.

The invariant matrix element $\mathcal{M}$ of this process is given by
\begin{align} \notag
  &(2\pi)^4 \delta^{(4)}(\sum_{\textrm{in}} p_{\textrm{in}} -  \sum_{\textrm{out}} k_{\textrm{out}})
  \mathcal{M}(\{\mathbf{r}_{\textrm{in}}, i_{\textrm{in}}, \mathbf{p}_{\textrm{in}} \}
  \to
  \{\mathbf{r}_{\textrm{out}}, j_{\textrm{out}}, \mathbf{k}_{\textrm{out}} \})
  \\
  &= \lim_{T\to \infty} {}_{\textrm{out}}
  \bra{\mathbf{r}'_{1},j_1, \mathbf{k}_{1};\cdots ;\mathbf{r}'_m,j_m, \mathbf{k}_m }
  e^{-i2 \mathcal{H}T}
  \ket{\mathbf{r}_1,i_1, \mathbf{p}_1;\cdots ;\mathbf{r}_n,i_n,\mathbf{p}_n}
  _{\textrm{in}},
\end{align}
where $\{\mathbf{r}_{\textrm{in}}, i_{\textrm{in}}, \mathbf{p}_{\textrm{in}} \}$ and $\{\mathbf{r}_{\textrm{out}}, j_{\textrm{out}}, \mathbf{k}_{\textrm{out}} \}$ formally represents the asymptotic states.
They are given by
\begin{align} \notag
  \ket{\mathbf{r}_1,i_1, \mathbf{p}_1;\cdots ;\mathbf{r}_n,i_n, \mathbf{p}_n}
  &\equiv a^{(\mathbf{r}_1)\dagger}_{i_1, \mathbf{p}_1}...a^{(\mathbf{r}_n) \dagger}_{i_n, \mathbf{p}_n}\ket{0},
  \\
  \ket{\mathbf{r}'_{1},j_1, \mathbf{k}_{1};\cdots ;\mathbf{r}'_m,j_m,\mathbf{k}_m }
  &\equiv a^{(\mathbf{r}'_1) \dagger}_{j_1, \mathbf{k}_1}...a^{(\mathbf{r}'_m)\dagger}_{j_m, \mathbf{k}_m} \ket{0},
\end{align}
where $a^{(\mathbf{r})\dagger}_{i, \mathbf{p}}$ is the particle creation operator of the $i$-th component (flavor) of 
$\phi_{\mathbf{r}}$ with momentum $\mathbf{p}$.\footnote{The initial and final state can contains antiparticles.
However, it just make the notation more complicated, and does not affect the result.
We assume the external lines are particles in this appendix.}
The flavor indices $i$ and $j$ are explicitly shown for our purpose.

If the model is invariant under a physical CP transformation given by $\phi_{\mathbf{r}} \to U_{\mathbf{r}} \phi_{\mathbf{r}}^{\mathrm{CP}}$ where $U_{\mathbf{r}}$ is a unitary matrix, the invariant matrix element is transformed as
\begin{align}
  \mathcal{M}(\{\mathbf{r}_{\textrm{in}}, i_{\textrm{in}}, \mathbf{p}_{\textrm{in}} \} \to \{\mathbf{r}_{\textrm{out}}, j_{\textrm{out}}, \mathbf{k}_{\textrm{out}} \} )
  &\to
  \mathcal{M}(\{\mathbf{r}^*_{\textrm{in}}, U_{\mathbf{r}}^{i_{\textrm{in}}i'} i', -\mathbf{p}_{\textrm{in}} \} \to \{\mathbf{r}^*_{\textrm{out}}, U_{\mathbf{r}}^{j_{\textrm{out}}j'} j', -\mathbf{k}_{\textrm{out}} \})
  \notag
  \\
  &\hspace{-10mm} =
  \prod_{i_{\textrm{in}}, j_{\textrm{out}}} U_{\mathbf{r}}^{i_{\textrm{in}}i'}
  U_{\mathbf{r}}^{j_{\textrm{out}}j'*}
  \mathcal{M}(\{\mathbf{r}^*_{\textrm{in}},  i', -\mathbf{p}_{\textrm{in}} \} \to \{\mathbf{r}^*_{\textrm{out}},  j', -\mathbf{k}_{\textrm{out}} \}),
  \label{eq:invariant_matrix}
\end{align}
under the physical CP transformation.
And hence, this process is CP conserved,
\begin{align}
  |\mathcal{M}(\{\mathbf{r}_{\textrm{in}}, i_{\textrm{in}}, \mathbf{p}_{\textrm{in}} \} \to \{\mathbf{r}_{\textrm{out}}, j_{\textrm{out}}, \mathbf{k}_{\textrm{out}} \})|^2
  -
  |\mathcal{M}(\{\mathbf{r}^*_{\textrm{in}}, U_{\mathbf{r}}^{i_{\textrm{in}}i'} i', -\mathbf{p}_{\textrm{in}} \} \to \{\mathbf{r}^*_{\textrm{out}}, U_{\mathbf{r}}^{j_{\textrm{out}}j'} j', -\mathbf{k}_{\textrm{out}} \})|^2
  =0,
\end{align}

To confirm physical CP violation by calculating an invariant matrix,
we must prove that there is no unitary matrix which satisfies Eq.~\eqref{eq:invariant_matrix}.
It might be difficult if we don't know the precise transformation rule for every irreducible representation.
Our goal is to establish an easy method to determine whether CP is violated without specifying physical CP transformation rules.
One easy way is to calculate
\begin{align} \notag
  \mathrm{tr}\, A_{i_{\textrm{in}} j_{\textrm{out}}} \equiv
  \sum_{i_{\textrm{in}}\in \mathbf{r}_{\textrm{in}}, j_{\textrm{out}}\in \mathbf{r}_{\textrm{out}}}&
   |\mathcal{M}(\{\mathbf{r}_{\textrm{in}}, i_{\textrm{in}}, \mathbf{p}_{\textrm{in}} \} \to \{\mathbf{r}_{\textrm{out}}, j_{\textrm{out}}, \mathbf{k}_{\textrm{out}} \} )|^2
   \\
  &- |\mathcal{M}(\{\mathbf{r}^*_{\textrm{in}}, i_{\textrm{in}}, -\mathbf{p}_{\textrm{in}} \} \to \{\mathbf{r}^*_{\textrm{out}}, j_{\textrm{out}}, -\mathbf{k}_{\textrm{out}} \})|^2,
\end{align}
where we take the sum of all initial and the final states.
If the model is physical CP invariant,
\begin{align} \notag
  \mathrm{tr}\, A_{i_{\textrm{in}} j_{\textrm{out}}}
  =&
  \sum_{i_{\textrm{in}}\in \mathbf{r}_{\textrm{in}}, j_{\textrm{out}}\in \mathbf{r}_{\textrm{out}}}
  |\prod_{i_{\textrm{in}}} U_{\mathbf{r}}^{i_{\textrm{in}}i'}
  \prod_{j_{\textrm{out}}} U_{\mathbf{r}}^{j_{\textrm{out}}j'*}
  \mathcal{M}(\{\mathbf{r}^*_{\textrm{in}},  i', -\mathbf{p}_{\textrm{in}} \} \to \{\mathbf{r}^*_{\textrm{out}},  j', -\mathbf{k}_{\textrm{out}} \})|^2
  \\ \notag
  &\hspace{5mm} - |\mathcal{M}(\{\mathbf{r}^*_{\textrm{in}}, i_{\textrm{in}}, -\mathbf{p}_{\textrm{in}} \} \to \{\mathbf{r}^*_{\textrm{out}}, j_{\textrm{out}}, -\mathbf{k}_{\textrm{out}} \})|^2
  \\ \notag
  =&
  \sum_{i'\in \mathbf{r}_{\textrm{in}}, j'\in \mathbf{r}_{\textrm{out}}}
  |\mathcal{M}(\{\mathbf{r}^*_{\textrm{in}},  i', -\mathbf{p}_{\textrm{in}} \} \to \{\mathbf{r}^*_{\textrm{out}},  j', -\mathbf{k}_{\textrm{out}} \})|^2
  \\ \notag
  &\hspace{5mm} - \sum_{i_{\textrm{in}}\in \mathbf{r}_{\textrm{in}}, j_{\textrm{out}}\in \mathbf{r}_{\textrm{out}}}
  |\mathcal{M}(\{\mathbf{r}^*_{\textrm{in}}, i_{\textrm{in}}, -\mathbf{p}_{\textrm{in}} \} \to \{\mathbf{r}^*_{\textrm{out}}, j_{\textrm{out}}, -\mathbf{k}_{\textrm{out}} \})|^2
  \\
  =&0.
\end{align}
Hence $\mathrm{tr}\, A_{ij} \neq 0$ is a sufficient condition for CP violation.
On the other hand, $\mathrm{tr}\, A_{ij} = 0$ implies the existence of generalized CP.

\end{document}